\documentclass[times,twocolumn]{aastex63}
\usepackage{amsmath}

\begin{document}
\title{Application of X-Ray Clumpy Torus Model (XCLUMPY) to 10 Obscured Active Galactic Nuclei Observed with Suzaku and NuSTAR}
\correspondingauthor{Atsushi Tanimoto}
\email{atsushi.tanimoto@phys.s.u-tokyo.ac.jp}
\author[0000-0002-0114-5581]{Atsushi Tanimoto}
\affiliation{Department of Astronomy, Kyoto University, Kyoto 606-8502, Japan}
\affiliation{Department of Physics, The University of Tokyo, Tokyo 113-0033, Japan}
\author[0000-0001-7821-6715]{Yoshihiro Ueda}
\affiliation{Department of Astronomy, Kyoto University, Kyoto 606-8502, Japan}
\author{Hirokazu Odaka}
\affiliation{Department of Physics, The University of Tokyo, Tokyo 113-0033, Japan}
\affiliation{Kavil IPMU, The University of Tokyo, Chiba 277-8583, Japan}
\author[0000-0002-5701-0811]{Shoji Ogawa}
\affiliation{Department of Astronomy, Kyoto University, Kyoto 606-8502, Japan}
\author[0000-0002-9754-3081]{Satoshi Yamada}
\affiliation{Department of Astronomy, Kyoto University, Kyoto 606-8502, Japan}
\author[0000-0002-3866-9645]{Toshihiro Kawaguchi}
\affiliation{Department of Economics, Management and Information Science, Onomichi City University, Hiroshima 722-8506, Japan}
\author[0000-0002-4377-903X]{Kohei Ichikawa}
\affiliation{Frontier Research Institute for Interdisciplinary Sciences, Tohoku University, Miyagi 980-8578, Japan}
\affiliation{Astronomical Institute, Tohoku University, Miyagi 980--8578, Japan}
\begin{abstract}
We apply XCLUMPY \citep{Tanimoto19}, an X-ray spectral model from a clumpy torus in an active galactic nucleus (AGN), to the broadband X-ray spectra of 10 obscured AGNs observed with both \textit{Suzaku} and \textit{NuSTAR}. The infrared spectra of these AGNs were analyzed by \cite{Ichikawa15} with the CLUMPY code \citep{Nenkova08a, Nenkova08b}. Since XCLUMPY adopts the same clump distribution as that in the CLUMPY, we can directly compare the torus parameters obtained from the X-ray spectra and those from the infrared ones. The torus angular widths determined from the infrared spectra ($\sigma_{\mathrm{IR}}$) are systematically larger than those from the X-ray data ($\sigma_{\mathrm{X}}$); the difference ($\sigma_{\mathrm{IR}}-\sigma_{\mathrm{X}}$) correlates with the inclination angle determined from the X-ray spectrum. These results can be explained by the contribution from dusty polar outflows to the observed infrared flux, which becomes more significant at higher inclinations (more edge-on views). The ratio of the hydrogen column density and V-band extinction in the line of sight absorber shows large scatter ($\simeq$1 dex) around the Galactic value, suggesting that a significant fraction of AGNs have dust-rich circumnuclear environments.
\end{abstract}

\section{Introduction}
The unification model of an active galactic nucleus (AGN: \citealt{Antonucci93, Urry95, Netzer15, Ramos17}) indicates the ubiquitous presence of an obscuring dusty, molecular gas region (so-called ``torus'') around the accreting supermassive black hole (SMBH). The torus is a key structure to understand the mechanisms of the coevolution between the SMBH and the host galaxy \citep{Kormendy13, Heckman14} because it is considered as a mass reservoir that feeds material onto the SMBH from the host galaxy. Nevertheless, the basic properties of the tori (e.g., spatial distribution of matter and the gas-to-dust ratio) are still unclear.

Many studies indicated that the torus consists of dusty clumps (clumpy torus: \citealt{Krolik88, Laor93, Honig06, Honig07, Nenkova08a, Nenkova08b, Honig12}). \cite{Nenkova08a, Nenkova08b} constructed an infrared spectral model from the clumpy torus called CLUMPY. They assumed a power law distribution of clumps in the radial direction and a Gaussian distribution in the elevation direction. This CLUMPY model has been widely used to analyze the infrared spectra of AGNs \citep[e.g.,][]{Ramos09, Ramos11b, Ramos11a, Ramos11c, Ramos14a, Ramos14b, Alonso11, Alonso12a, Alonso12b, Alonso13, Ichikawa15, Garcia15, Garcia19, Fuller16, Fuller19, Audibert17, Mateos17, Lopez-Rodriguez18}.

X-ray spectra of AGNs provide a powerful tool to study the properties of the tori. This is because X-rays (in particular hard X-rays above 10 keV) have strong penetrating power against absorption and can trace all material including gas and dust in an unbiased manner, unlike the infrared continuum emission that is sensitive only to dust. The torus produces a line of sight absorption of the primary emission and a reflected spectrum accompanied by fluorescence lines. These signals carry important information on the torus parameters (e.g., the hydrogen column density and the covering factor). In many previous studies, analytic reflection models, such as the pexrav model \citep{Magdziarz95}, were used to approximate the torus-reflection component, although the assumed geometry and condition are too simple. To consider the complex geometry of a torus, several Monte-Carlo based numerical models were developed (e.g., MYTorus model \citealt{Murphy09}; Ikeda model \citealt{Ikeda09}; borus02 model \citealt{Balokovic18}). These models, however, assumed a uniform density torus (``smooth torus''), which is not realistic, as we mentioned above.

Following the earlier works by \citet{Liu14} and \citet{Furui16}, \cite{Tanimoto19} have made a new X-ray spectral model from a clumpy torus called XCLUMPY\footnote{More recently, \citet{Buchner19} have also published a similar model called UXCLUMPY.}, utilizing the Monte Carlo simulation for astrophysics and cosmology (MONACO: \citealt{Odaka11, Odaka16}) framework. XCLUMPY assumes the same torus geometry of the clump distribution as that of the CLUMPY model \citep{Nenkova08a, Nenkova08b}. This enables us to directly compare the torus parameters obtained from the X-ray spectra and those from the infrared ones.

Up to present, XCLUMPY has been applied to the X-ray spectra of two Seyfert 1 galaxies (IC 4329A and NGC 7469: \citealt{Ogawa19}) and one Compton-thick, Seyfert 2 galaxy (the Circinus galaxy: \citealt{Tanimoto19}). Interpretation of the results is somewhat puzzling, however. \cite{Tanimoto19} found that the ratio of the hydrogen column density to the V-band extinction along the line of sight ($N_{\mathrm{H}}^{\mathrm{LOS}}/A_{\mathrm{V}}^{\mathrm{LOS}}$) in the Circinus galaxy is $\simeq$10 times larger than that of Galactic interstellar medium (ISM), while \cite{Ogawa19} showed that they are $\simeq$2--70 times smaller than the Galactic value in IC 4329A and NGC 7469. To have an overview of AGN torus properties, we need to increase the number of AGNs whose spectra are analyzed with the XCLUMPY model.

This paper presents the results of applications of XCLUMPY to the broadband X-ray spectra of 10 obscured AGNs observed with both \textit{Suzaku} and \textit{NuSTAR}. The structure of this paper is organized as follows. Section 2 describes our sample and the data reduction. Section 3 presents the X-ray spectral analysis using XCLUMPY. Section 4 summarizes the results. In Section 5, we compare the torus parameters obtained from the X-ray spectra and those from the infrared data. We assume the solar abundances by \citet{Anders89}. To estimate the luminosity, we adopt the cosmological parameters of $H_{0} = 70.0$ km s$^{-1}$ Mpc$^{-1}$, $\Omega_{\mathrm{m}} = 0.30$, $\Omega_{\lambda} = 0.70$. The error on a spectral parameter corresponds to the 90\% confidence limit for a single parameter estimated by the Markov chain Monte Carlo method.

\begin{deluxetable*}{llllll}
\tablecaption{Information on Objects}
\tablehead{
Galaxy Name					                                & Classification                                            &
RA                                                          & Dec                                                       &
Redshift                                                    & $N_{\mathrm{H}}^{\mathrm{Gal}}$                           \\
(1)                                                         & (2)							                            &
(3)                                                         & (4)							                            &
(5)                                                         & (6)                                                       }
\startdata
IC 5063                                                     & Seyfert 2.0                                               &
20h52m02.34s                                                & --57d04m07.6s			                                    & 
0.01135                                                     & 0.07390                                                   \\
NGC 2110				                                    & Seyfert 2.0                                               &
05h52m11.38s                                                & --07d27m22.4s                                             &
0.00779                                                     & 0.29800                                                   \\
NGC 3227				                                    & Seyfert 2.0                                               &
10h23m30.58s                                                & +19d51m54.2s                                              &
0.00386                                                     & 0.02130                                                   \\
NGC 3281			                                        & Seyfert 2.0                                               &
10h31m52.09s                                                & --34d51m13.3s			                                    & 
0.01067                                                     & 0.08940                                                   \\
NGC 5506				                                    & Seyfert 1.9                                               &
14h13m14.89s						                        & --03d12m27.3s			                                    & 
0.00618                                                     & 0.04890                                                   \\
NGC 5643			                                        & Seyfert 2.0                                               &
14h32m40.74s                                                & --44d10m27.8s			                                    & 
0.00400                                                     & 0.12400                                                   \\
NGC 5728				                                    & Seyfert 2.0                                               &
14h42m23.89s						                        & --17d15m11.1s			                                    & 
0.00935                                                     & 0.10000                                                   \\
NGC 7172				                                    & Seyfert 2.0                                               &
22h02m01.89s						                        & --31d52m10.8s			                                    &
0.00868                                                     & 0.02120                                                   \\
NGC 7582				                                    & Seyfert 2.0                                               &
23h18m23.50s                                                & --42d22m14.0s			                                    & 
0.00525                                                     & 0.01390                                                   \\
NGC 7674				                                    & Seyfert 2.0                                               &
23h27m56.72s						                        & +08d46m44.5s                                              &
0.02892                                                     & 0.05200                                                   \\
\enddata
\tablecomments{Column (1): galaxy name. Column (2): optical classification from the NASA/IPAC extragalactic database (NED). Column (3): right ascension from the NED. Column (4): declination from the NED. Column (5) redshift from the NED. Column (6): total Galactic $HI$ and $H_{2}$ values in units of 10$^{22}$ cm$^{-2}$ \citep{Willingale13}.}
\end{deluxetable*}
\begin{deluxetable*}{llllllll}
\tablecaption{Summary of Observations}
\tablehead{
Galaxy Name					                                & Observatory				                                &
Observation ID		                                        & Start Date                                                &
End Date				                                    & Exposure                                                  &
Binning                                                     & Reference						                            \\
(1)						                                    & (2)							                            &
(3)					                                        & (4)							                            &
(5)						                                    & (6)							                            &
(7)					                                        & (8)                                                       }
\startdata
IC 5063					                                    & \textit{Suzaku}				                            &
704010010				                                    & 2009 Apr 24                                               &
2009 Apr 25                                                 & 45                                                        &
50					                                        & (01)							                            \\
						                                    & \textit{NuSTAR}                                           &
60061302002				                                    & 2013 Jul 08                                               &
2013 Jul 08                                                 & 18                                                        &
50                                                          & (02)							                            \\
NGC 2110				                                    & \textit{Suzaku}				                            &
707034010			                                        & 2012 Aug 31                                               &
2012 Sep 02                                                 & 103                                                       &
100                                                         & (03) (04)     				                            \\
                            							    & \textit{NuSTAR}				                            &
60061061002			                                        & 2012 Oct 05                                               &
2012 Oct 05                                                 & 16                                                        &
100                                                         & (02) (05)  					                            \\
NGC 3227				                                    & \textit{Suzaku}				                            &
703022050				                                    & 2008 Nov 27                                               &
2008 Nov 29                                                 & 79                                                        &
50                                                          & (06)      					                            \\
						                                    & \textit{NuSTAR}				                            &
60202002002			                                        & 2016 Nov 09 					                            &
2016 Nov 10                                                 & 49                                                        &
50					                                        & \nodata						                            \\
NGC 3281                                                    & \textit{Suzaku}				                            &
703033010				                                    & 2008 May 21                                               &
2008 May 23                                                 & 46                                                        & 
50					                                        & \nodata							                        \\
						                                    & \textit{NuSTAR}			                                &
60061201002                                                 & 2016 Jan 22                                               &
2016 Jan 23                                                 & 22                                                        &
50                                                          & \nodata						                            \\
NGC 5506				                                    & \textit{Suzaku}			                                &
701030030				                                    & 2007 Jan 31                                               &
2007 Feb 01                                                 & 57                                                        &
100					                                        & (04)						                                \\
						                                    & \textit{NuSTAR}				                            &
60061323002			                                        & 2014 Apr 01					                            &
2014 Apr 03                                                 & 56                                                        &
100                                                         & (07)							                            \\
NGC 5643			                                        & \textit{Suzaku}				                            &
702010010				                                    & 2007 Aug 19                                               &
2007 Aug 20                                                 & 43                                                        & 
50					                                        & (08)                                                      \\
                                                            & \textit{NuSTAR}				                            &
60061362002                                                 & 2014 May 24                                               &
2014 May 25                                                 & 22                                                        &
50                                                          & (09)						                                \\
NGC 5728				                                    & \textit{Suzaku}				                            &
701079010				                                    & 2006 Jun 19                                               &
2006 Jun 20                                                 & 41                                                        &
50					                                        & (10)                                                      \\
						                                    & \textit{NuSTAR}			                            	&
60061256002			                                        & 2013 Jan 02					                            &
2013 Jan 02                                                 & 24                                                        &
50                                                          & (09)							                            \\
NGC 7172				                                    & \textit{Suzaku}				                            &
703030010				                                    & 2008 May 25                                               &
2008 May 26                                                 & 82                                                        &
100					                                        & (04)						                                \\
						                                    & \textit{NuSTAR}				                            &
60061308002			                                        & 2014 Oct 07                                               &
2014 Oct 08                                                 & 32                                                        &
100                                                         & \nodata						                            \\
NGC 7582				                                    & \textit{Suzaku}				                            &
702052040				                                    & 2007 Nov 16                                               &
2007 Nov 16                                                 & 32                                                        &
50					                                        & (10) (11)                                                 \\
						                                    & \textit{NuSTAR}				                            &
60201003002			                                        & 2016 Apr 28					                            &
2016 Apr 29                                                 & 47                                                        &
50                                                          & (02) (09)						                            \\
NGC 7674				                                    & \textit{Suzaku}				                            &
708023010				                                    & 2013 Dec 08                                               &
2013 Dec 10                                                 & 52                                                        &
50					                                        & (12)  						                            \\
						                                    & \textit{NuSTAR}				                            &
60001151002			                                        & 2014 Sep 30					                            &
2014 Oct 01                                                 & 50                                                        &
50                                                          & (12)                                                      \\
\enddata
\tablecomments{Column (1): galaxy name. Column (2): observatory. Column (3): observation identification number. Column (4): start date in units of ymd. Column (5): end date in units of ymd. Column (6): exposure in units of ksec. Here we adopt \textit{Suzaku}/XIS0 and \textit{NuSTAR}/FPMA exposures. Column (7): binning. Column (8): references of the previous work.}
\tablerefs{(01) \cite{Tazaki11}. (02) \cite{Balokovic18}. (03) \cite{Rivers14}. (04) \cite{Kawamuro16a}. (05) \cite{Marinucci15}. (06) \cite{Noda14}. (07) \cite{Matt15}. (08) \cite{Kawamuro16b}. (09) \cite{Marchesi19a}. (10) \cite{Tanimoto18}. (11) \cite{Bianchi09}. (12) \cite{Gandhi17}.}
\end{deluxetable*}

\section{Sample and Data Analysis}
\subsection{Sample}
Our sample is taken from that of \cite{Ichikawa15}\footnote{The analysis of broadband X-ray spectra including the new sample of \cite{Garcia19} will be reported in a forthcoming paper \citep{Ogawa20}.}. They compiled high spatial resolution mid-infrared N-band spectroscopy, Q-band imaging, and nuclear near- and mid-infrared photometries from \cite{Alonso11} and \cite{Gonzalez13} for 21 nearby AGNs. By applying the CLUMPY model to the infrared spectra, \cite{Ichikawa15} examined the torus properties such as the V-band extinction of the torus ($A_{\mathrm{V}}$) along the equatorial plane, the torus angular width ($\sigma_{\mathrm{IR}}$), and the inclination angle ($i_{\mathrm{IR}}$).

In this paper, we analyze the broadband X-ray spectra of 10 obscured AGNs ($\log N_{\mathrm{H}}/\mathrm{cm}^{-2} \geq 22$) observed with both \textit{Suzaku} and \textit{NuSTAR} among the 21 objects in \cite{Ichikawa15}. For later discussion, we also include the Circinus galaxy and NGC 5135, for which \citet{Tanimoto19} and \citet{Yamada20} published the X-ray analysis results utilizing XCLUMPY, respectively. We have excluded 3 objects from the \cite{Ichikawa15} sample that show too complex X-ray spectra: (1) NGC 1068, a heavily Compton-thick AGN whose X-ray spectrum is dominated by photoionized plasma emission \citep[e.g.,][]{Kallman14}, (2) NGC 1386, which exhibited strong spectral variability between the \textit{Suzaku} and \textit{NuSTAR} observations according to our analysis, and (3) Cen A, which may contain a jet component \citep[e.g.,][]{Fukazawa11}. We focus on obscured AGNs because the line of sight absorption can be used to constrain the torus parameters (unless the absorption by the host galaxy is significant). High quality broadband X-ray spectra, like those of \textit{Suzaku} and \textit{NuSTAR}, are essential for separating and characterizing the torus-reflection component. Tables 1 and 2 summarize our sample and the observations, respectively.

\subsection{Data Analysis}
\subsubsection{\textit{Suzaku}}
\textit{Suzaku} (2005--2015) is the fifth Japanese X-ray astronomical satellite \citep{Mitsuda07}. It carried four CCD cameras called the X-ray imaging spectrometers (XIS0, XIS1, XIS2, XIS3: \citealt{Koyama07}) and collimated hard X-ray instrument called the hard X-ray detector (HXD: \citealt{Takahashi07}). XIS1 is the back-illuminated CCD (BIXIS) sensitive to 0.2--12.0 keV photons, and XIS0, XIS2, and XIS3 are front-side-illuminated ones (FIXIS) sensitive to 0.4--12.0 keV photons. HXD consists of the PIN photodiodes (PIN) covering the 10--70 keV band and the gadolinium silicon oxide (GSO) scintillation counters covering the 40--600 keV band \citep{Kokubun07}.

We analyzed the XIS and HXD-PIN data with the HEAsoft 6.26 and the calibration database (CALDB) released on 2018 October 10 (XIS) and 2011 September 13 (HXD). The XIS and HXD data were reprocessed by using \textsf{aepipeline}. We extracted the source spectrum of the XIS from the 1-arcmin radius circular region centered on the source peak and the background from a 1-arcmin radius source-free region. We generated the redistribution matrix files (RMF) with \textsf{xisrmfgen} and the ancillary response files (ARF) with \textsf{xissimarfgen} \citep{Ishisaki07}. The source spectrum, the background spectrum, the RMF, and the ARF of FIXIS were combined with \textsf{addascaspec}. We made the HXD/PIN spectrum with \textsf{hxdpinxbpi}. We utilized the tuned background files \citep{Fukazawa09} to reproduce the non X-ray background (NXB). The simulated spectrum of the cosmic X-ray background (CXB) was added to the NXB.

\subsubsection{\textit{NuSTAR}}
\textit{NuSTAR} (2012--) is the first imaging satellite in the hard X-ray band above 10 keV \citep{Harrison13}. It carries two co-aligned grazing incidence telescopes coupled with two focal plane modules (FPMs) and covers the energy band of 3--79 keV. We analyzed the FPM data with the HEAsoft 6.26 and the CALDB released on 2019 April 10. The FPM data were reprocessed by using \textsf{nupipeline}. We extracted the source spectrum from a 1-arcmin radius circular region centered on the source peak and the background from a 1-arcmin radius source-free region, using the \textsf{nuproducts} script. The source spectrum, the background spectrum, the RMF, and the ARF were combined with \textsf{addascaspec}.

\section{Spectral Analysis}
We employ the XCLUMPY model to reproduce the reflection spectra from the torus. The torus geometry of the clump distribution is the same as that of the CLUMPY model \citep{Nenkova08a, Nenkova08b} i.e., a power law distribution in the radial direction and a normal distribution in the elevation direction. The number density function $d(r,\theta,\phi)$ (in units of pc$^{-3}$) is represented in the spherical coordinate system (where $r$ is radius, $i$ is inclination angle measured from the rotation axis, and $\phi$ is azimuth) as:
\begin{equation}
d(r,\theta,\phi) = N \left(\frac{r}{r_{\mathrm{in}}}\right)^{-1/2} \exp{\left(-\frac{(i-\pi/2)^2}{\sigma^2}\right)}.
\end{equation}
where $N$ is the normalization, $r_{\mathrm{in}}$ is the inner radius of the torus, and $\sigma$ is the torus angular width around the mid-plane \citep{Tanimoto19}. The inner and outer radii of the torus and the radius of each clump is set to be 0.05 pc, 1.00 pc, and 0.002 pc\footnote{The absolute numbers of these 3 parameters are arbitrary and only their ratios are important, since a self similar geometry produces the identical results.}, respectively. This model has five free parameters: (1) hydrogen column density along the equatorial plane ($N_{\mathrm{H}}^{\mathrm{Equ}}$: $10^{23}$--$10^{26}$ cm$^{-2}$), (2) torus angular width ($\sigma$: 10\degr--70\degr), (3) inclination angle ($i$: 20\degr--87\degr), (4) photon index ($\Gamma$: 1.5--2.5), and (5) cutoff energy ($E_{\mathrm{cut}}$: $10^1$--$10^3$ keV).

For each object, we perform simultaneous fitting to the \textit{Suzaku}/BIXIS (0.5--8.0 keV), \textit{Suzaku}/FIXIS (2--10 keV), \textit{Suzaku}/HXD (16--40 keV: the widest case), and \textit{NuSTAR}/FPM (8--60 keV: the widest case) spectra. Our model is represented as follows in the XSPEC \citep{Arnaud96} terminology:
\begin{align}
&	\mathbf{const1*phabs}\nonumber\\
& *(\mathbf{const2*zphabs*cabs*zcutoffpl}\nonumber\\
& +	\mathbf{const3*zcutoffpl+atable\{xclumpy\_R.fits\}}\nonumber\\
& + \mathbf{const4*atable\{xclumpy\_L.fits\}+apec})
\end{align}

This model consists of six components:
\begin{enumerate}
\item \textbf{const1*phabs}. The \textbf{const1} term is a cross-normalization constant to adjust small differences in the absolute flux calibration among different instruments. We set those of \textit{Suzaku}/FIXIS and \textit{NuSTAR}/FPM to unity as references. That of \textit{Suzaku}/HXD is set to 1.16 (for the XIS-nominal pointing position) or 1.18 (HXD nominal). We leave that of \textit{Suzaku}/BIXIS ($C_{\mathrm{BIXIS}}$) as a free parameter. The \textbf{phabs} term represents the Galactic absorption. We fix the hydrogen column density to the total Galactic $HI$ and $H_{2}$ values provided by \citet{Willingale13}.

\item \textbf{const2*zphabs*cabs*zcutoffpl}. This component represents the transmitted continuum through the torus. The \textbf{const2} term ($C_{\mathrm{Time}}$) is a constant to consider time variability between the \textit{Suzaku} and \textit{NuSTAR} observations. We do not multiply this constant to the scattered component and the reflection component. This is because the sizes of the scatterer and reflector are most likely parsec or larger scales and hence little time variability is expected. We limit $C_{\mathrm{Time}}$ value within a range of 0.10--10.0 to avoid unrealistic results \citep[e.g.][]{Kawamuro16a, Tanimoto18}. The \textbf{zphabs} and \textbf{cabs} terms represent the photoelectric absorption\footnote{The differences in the absorption cross section between the \textbf{zphabs} model and that utilized in XCLUMPY (xraylib: \citealt{Schoonjans11}) are almost ignorable at energies above 1 keV.} and Compton scattering by the torus, respectively. The hydrogen column density along the line of sight ($N_{\mathrm{H}}^{\mathrm{LOS}}$) is determined according to equation (3) (see below). The \textbf{zcutoffpl} is the intrinsic continuum modeled by a power law with an exponential cutoff. Since it is difficult to determine the cutoff energy, we fix this value at a typical value ($E_{\mathrm{cut}} = 370$ keV: \citealt{Ricci18}).

\item \textbf{const3*zcutoffpl}. This represents the scattered component, where \textbf{const3} is the scattering fraction ($f_{\mathrm{scat}}$). We link the photon index ($\Gamma$), the cutoff energy ($E_{\mathrm{cut}}$), and the normalization ($N_{\mathrm{Dir}}$) to those of the intrinsic continuum.

\item \textbf{xclumpy\_R.fits}. This component represents the reflection continuum from the torus based on XCLUMPY. XCLUMPY has six free parameters: $N_{\mathrm{H}}^{\mathrm{Equ}}$, $\sigma$, $i$, $\Gamma$, $E_{\mathrm{cut}}$, and $N_{\mathrm{Dir}}$. We link $\Gamma$, $E_{\mathrm{cut}}$, and $N_{\mathrm{Dir}}$ to those of the intrinsic continuum. The line-of-sight absorption $N_{\mathrm{H}}^{\mathrm{LOS}}$ is related to torus parameters as follows:
\begin{equation}
N_{\mathrm{H}}^{\mathrm{LOS}} = \alpha N_{\mathrm{H}}^{\mathrm{Equ}} \exp{\left(-\frac{(i-\pi/2)^2}{\sigma^2}\right)}.
\end{equation}
Here the dimensionless factor $\alpha$ is introduced to take into account a possible statistical fluctuation in the number of clumps along the line of sight. To avoid unrealistic solutions, we limit $\alpha$ within a range of 0.5--2.0 (i.e., a factor of 2); this is because a typical clump number along the line of sight is found to be $\simeq$4 in our analysis, whose fractional standard deviation assuming the Poisson distribution is $\simeq$50\%.  When the error of the inclination angle is greater than 30\degr, we fix it to the value obtained from the infrared data (for NGC 3227, NGC 5643 and NGC 5728).

\item \textbf{const4*xclumpy\_L.fits}. This component represents fluorescence lines from the torus based on XCLUMPY. The \textbf{const4} term is a relative normalization ($N_{\mathrm{Line}}$ to consider possible systematic uncertainties. For instance, recent studies implied contribution from spatially extended fluorescence lines \citep{Arevalo14, Bauer15, Fabbiano17, Kawamuro19}. We link $N_{\mathrm{H}}^{\mathrm{Equ}}$, $\sigma$, $i$, $\Gamma$, and $E_{\mathrm{cut}}$ to those of the reflection continuum.

\item \textbf{apec}. This component represents emission from an optically thin thermal plasma in the host galaxy. We adopt it when the improvement of the fit by adding this component is significant at a $>$99\% confidence level with the F-test\footnote{Note that this approach is an approximation because the F-test is known to be invalid when the simpler model is at the border of the parameter space of the more complex model \cite{Protassov02}}.
\end{enumerate}

\begin{figure*}
\gridline{\fig{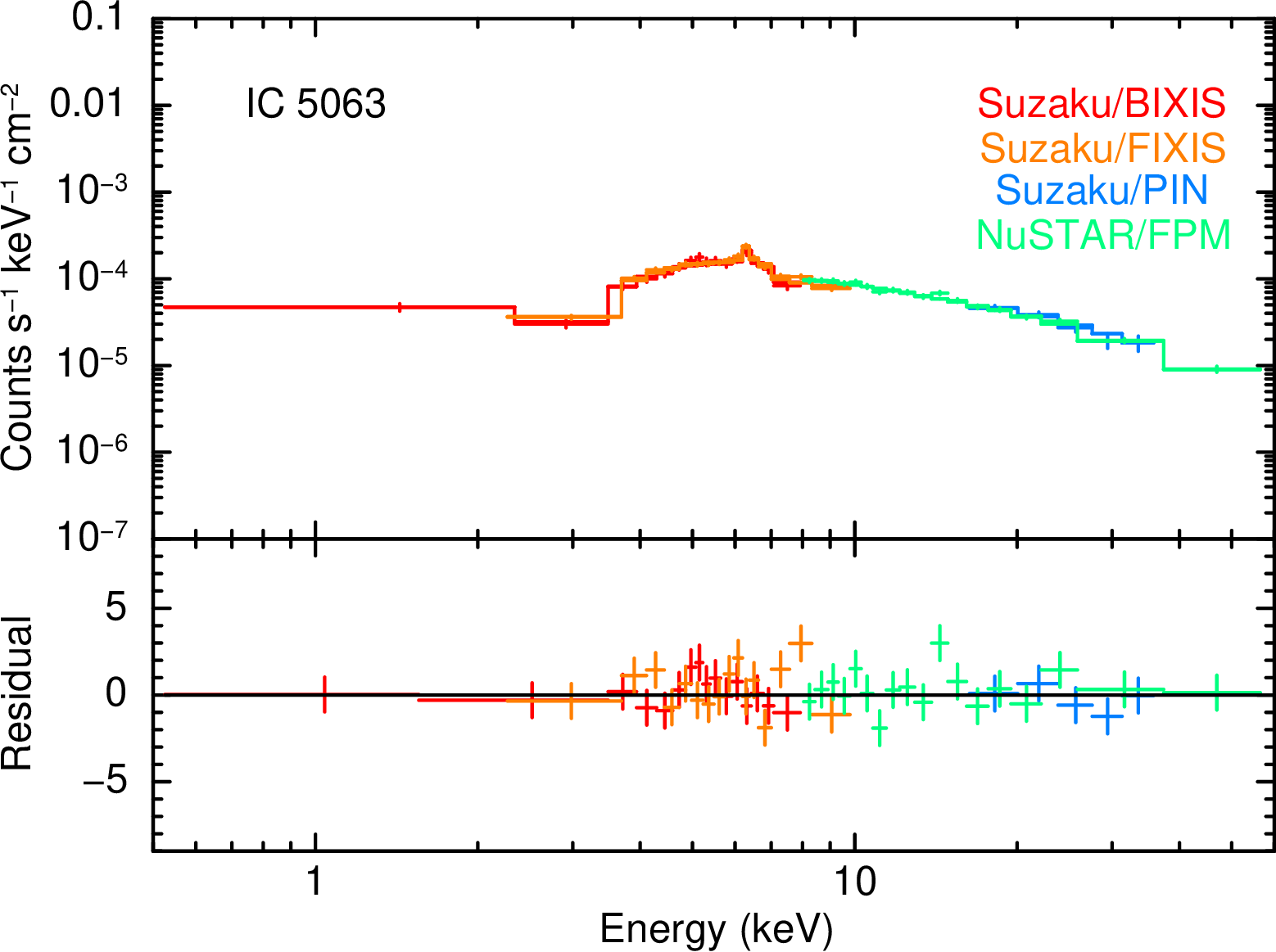}{0.45\textwidth}{}\fig{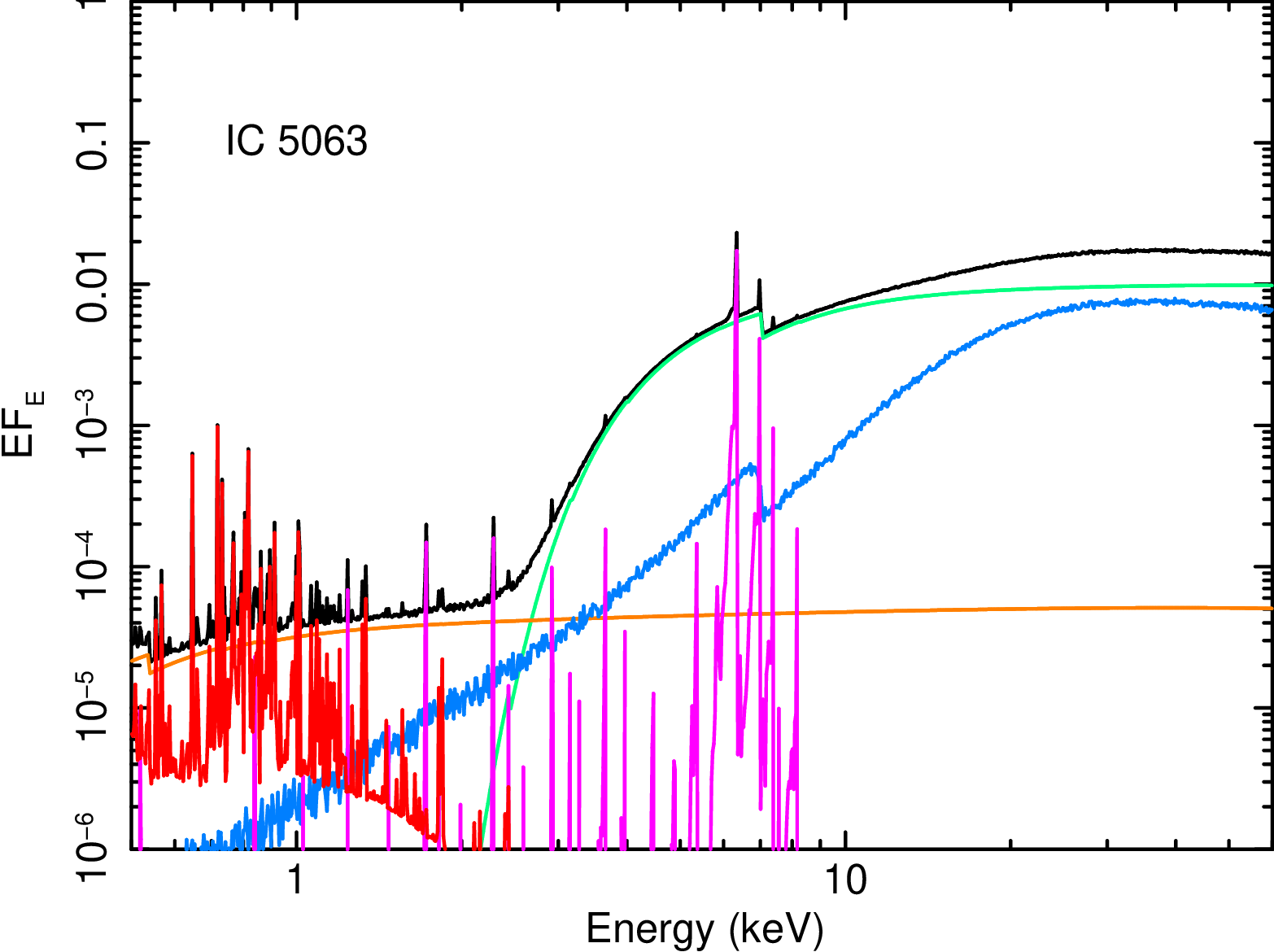}{0.45\textwidth}{}}
\gridline{\fig{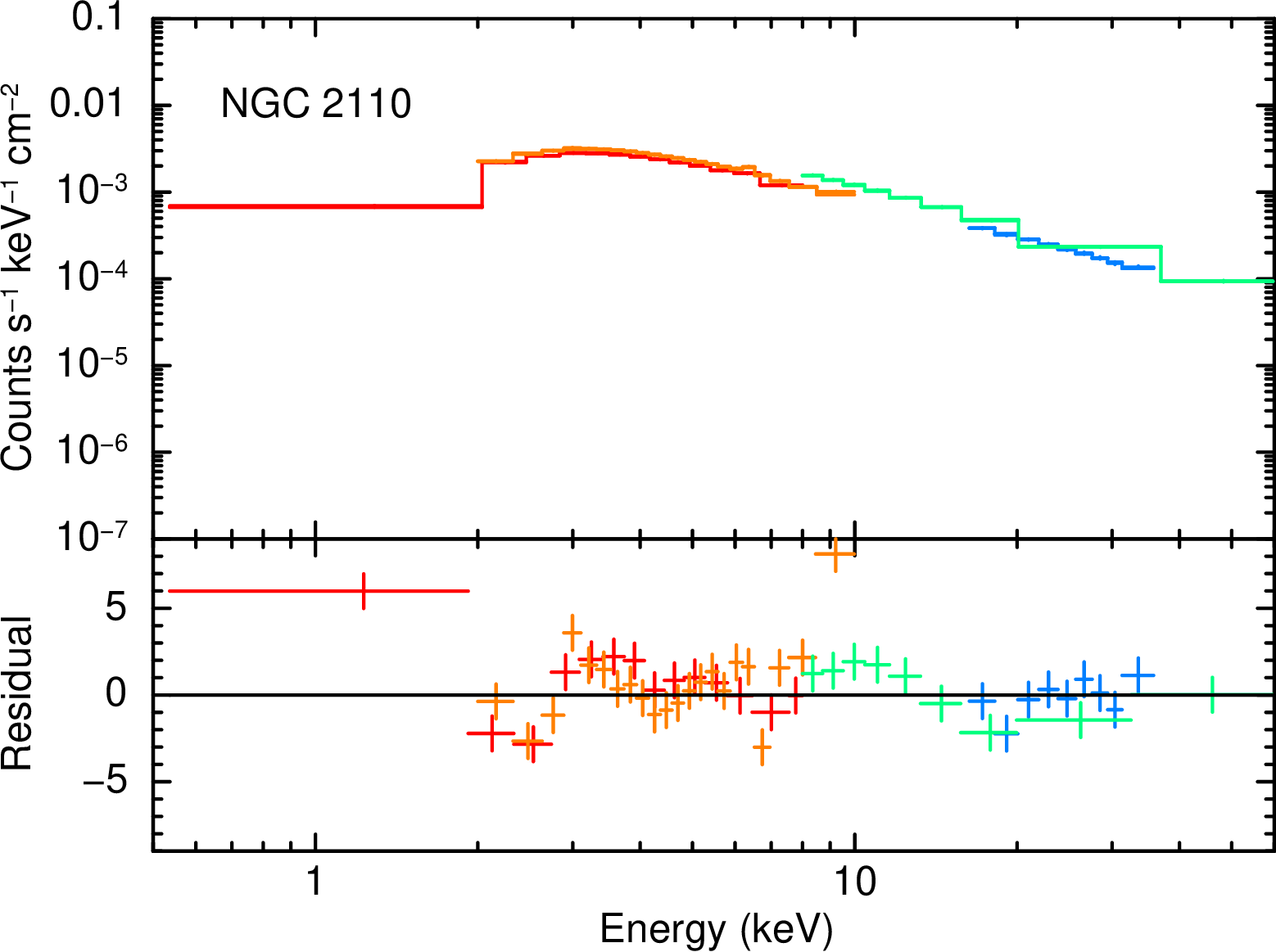}{0.45\textwidth}{}\fig{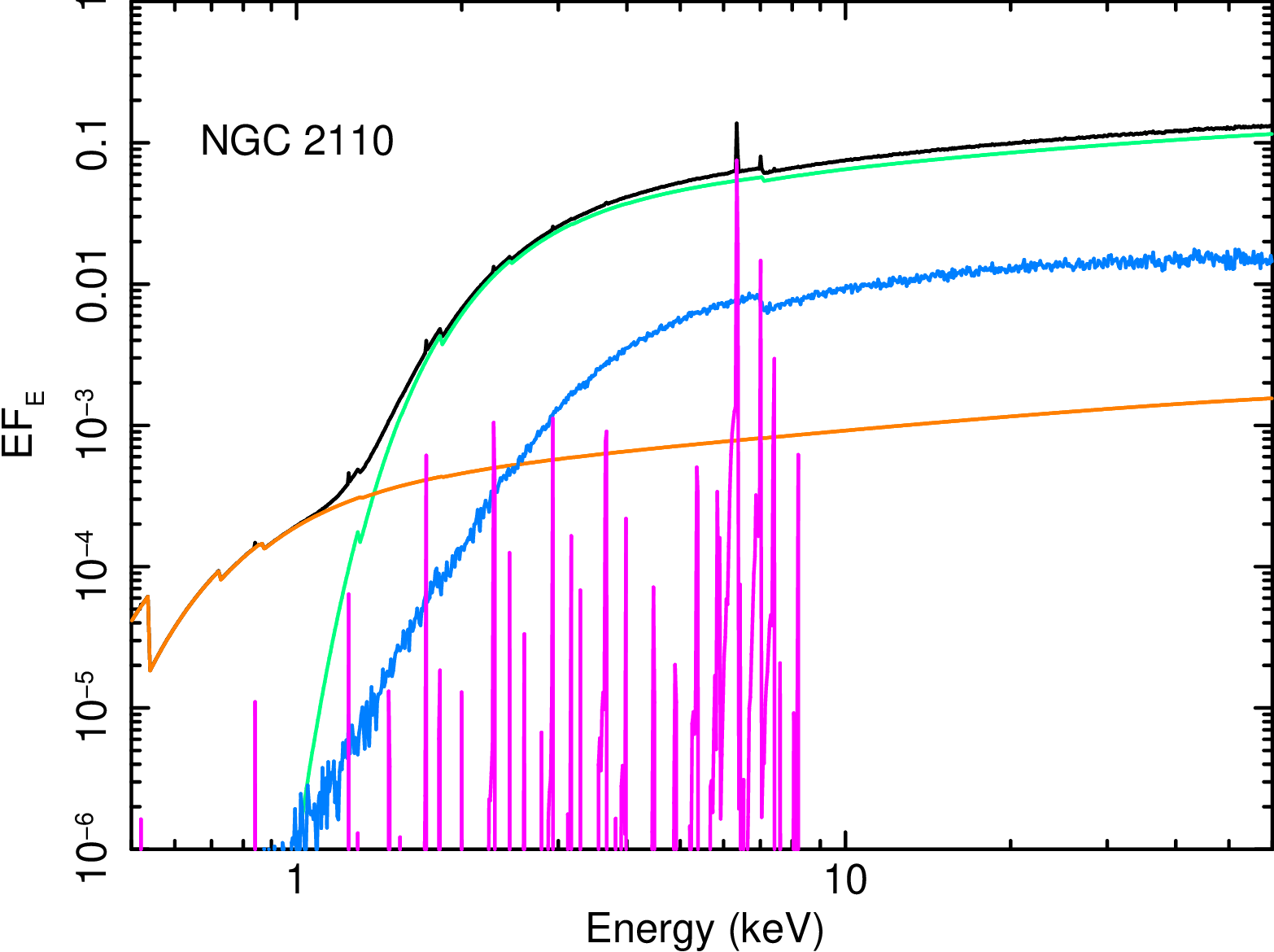}{0.45\textwidth}{}}
\gridline{\fig{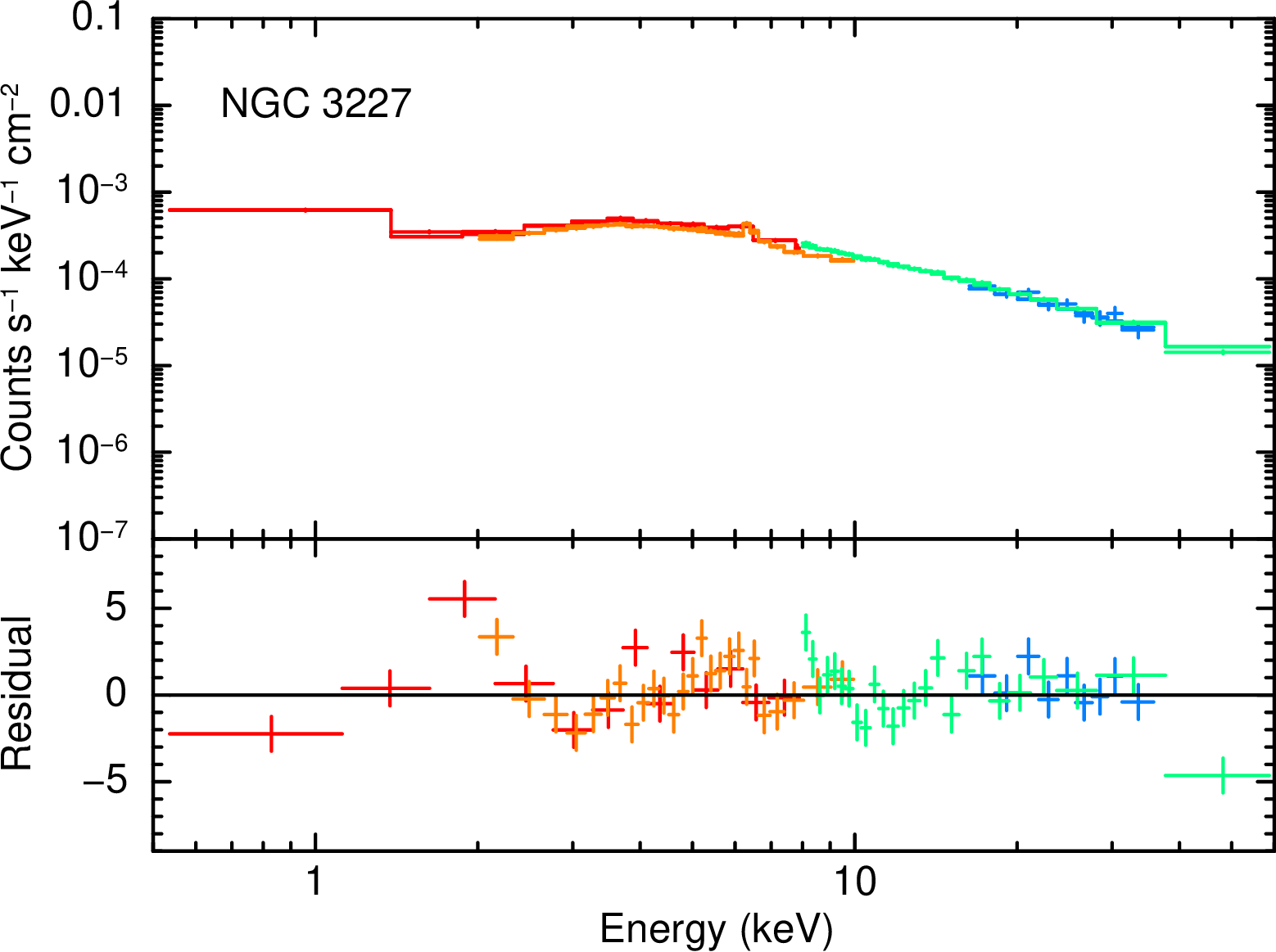}{0.45\textwidth}{}\fig{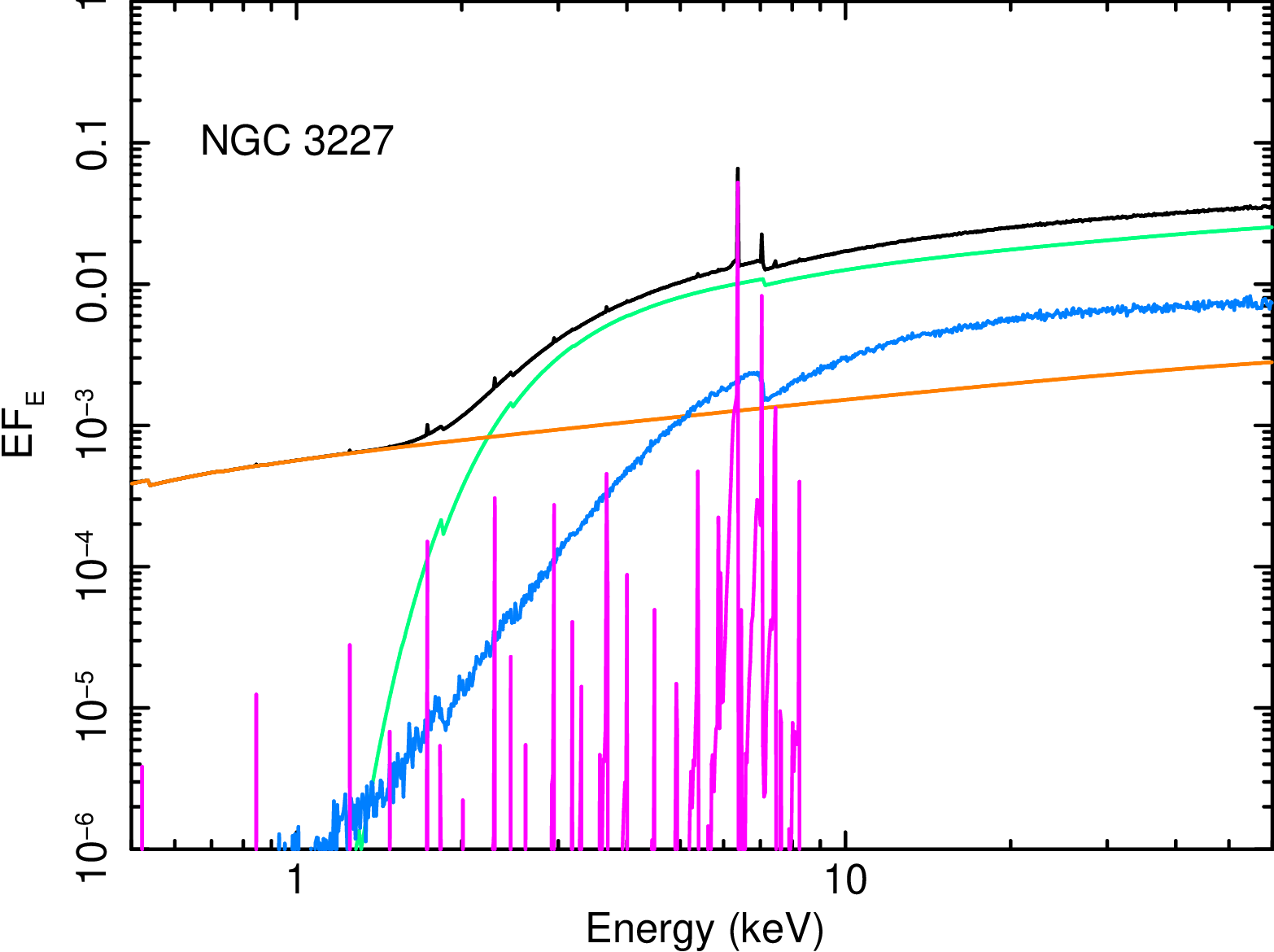}{0.45\textwidth}{}}
\caption{Left: The folded X-ray spectra fitted with the XCLUMPY. Red crosses: \textit{Suzaku}/BIXIS. Orange crosses: \textit{Suzaku}/FIXIS. Blue crosses: \textit{Suzaku}/PIN. Green crosses: \textit{NuSTAR}/FPM. Solid curves: the best fitting model. Lower panel: the residuals. Right: The best fitting models. Black line: total. Red line: thermal emission from optically thin plasma. Orange line: scattered component. Green line: direct component. Blue line: reflection continuum from the torus. Magenta line: emission lines from the torus.}
\end{figure*}
\begin{figure*}
\gridline{\fig{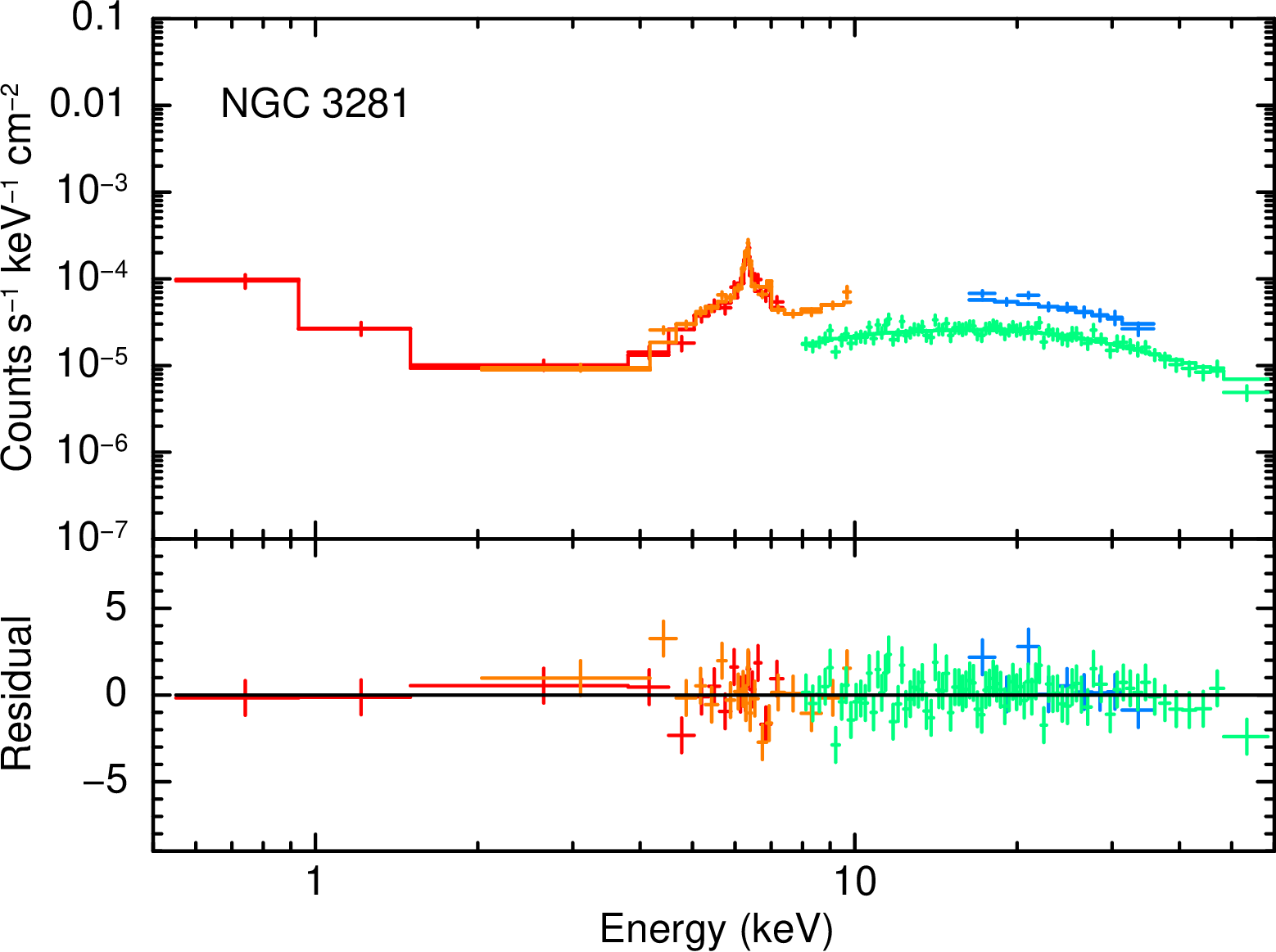}{0.45\textwidth}{}\fig{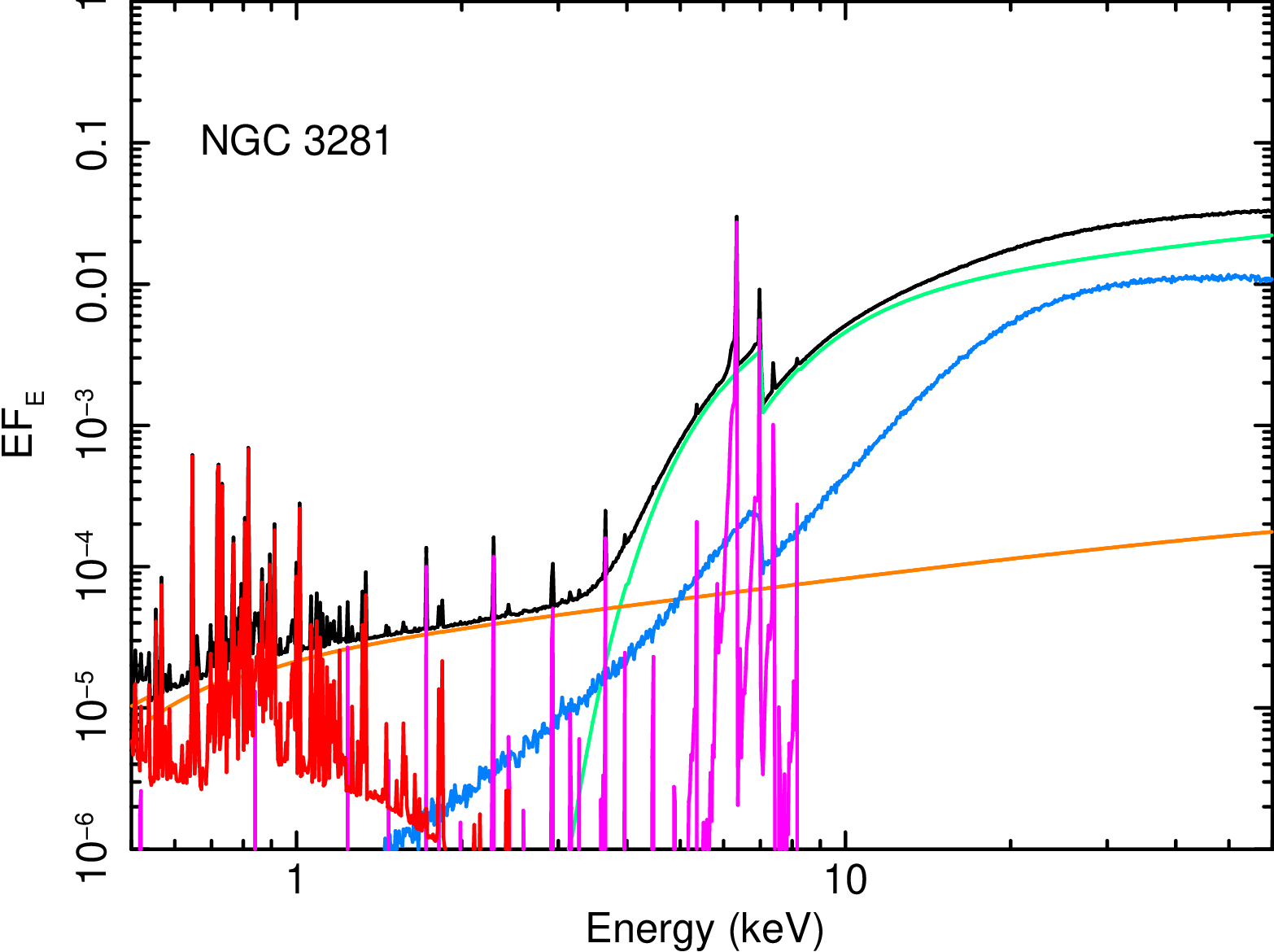}{0.45\textwidth}{}}
\gridline{\fig{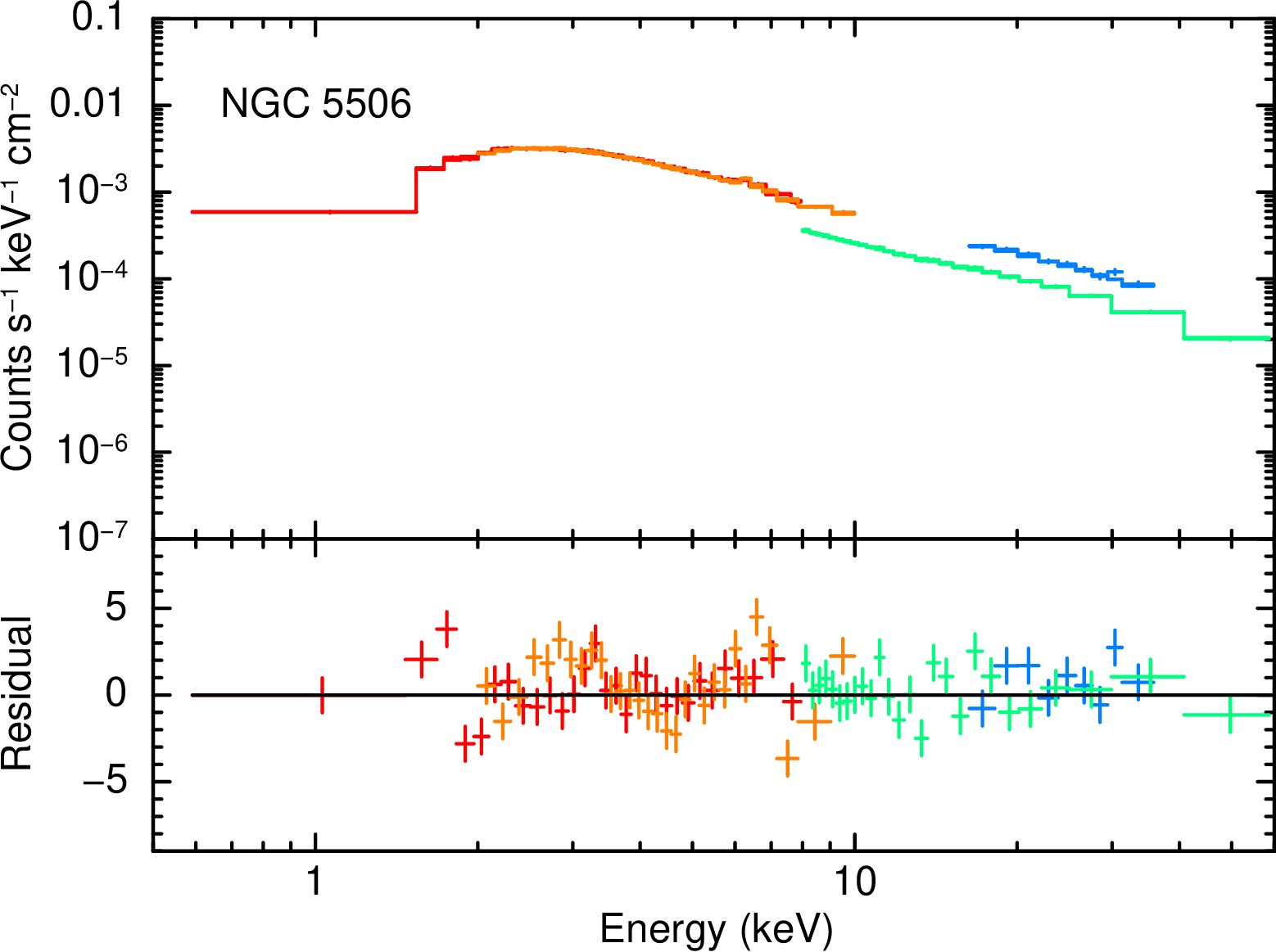}{0.45\textwidth}{}\fig{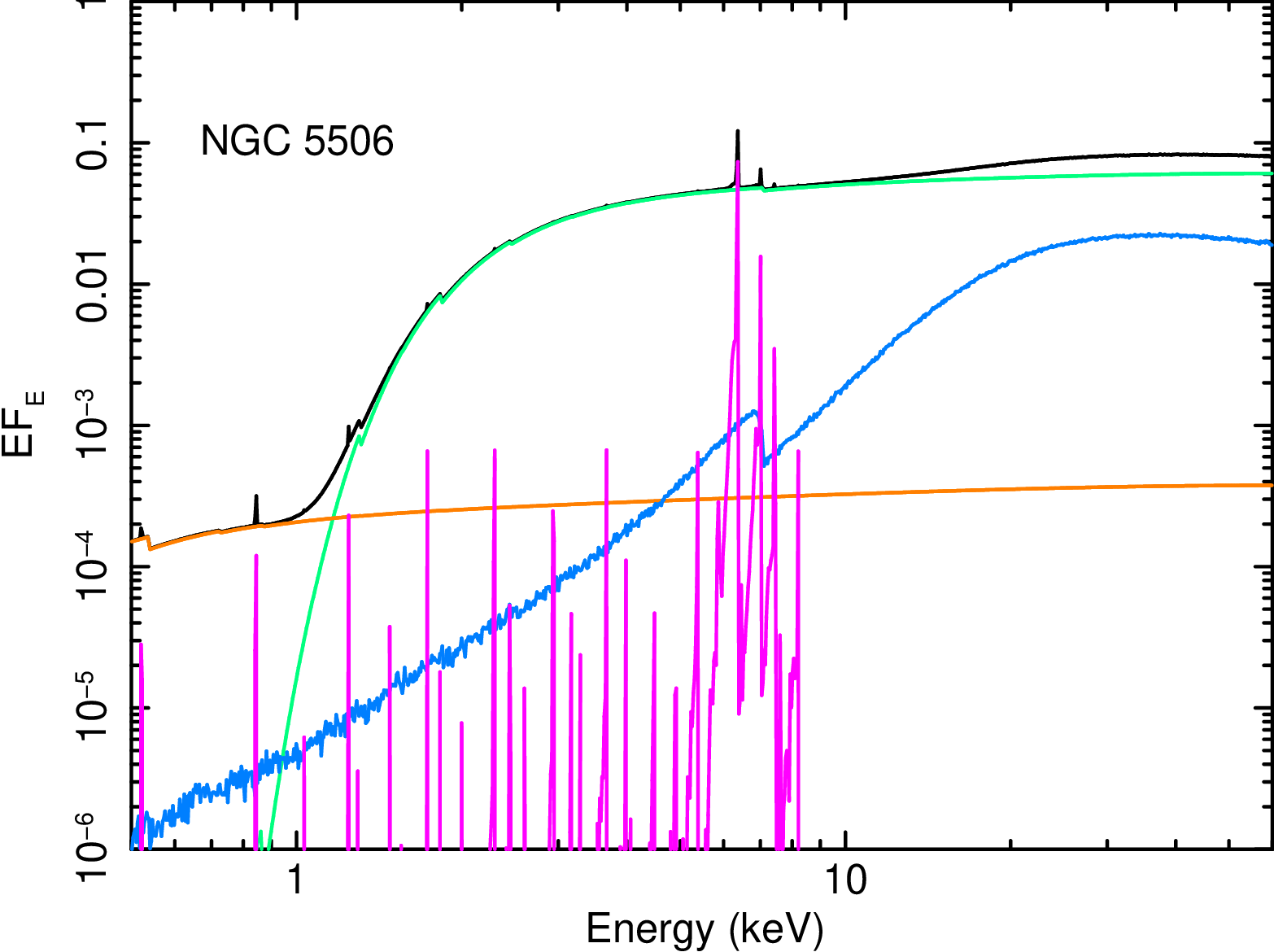}{0.45\textwidth}{}}
\gridline{\fig{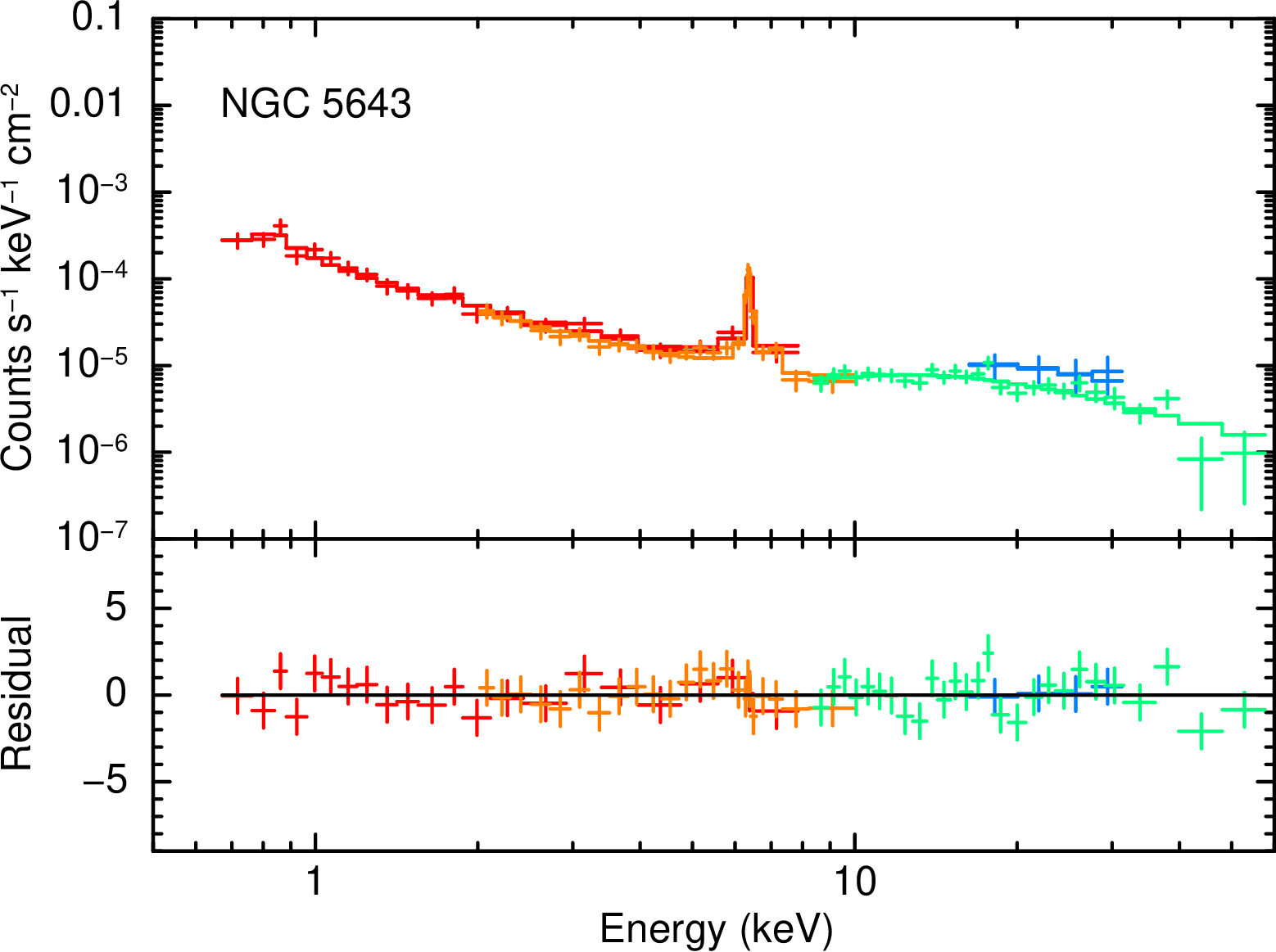}{0.45\textwidth}{}\fig{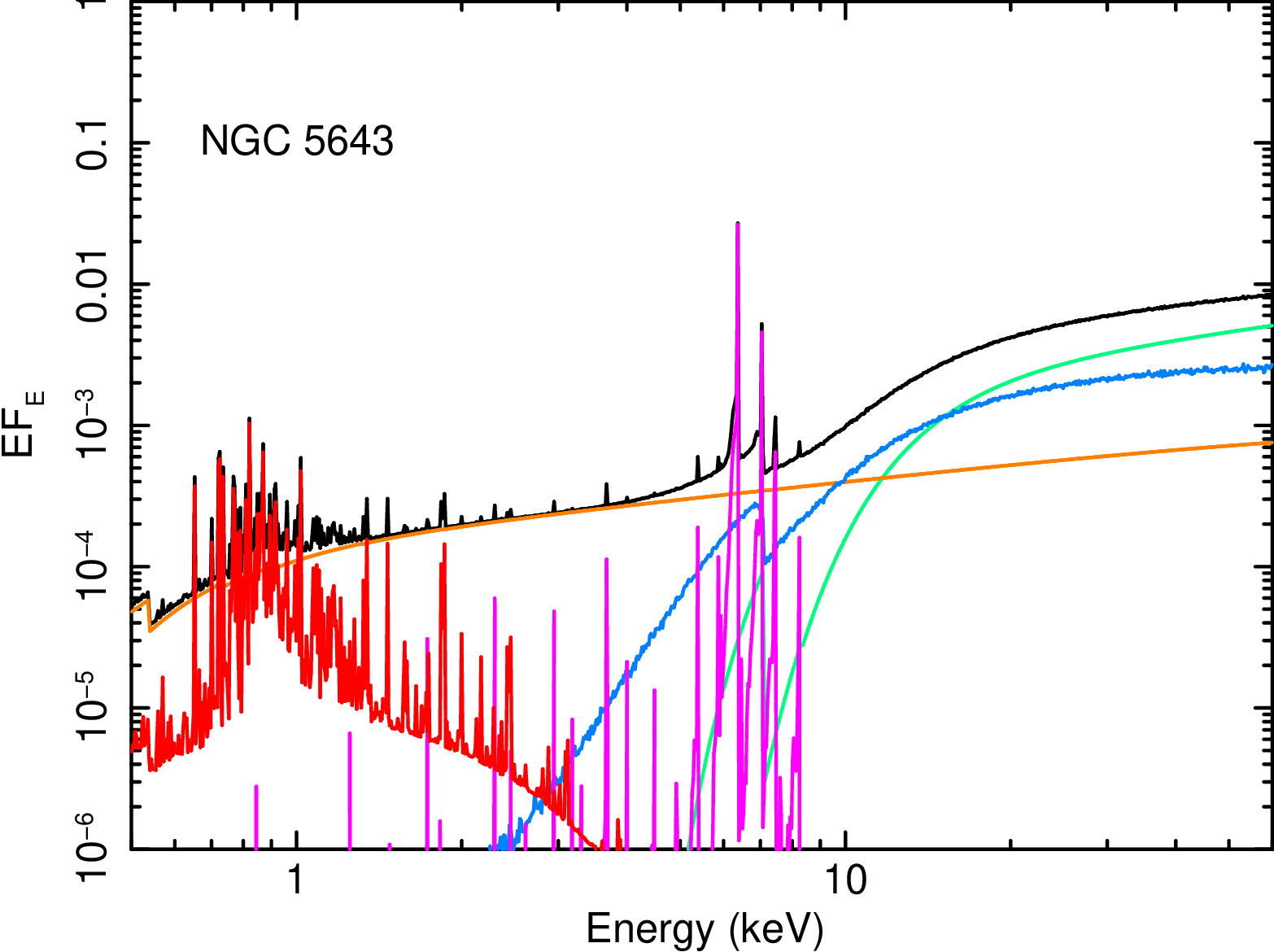}{0.45\textwidth}{}}
\setcounter{figure}{0}
\caption{Continued.}
\end{figure*}
\begin{figure*}
\gridline{\fig{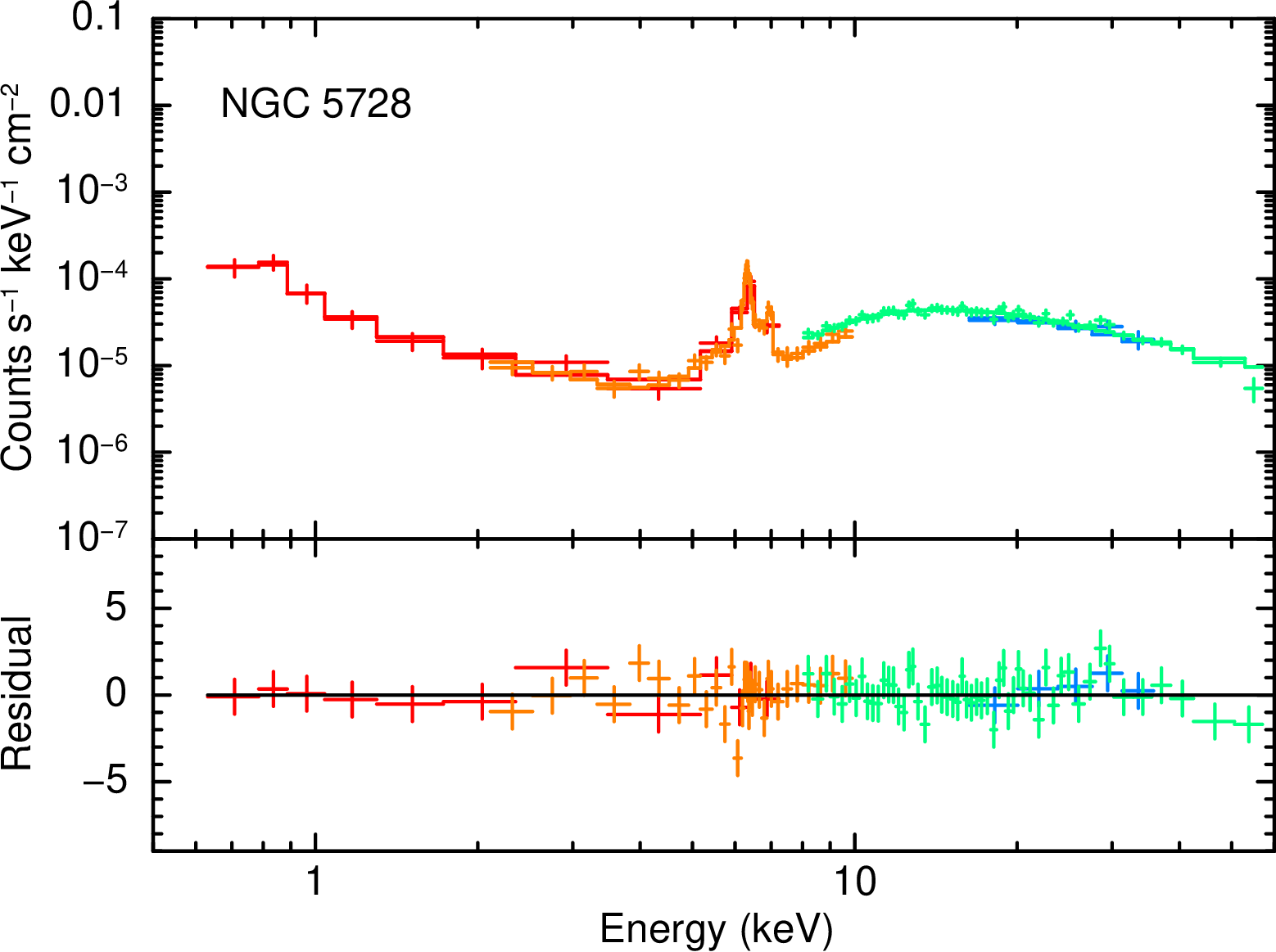}{0.45\textwidth}{}\fig{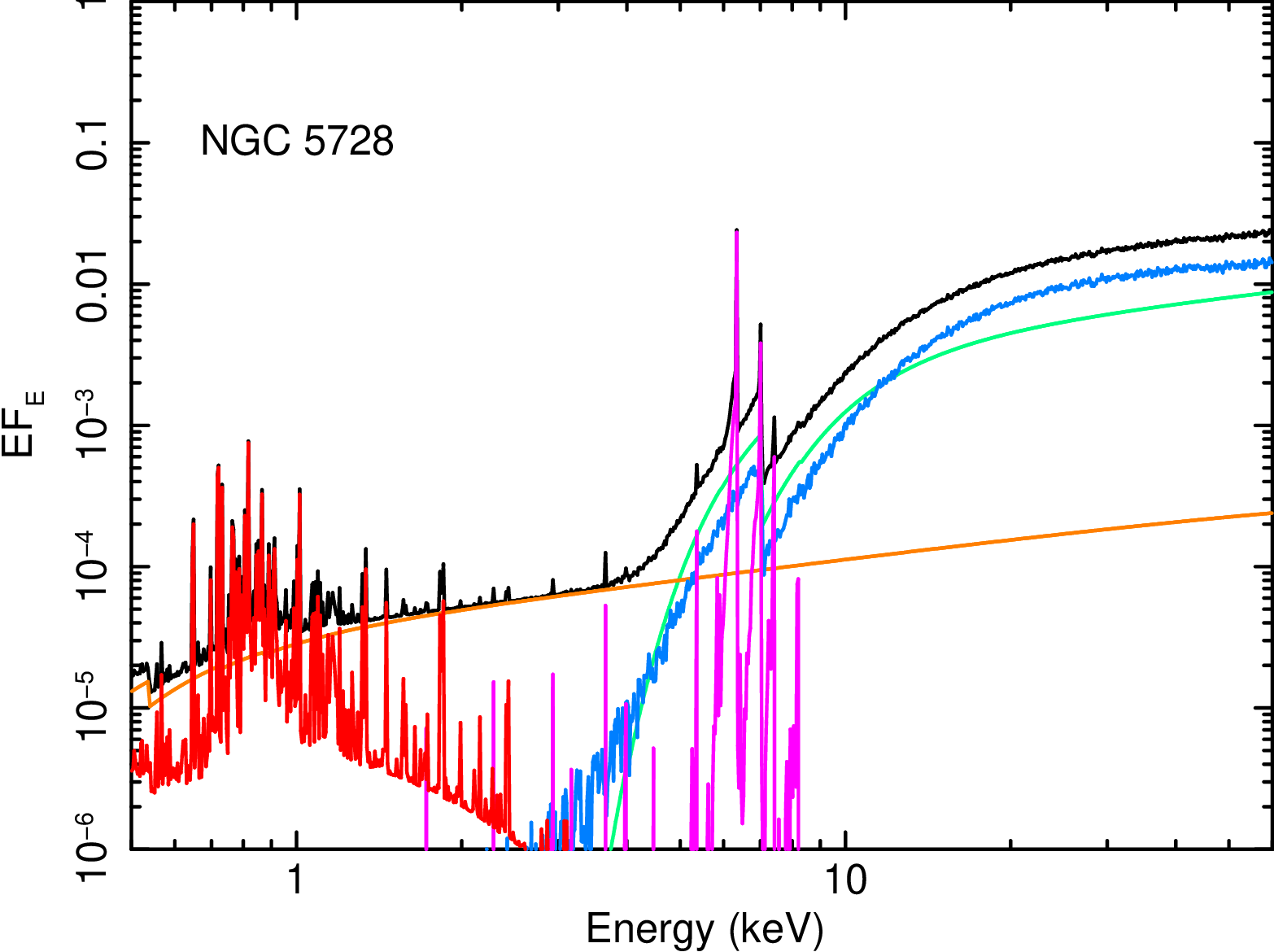}{0.45\textwidth}{}}
\gridline{\fig{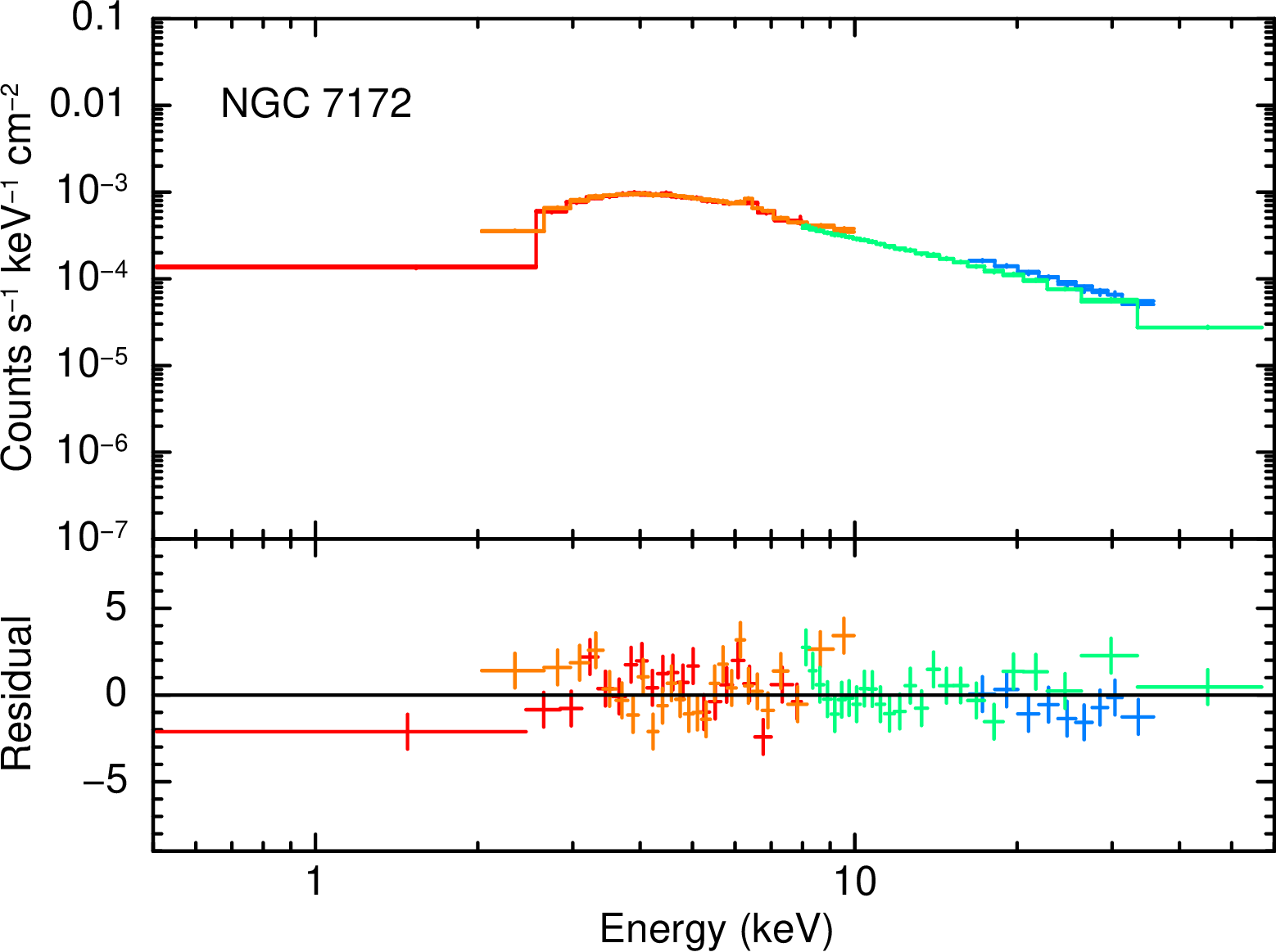}{0.45\textwidth}{}\fig{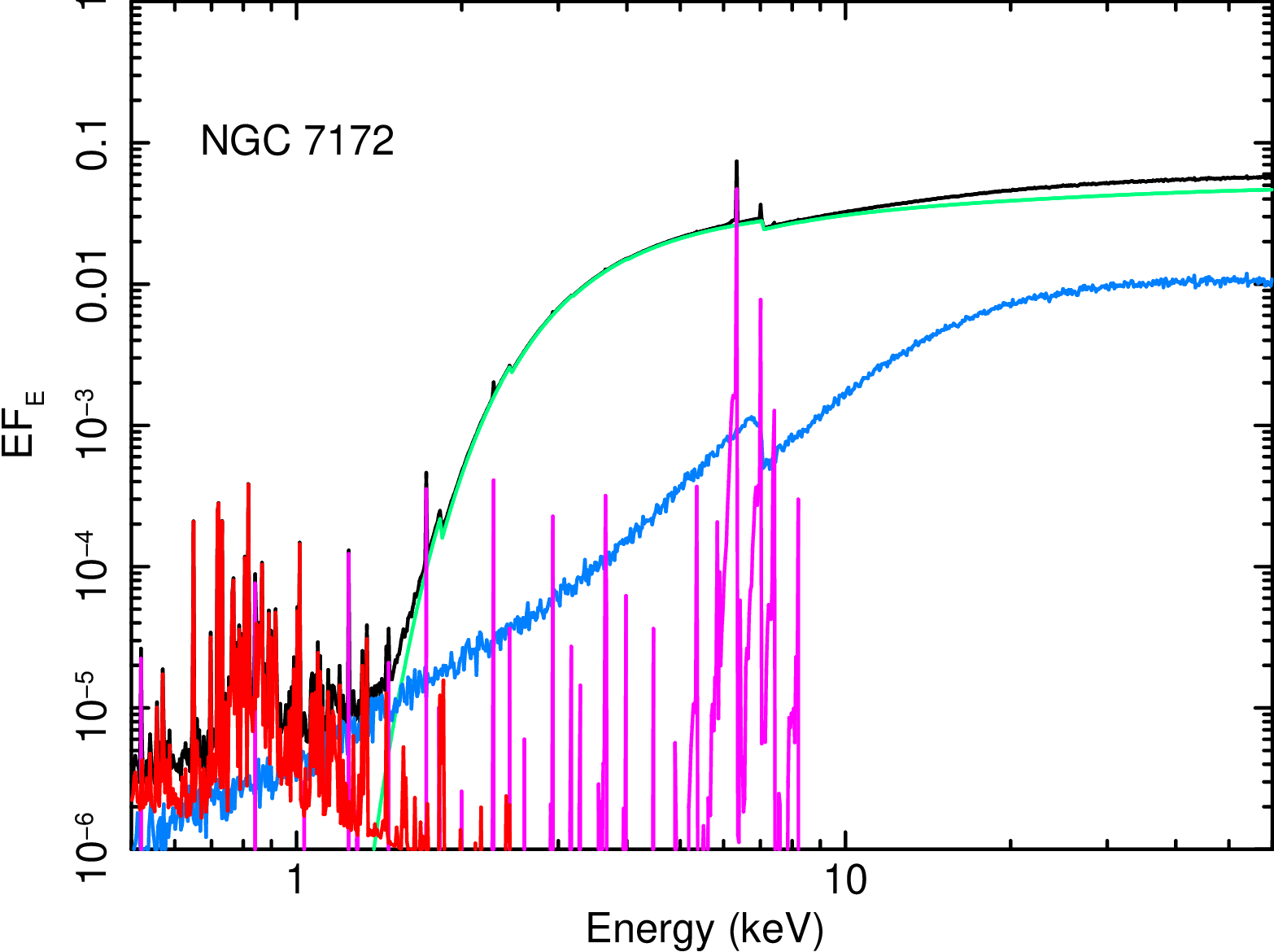}{0.45\textwidth}{}}
\gridline{\fig{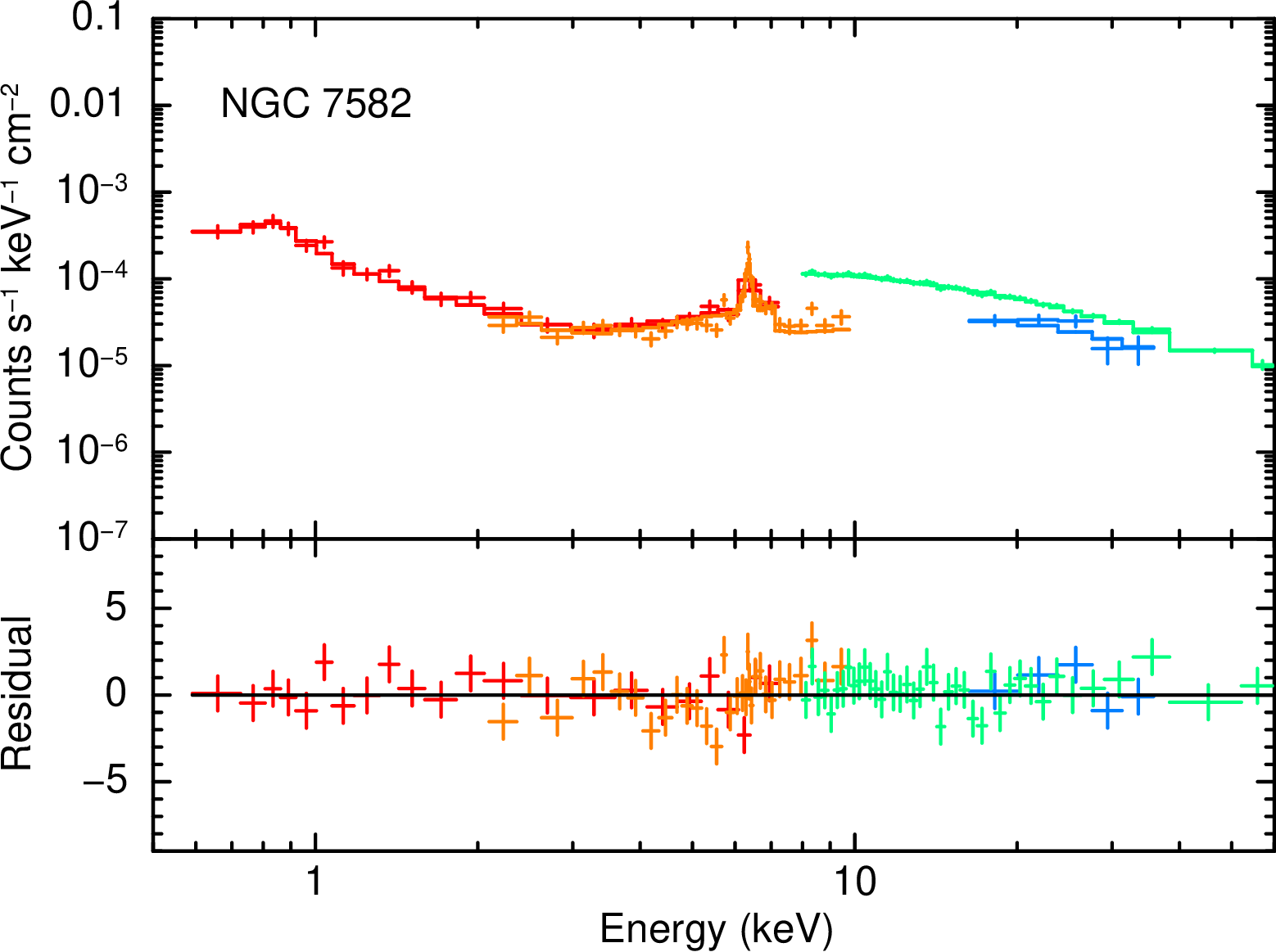}{0.45\textwidth}{}\fig{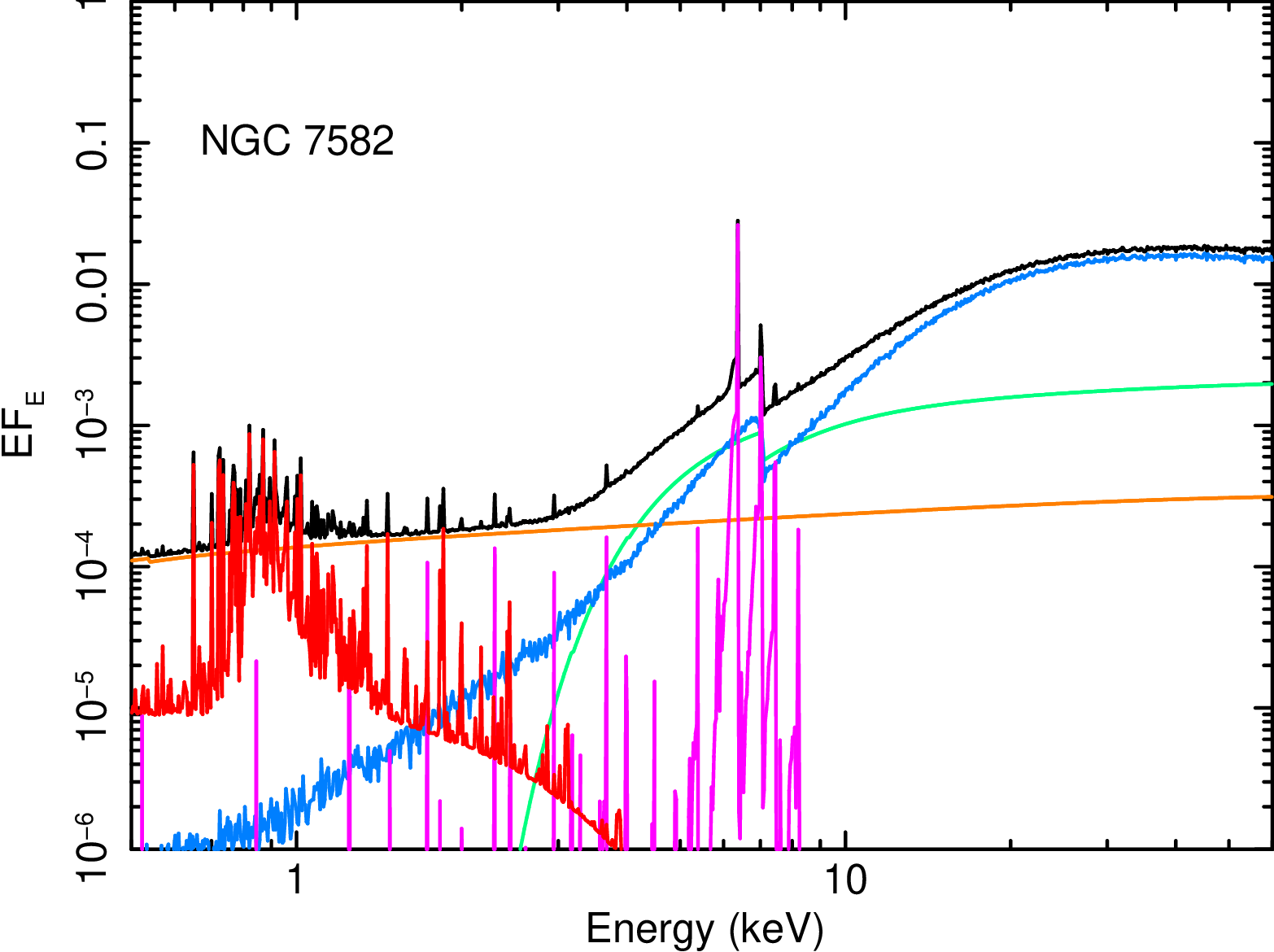}{0.45\textwidth}{}}
\setcounter{figure}{0}
\caption{Continued.}
\end{figure*}
\begin{figure*}
\gridline{\fig{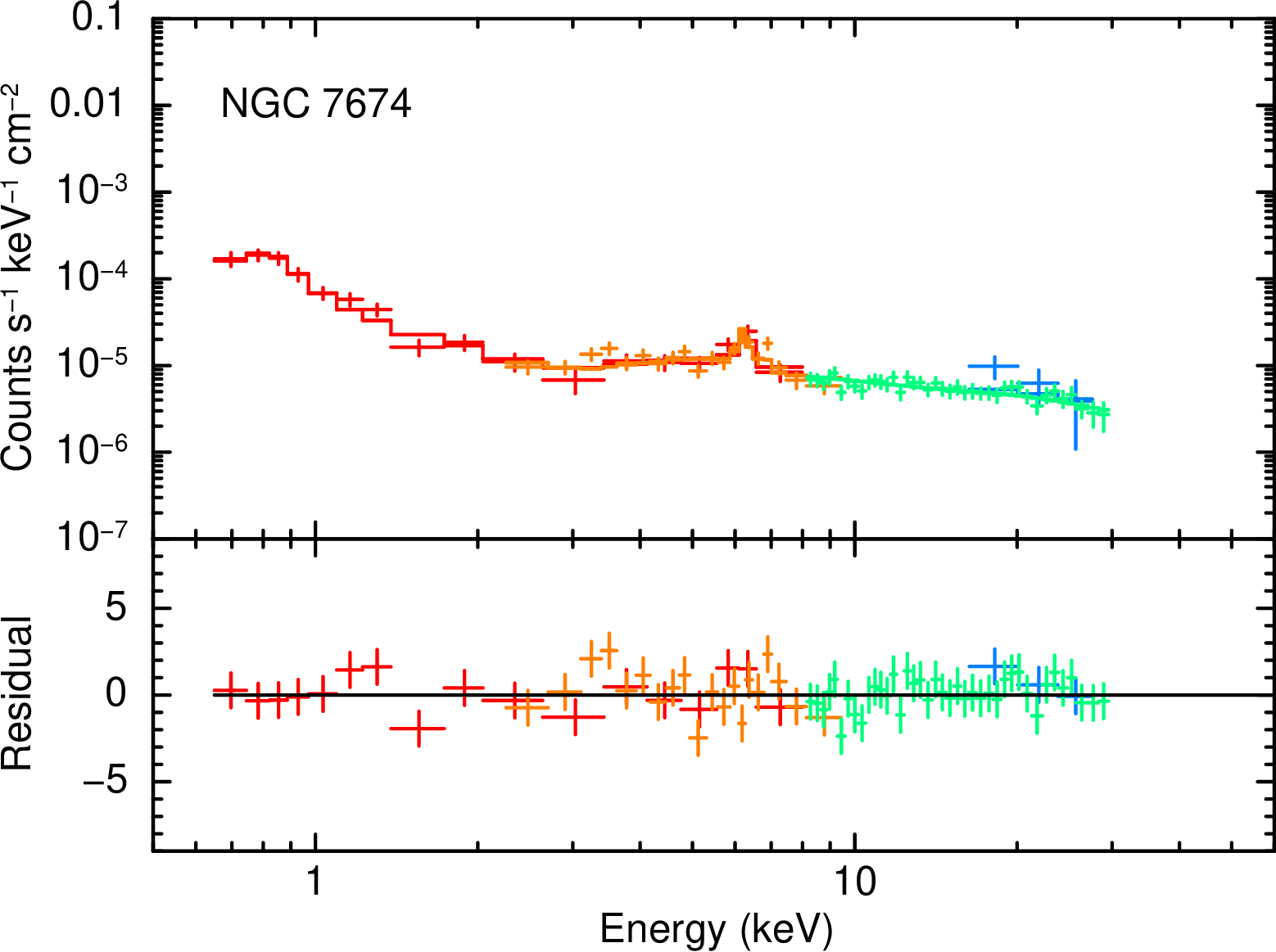}{0.45\textwidth}{}\fig{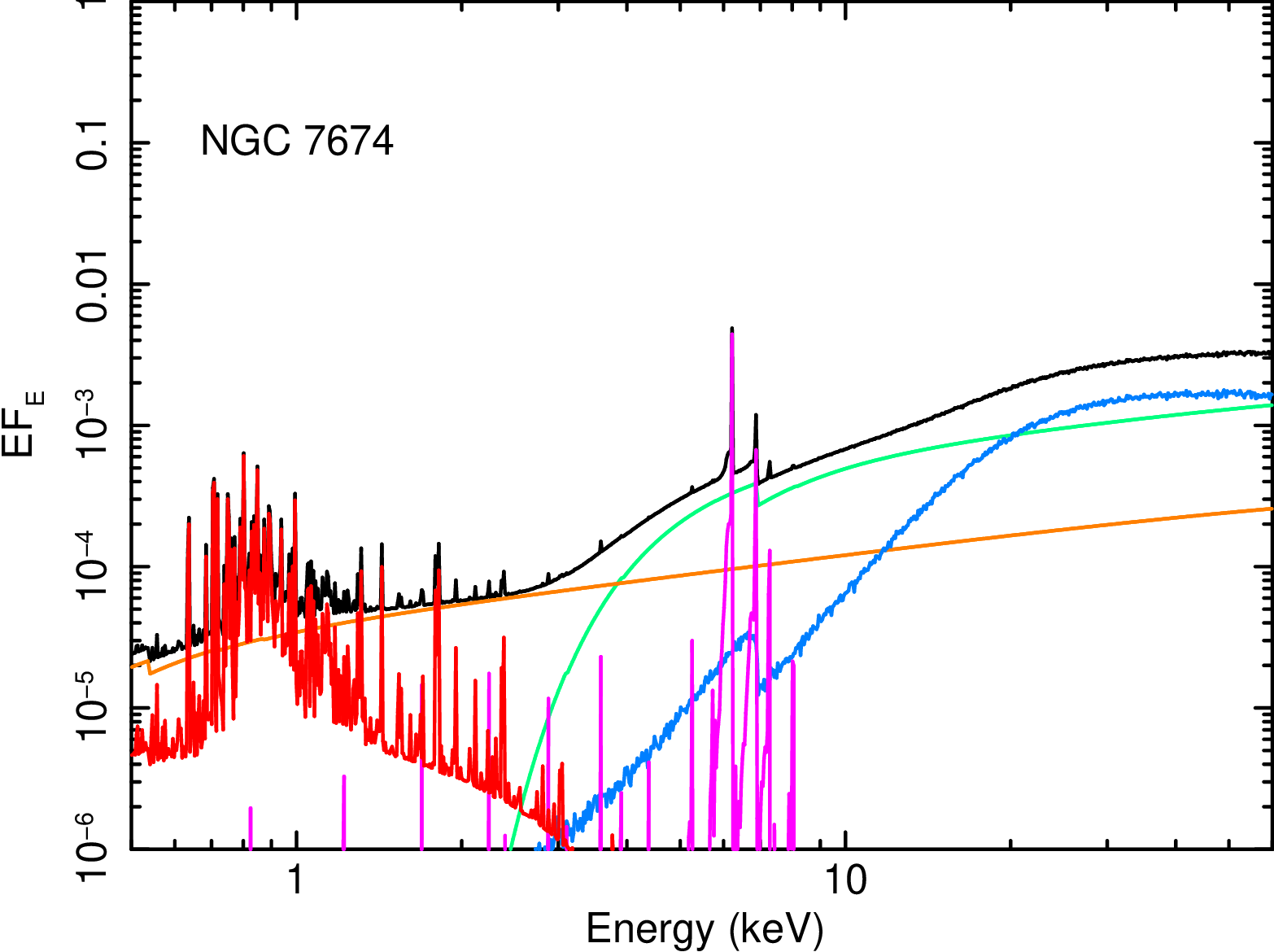}{0.45\textwidth}{}}
\setcounter{figure}{0}
\caption{Continued.}
\end{figure*}
\begin{deluxetable*}{lllllllll}
\tablecaption{Best Fitting Parameters}
\tablehead{
Galaxy Name                                                 & $C_{\mathrm{BIXIS}}$                                      &
$C_{\mathrm{Time}}$                                         & $\Gamma$                                                  &
$N_{\mathrm{Dir}}$                                          & $f_{\mathrm{scat}}$                                       &
$N_{\mathrm{H}}^{\mathrm{LOS}}$                             & $\alpha$                                                  \\
(01)		                                                & (02)				    		                            &
(03)				                                        & (04)					                                    &
(05)				                                        & (06)						                                &
(07)                                                        & (08)                                                      \\
                                                            & $N_{\mathrm{H}}^{\mathrm{Equ}}$                           &
$\sigma$		                                            & $i$	                                                    & 
$N_{\mathrm{Line}}$                                         & $k_{\mathrm{B}}T$			                                &
$N_{k_B T}$	                                                & $\chi^2_{\mathrm{red}}$	                                \\
                                                            & (09)		    				                            &
(10)					                                    & (11)					    	                            &
(12)					                                    & (13)                                                      &
(14)                                                        & (15)                                                      }
\startdata
IC 5063                                                     & $0.96_{-0.04}^{+0.04}$                                    &
$0.87_{-0.04}^{+0.04}$                                      & $1.89_{-0.09}^{+0.10}$                                    &
$1.06_{-0.20}^{+0.29}$                                      & $0.34_{-0.12}^{+0.13}$                                    &
$0.26_{-0.14}^{+0.49}$                                      & $0.67_{-0.13}^{+1.33}$                                    \\
                                                            & $10.0_{-5.26}^{+19.1}$                                    &
$22.4_{-4.13}^{+5.30}$                                      & $49.5_{-8.40}^{+10.0}$                                    &
$1.21_{-0.34}^{+0.69}$                                      & $0.75_{-0.49}^{+0.28}$                                    &
$0.15_{-0.08}^{+0.08}$                                      & $0.97$                                                    \\
NGC 2110                                                    & $0.90_{-0.01}^{+0.01}$                                    &
$0.64_{-0.01}^{+0.01}$                                      & $1.63_{-0.01}^{+0.01}$                                    &
$5.40_{-0.11}^{+0.10}$                                      & $0.84_{-0.05}^{+0.05}$                                    &
$0.04_{-0.01}^{+0.01}$                                      & $0.50_{-0.00}^{+0.26}$                                    \\
                                                            & $0.18_{-0.01}^{+0.01}$                                    &
$63.8_{-12.9}^{+6.20}$                                      & $31.7_{-11.7}^{+2.50}$                                    &
$0.36_{-0.03}^{+0.04}$                                      & \nodata                                                   &
\nodata                                                     & $1.03$                                                    \\
NGC 3227                                                    & $1.13_{-0.02}^{+0.02}$                                    &
$0.73_{-0.02}^{+0.02}$                                      & $1.58_{-0.02}^{+0.03}$                                    &
$0.69_{-0.04}^{+0.03}$                                      & $7.74_{-0.45}^{+0.46}$                                    &
$0.07_{-0.02}^{+0.01}$                                      & $0.77_{-0.27}^{+0.87}$                                    \\
                                                            & $0.56_{-0.12}^{+0.07}$                                    &
$51.6_{-8.00}^{+12.0}$                                      & $20.0$ (fixed)                                            &
$0.66_{-0.06}^{+0.07}$                                      & \nodata                                                   &
\nodata                                                     & $1.22$                                                    \\
NGC 3281                                                    & $0.92_{-0.07}^{+0.07}$                                    &
$3.00_{-0.27}^{+0.28}$                                      & $1.50_{-0.00}^{+0.05}$                                    &
$0.19_{-0.01}^{+0.04}$                                      & $1.54_{-0.31}^{+0.31}$                                    &
$0.66_{-0.17}^{+0.38}$                                      & $2.00_{-1.21}^{+0.00}$                                    \\
                                                            & $9.79_{-2.56}^{+5.70}$                                    &
$38.0_{-6.50}^{+9.10}$                                      & $20.1_{-0.10}^{+6.40}$                                    &
$2.63_{-0.51}^{+0.67}$                                      & $0.37_{-0.14}^{+0.32}$                                    &
$0.34_{-0.18}^{+0.31}$                                      & $1.23$                                                    \\
NGC 5506                                                    & $1.02_{-0.01}^{+0.01}$                                    &
$2.02_{-0.03}^{+0.04}$                                      & $1.84_{-0.01}^{+0.02}$                                    &
$1.88_{-0.06}^{+0.06}$                                      & $1.23_{-0.12}^{+0.12}$                                    &
$0.03_{-0.01}^{+0.01}$                                      & $0.50_{-0.00}^{+1.50}$                                    \\
                                                            & $12.5_{-3.75}^{+3.70}$                                    &
$24.6_{-2.58}^{+2.70}$                                      & $33.4_{-10.6}^{+2.77}$                                    &
$1.78_{-0.36}^{+0.30}$                                      & \nodata                                                   &
\nodata                                                     & $1.08$                                                    \\
NGC 5643                                                    & $1.19_{-0.11}^{+0.12}$                                    &
$2.57_{-1.63}^{+2.50}$                                      & $1.68_{-0.17}^{+0.18}$                                    &
$0.23_{-0.12}^{+0.53}$                                      & $6.30_{-3.94}^{+5.48}$                                    &
$2.36_{-0.58}^{+2.24}$                                      & $2.00_{-0.53}^{+0.00}$                                    \\
                                                            & $1.50_{-0.36}^{+1.40}$                                    &
$34.8_{-18.5}^{+35.2}$                                      & $74.0$ (fixed)                                            &
$2.70_{-0.51}^{+0.81}$                                      & $0.25_{-0.06}^{+0.12}$                                    &
$1.17_{-0.59}^{+0.83}$                                      & $0.99$                                                    \\
NGC 5728                                                    & $0.99_{-0.11}^{+0.12}$                                    &
$0.56_{-0.07}^{+0.08}$                                      & $1.50_{-0.00}^{+0.05}$                                    &
$0.45_{-0.03}^{+0.13}$                                      & $0.83_{-0.28}^{+0.14}$                                    &
$0.95_{-0.19}^{+0.28}$                                      & $0.50_{-0.00}^{+0.20}$                                    \\
                                                            & $2.00_{-0.40}^{+0.58}$                                    &
$60.0_{-13.5}^{+10.0}$                                      & $80.0$ (fixed)                                            &
$1.20_{-0.32}^{+0.08}$                                      & $0.56_{-0.27}^{+0.17}$                                    &
$0.31_{-0.10}^{+0.30}$                                      & $1.00$                                                    \\
NGC 7172                                                    & $0.98_{-0.01}^{+0.01}$                                    &
$1.08_{-0.02}^{+0.02}$                                      & $1.76_{-0.02}^{+0.02}$                                    &
$2.05_{-0.10}^{+0.11}$                                      & $0.00_{-0.00}^{+0.08}$                                    &
$0.09_{-0.02}^{+0.18}$                                      & $0.50_{-0.00}^{+1.50}$                                    \\
                                                            & $5.26_{-1.40}^{+10.3}$                                    &
$12.6_{-2.60}^{+3.00}$                                      & $66.9_{-6.44}^{+3.51}$                                    &
$1.13_{-0.16}^{+0.50}$                                      & $0.59_{-0.27}^{+0.48}$                                    &
$0.10_{-0.04}^{+0.08}$                                      & $1.02$                                                    \\
NGC 7582                                                    & $1.10_{-0.09}^{+0.09}$                                    &
$0.10_{-0.00}^{+0.01}$                                      & $1.77_{-0.07}^{+0.08}$                                    &
$1.05_{-0.22}^{+0.32}$                                      & $1.20_{-0.25}^{+0.27}$                                    &
$0.31_{-0.13}^{+0.16}$                                      & $1.47_{-0.97}^{+0.53}$                                    \\
                                                            & $7.29_{-3.00}^{+3.64}$                                    &
$25.8_{-3.60}^{+10.5}$                                      & $41.4_{-5.80}^{+7.10}$                                    &
$0.64_{-0.13}^{+0.20}$                                      & $0.76_{-0.11}^{+0.09}$                                    &
$0.47_{-0.12}^{+0.13}$                                      & $1.11$                                                    \\
NGC 7674                                                    & $1.01_{-0.12}^{+0.13}$                                    &
$0.96_{-0.16}^{+0.20}$                                      & $1.50_{-0.00}^{+0.07}$                                    &
$0.03_{-0.01}^{+0.01}$                                      & $15.1_{-2.98}^{+3.02}$                                    &
$0.24_{-0.10}^{+0.22}$                                      & $0.51_{-0.01}^{+1.49}$                                    \\
                                                            & $10.0_{-4.10}^{+9.20}$                                    &
$39.9_{-8.80}^{+5.70}$                                      & $20.0_{-0.00}^{+8.00}$                                    &
$2.76_{-0.98}^{+0.88}$                                      & $0.71_{-0.12}^{+0.08}$                                    &
$0.34_{-0.07}^{+0.08}$                                      & $1.20$                                                    \\
\enddata
\tablecomments{Column (01): galaxy name. Column (02): relative normalization of \textit{Suzaku}/BIXIS to \textit{Suzaku}/FIXIS. Column (03): time variability constant. Column (04): photon index. Column (05): normalization of the direct component in units of 10$^{-2}$ photons keV$^{-1}$ cm$^{-2}$ s$^{-1}$. Column (06): scattering fraction in units of percent. Column (07): hydrogen column density along the equatorial plane in units of 10$^{24}$ cm$^{-2}$. Column (08): correction factor of the line-of-sight column density. Column (09): hydrogen column density along the line of sight in units of 10$^{24}$ cm$^{-2}$. Column (10): torus angular width in units of degree. Column (11): inclination angle in units of degree. Column (12): relative normalization of the emission lines to the reflection component. Column (13): temperature of the \textsf{apec} model in units of keV. Column (14): normalization of the \textsf{apec} model in units of 10$^{-18}/4\pi [D_{\mathrm{A}}(1+z)]^2 \int n_{\mathrm{e}} n_{\mathrm{H}}dV$, where $D_{\mathrm{A}}$ is the angular diameter distance to the source in units of cm, $n_{\mathrm{e}}$ and $n_{\mathrm{H}}$ are the electron and hydrogen densities in units of cm$^{-3}$. Column (15): reduced $\chi^2$.}
\end{deluxetable*}
\begin{deluxetable*}{llllllll}
\tablecaption{Fluxes and Luminosities}
\tablehead
{
Galaxy Name                                                 & $\log F_{2-10}$ 				                            &
$\log L_{2-10}$				                                & $\log M_{\mathrm{BH}}/M_{\sun}$                           &
$\log \lambda_{\mathrm{Edd}}$ 							    & $M_{\mathrm{BH}}$ Reference                               \\
(1)                                                         & (2)                                                       &
(3)					                                        & (4)							                            &
(5)						                                    & (6)                                                       }
\startdata
IC 5063                                                     & $-11.1$						                            &
$ 43.0$						                                & $ 8.45$                                                   &
$-2.29$                                                     & (1)                                                       \\
NGC 2110                                                    & $-9.90$                                                   &
$ 43.5$						                                & $ 9.25$                                                   &
$-2.53$			                    					    & (2)                                                       \\
NGC 3227                                                    & $-10.7$						                            &
$ 42.0$		                                                & $ 7.18$                                                   &
$-1.94$		                                                & (2)                                                       \\
NGC 3281			                                        & $-11.4$						                            &
$ 42.4$						                                & $ 8.00$                                                   &
$-2.38$				                                        & (3)                                                       \\
NGC 5506				                                    & $-10.0$						                            &
$ 42.7$						                                & $ 7.87$                                                   &
$-1.96$                                                     & (4)                                                       \\
NGC 5643		                                            & $-11.9$						                            &
$ 41.4$						                                & $ 7.05$                                                   &
$-2.41$                                                     & (1)                                                       \\
NGC 5728				                                    & $-11.8$						                            &
$ 42.7$						                                & $ 8.07$                                                   &
$-2.18$                                                     & (2)                                                       \\
NGC 7172				                                    & $-10.4$						                            &
$ 43.1$						                                & $ 8.45$                                                   &
$-2.15$				                                        & (2)                                                       \\
NGC 7582				                                    & $-11.6$						                            &
$ 42.3$						                                & $ 7.74$                                                   &
$-2.19$				        							    & (2)                                                       \\
NGC 7674				                                    & $-12.1$                                                   &
$ 42.4$						                                & $ 8.50$                                                   &
$-2.87$                                                     & (5)
\enddata
\tablecomments{Column (1): galaxy name. Column (2): logarithmic observed flux in the 2--10 keV (\textit{Suzaku}/FIXIS). Column (3): logarithmic intrinsic luminosity in the 2--10 keV. Column (4): logarithmic black hole mass. Column (5): logarithmic Eddington ratio ($\lambda_{\mathrm{Edd}} = L_{\mathrm{bol}}/L_{\mathrm{Edd}}$). Here we obtained the bolometric luminosity as $L_{\mathrm{bol}} = 20 L_{2-10}$ and defined the Eddington luminosity as $L_{\mathrm{Edd}} = 1.25 \times 10^{38} M_{\mathrm{BH}}/M_{\sun}$. Column (6) reference of the black hole mass.}
\tablerefs{(1) \cite{Bosch16}. (2) \cite{Koss17}. (3) \cite{Panessa15}. (4) \cite{Izumi16}. (5) \cite{Haan11}.}
\end{deluxetable*}

\section{Results}
Figure~1 shows the folded X-ray spectra and the best fitting models. Table 3 summarizes the best fitting parameters. Table 4 gives the observed fluxes, the intrinsic luminosities, and the Eddington ratios. Here we estimate the bolometric luminosity as $L_{\mathrm{Bol}} = 20 L_{2-10\mathrm{keV}}$ where $L_{2-10 \mathrm{keV}}$ is the intrinsic 2--10 keV luminosity, and define the Eddington luminosity as $L_{\mathrm{Edd}} = 1.25 \times 10^{38} M_{\mathrm{BH}}/M_{\sun}$ where $M_{\mathrm{BH}}$ is the black hole mass. Below, we compare our results with previous studies where different reflection models were adopted. To focus on differences in the spectral models, not in the data, here we only refer to previous works that utilized \textit{Suzaku} or \textit{NuSTAR} data.

\subsection{IC 5063}
The model with an apec component well reproduces the broadband (0.50--60.0 keV) X-ray spectrum ($\chi_\mathrm{red}^2 = 0.97$). We obtain $N_{\mathrm{H}}^{\mathrm{LOS}} = 0.26_{-0.14}^{+0.49} \times 10^{24}$ cm$^{-2}$ and $\Gamma = 1.89_{-0.09}^{+0.10}$. Our best fitting parameters are consistent with the \textit{Suzaku} results \citep{Tazaki11} and \textit{NuSTAR} results \citep{Balokovic18}. \cite{Tazaki11} estimated $N_{\mathrm{H}}^{\mathrm{LOS}} = 0.25_{-0.01}^{+0.10} \times 10^{24}$ cm$^{-2}$ and $\Gamma = 1.82_{-0.11}^{+0.08}$ with the Ikeda model. \cite{Balokovic18} obtained $N_{\mathrm{H}}^{\mathrm{LOS}} = 0.21 \times 10^{24}$ cm$^{-2}$ and $\Gamma = 1.75$ with the borus02 model.

\subsection{NGC 2110}
The model without an apec component is able to reproduce the broadband (0.50--60.0 keV) X-ray spectrum ($\chi_\mathrm{red}^2 = 1.03$). Our best fitting parameters are $N_{\mathrm{H}}^{\mathrm{LOS}} = 0.04_{-0.01}^{+0.01} \times 10^{24}$ cm$^{-2}$ and $\Gamma = 1.63_{-0.01}^{+0.01}$. Our results agree with the \textit{Suzaku} results \citep{Rivers14, Kawamuro16a} and \textit{NuSTAR} results \citep{Marinucci15, Balokovic18}. Utilizing the pexrav model \citep{Magdziarz95} for the reflection component, \cite{Rivers14} obtained $N_{\mathrm{H}}^{\mathrm{LOS}} = 0.05_{-0.01}^{+0.01} \times 10^{24}$ cm$^{-2}$ and $\Gamma = 1.66_{-0.01}^{+0.01}$, and \cite{Kawamuro16a} obtained $N_{\mathrm{H}}^{\mathrm{LOS}} = 0.02_{-0.01}^{+0.01} \times 10^{24}$ cm$^{-2}$ and $\Gamma = 1.65_{-0.01}^{+0.01}$. \cite{Marinucci15} obtained $N_{\mathrm{H}}^{\mathrm{LOS}} = 0.04_{-0.01}^{+0.01} \times 10^{24}$ cm$^{-2}$ and $\Gamma = 1.64_{-0.03}^{+0.03}$ with the MYTorus model, and \cite{Balokovic18} obtained $N_{\mathrm{H}}^{\mathrm{LOS}} = 0.04 \times 10^{24}$ cm$^{-2}$ and $\Gamma = 1.63$ with the borus02 model.

\subsection{NGC 3227}
The \textit{NuSTAR} data are reported for the first time. The model without an apec component provides an adequate fit ($\chi_\mathrm{red}^2 = 1.22$). We obtain $N_{\mathrm{H}}^{\mathrm{LOS}} = 0.07_{-0.02}^{+0.01} \times 10^{24}$ cm$^{-2}$ and $\Gamma = 1.58_{-0.02}^{+0.03}$. These are consistent with the \textit{Suzaku} results by \cite{Noda14} ($N_{\mathrm{H}}^{\mathrm{LOS}} = 0.10_{-0.01}^{+0.01} \times 10^{24}$ cm$^{-2}$ and $\Gamma = 1.67_{-0.06}^{+0.06}$) utilizing the pexrav model\footnote{Among the total six \textit{Suzaku} observations analyzed by \cite{Noda14}, we analyze the data of the fifth observation that has the longest exposure.}.

\subsection{NGC 3281}
The \textit{NuSTAR} data are reported for the first time. The model with an apec component provides an adequate fit ($\chi_\mathrm{red}^2 = 1.23$). Our best fitting parameters are $N_{\mathrm{H}}^{\mathrm{LOS}} = 0.66_{-0.17}^{+0.38}
\times 10^{24}$ cm$^{-2}$ and $\Gamma = 1.50^{+0.05}$. We note that $\alpha$ is pegged at the upper boundary (2.0).

\subsection{NGC 5506}
The model with an apec component well reproduces the broadband (0.60--60.0 keV) X-ray spectrum ($\chi_\mathrm{red}^2 = 1.08$). We obtain $N_{\mathrm{H}}^{\mathrm{LOS}} = 0.03_{-0.01}^{+0.01} \times 10^{24}$ cm$^{-2}$ and $\Gamma = 1.84_{-0.01}^{+0.02}$. Our photon index is slightly smaller from those of \textit{Suzaku} \citep{Kawamuro16a} and \textit{NuSTAR} \citep{Matt15}, while the column density is consistent with their results. \cite{Kawamuro16a} estimated $N_{\mathrm{H}}^{\mathrm{LOS}} = 0.03_{-0.01}^{+0.01} \times 10^{24}$ cm$^{-2}$ and $\Gamma = 1.95_{-0.01}^{+0.01}$ by applying the pexrav model for the reflection continuum. \cite{Matt15} obtained $N_{\mathrm{H}}^{\mathrm{LOS}} = 0.03_{-0.01}^{+0.01} \times 10^{24}$ cm$^{-2}$ and $\Gamma = 1.91_{-0.03}^{+0.03}$ with the xillver model \citep{Garcia13}, which represents a reflection component from an illuminated accretion disk. We interpret that this is because XCLUMPY contains more unabsorbed (hence softer) reflected continuum than the pexrav model (see \citealt[][Section 4.3]{Tanimoto19}), resulting in a harder intrinsic continuum.

\subsection{NGC 5643}
The model with an apec component well fit the broadband (0.70--55.0 keV) X-ray spectrum ($\chi_\mathrm{red}^2 = 0.99$). Our best fitting parameters are $N_{\mathrm{H}}^{\mathrm{LOS}} = 2.36_{-0.58}^{+2.24} \times 10^{24}$ cm$^{-2}$ and $\Gamma = 1.68_{-0.17}^{+0.18}$. Our results agree with the \textit{Suzaku} results \citep{Kawamuro16b} and \textit{NuSTAR} results \citep{Marchesi19a}. \cite{Kawamuro16b} estimated $N_{\mathrm{H}}^{\mathrm{LOS}} = 0.94_{-0.32}^{+0.61} \times 10^{24}$ cm$^{-2}$ and $\Gamma = 1.57_{-0.31}^{+0.37}$ by employing the pexrav model. \cite{Marchesi19a} obtained $N_{\mathrm{H}}^{\mathrm{LOS}} = 2.69_{-0.65}^{+1.88} \times 10^{24}$ cm$^{-2}$ and $\Gamma = 1.55_{-0.15}^{+0.13}$ with the borus02 model.

\subsection{NGC 5728}
The model with an apec component well reproduces the broadband (0.60--55.0 keV) X-ray spectrum ($\chi_\mathrm{red}^2 = 1.00$). We obtain $N_{\mathrm{H}}^{\mathrm{LOS}} = 0.95_{-0.19}^{+0.29} \times 10^{24}$ cm$^{-2}$ and $\Gamma = 1.50^{+0.05}$. Our photon index is slightly smaller than those of \textit{Suzaku} \citep{Tanimoto18} and \textit{NuSTAR} \citep{Marchesi19a}, whereas the hydrogen column density is consistent with their results. \cite{Tanimoto18} obtained $N_{\mathrm{H}}^{\mathrm{LOS}} = 1.69_{-0.53}^{+1.45} \times 10^{24}$ cm$^{-2}$ and $\Gamma = 1.69_{-0.14}^{+0.14}$ by applying the Ikeda model\footnote{\cite{Tanimoto18} considered two models, ``Ikeda1'' and ``Ikeda2'', where the line of sight absorption is linked and not-linked to the torus parameters, respectively. Here we refer to the results of ``Ikeda2''.} and \cite{Marchesi19a} obtained $N_{\mathrm{H}}^{\mathrm{LOS}} = 0.96_{-0.03}^{+0.05} \times 10^{24}$ cm$^{-2}$ and $\Gamma = 1.81_{-0.04}^{+0.07}$ with the borus02 model. This trend is the same as the case of NGC 5506. It can be explained by a large unabsorbed reflection-continuum flux in the XCLUMPY model.

\subsection{NGC 7172}
The \textit{NuSTAR} data are reported for the first time. The model with an apec component well replicate the broadband (0.50--60.0 keV) X-ray spectrum ($\chi_\mathrm{red}^2 = 1.02$). Our best fitting parameters are $N_{\mathrm{H}}^{\mathrm{LOS}} = 0.09_{-0.02}^{+0.18} \times 10^{24}$ cm$^{-2}$ and $\Gamma = 1.76_{-0.02}^{+0.02}$. Our results agree with the \textit{Suzaku} results by \cite{Kawamuro16a}, who obtained $N_{\mathrm{H}}^{\mathrm{LOS}} = 0.09_{-0.01}^{+0.01} \times 10^{24}$ cm$^{-2}$ and $\Gamma = 1.74_{-0.02}^{+0.01}$ by applying the pexrav model for the reflection continuum.

\subsection{NGC 7582}
The model with an apec component well fit the broadband (0.60--60.0 keV) X-ray spectrum ($\chi_\mathrm{red}^2 = 1.11$). We obtain $N_{\mathrm{H}}^{\mathrm{LOS}} = 0.31_{-0.13}^{+0.16} \times 10^{24}$ cm$^{-2}$ and $\Gamma = 1.77_{-0.07}^{+0.08}$. Analyzing the same \textit{Suzaku} data with the Ikeda model, \cite{Tanimoto18} obtained $N_{\mathrm{H}}^{\mathrm{LOS}} = 0.71_{-0.15}^{+0.67} \times 10^{24}$ cm$^{-2}$ and $\Gamma = 1.80_{-0.10}^{+0.09}$ (``Ikeda1'' model) or $N_{\mathrm{H}}^{\mathrm{LOS}} = 3.23_{-1.78}^{+1.33} \times 10^{24}$ cm$^{-2}$ and $\Gamma = 1.88_{-0.12}^{+0.11}$ (``Ikeda2'' model). Our XCLUMPY results prefer the former model for this object (i.e., a Compton-thin AGN). \cite{Balokovic18} derived $N_{\mathrm{H}}^{\mathrm{LOS}} = 0.44 \times 10^{24}$ cm$^{-2}$ and $\Gamma = 1.67$ with the NuSTAR data by applying the borus02 model, which are similar to our results. We note that the time variability constant ($C_{\mathrm{Time}}$) is pegged at 0.1, implying a large time variability in the transmitted component between the Suzaku and NuSTAR observations. It may be a result by a change in the line-of-sight absorption as reported by \cite{Bianchi09}.

\subsection{NGC 7674}
The model with an apec component gives an adequate fit ($\chi_\mathrm{red}^2 = 1.20$). The best fitting parameters are $N_{\mathrm{H}}^{\mathrm{LOS}} = 0.24_{-0.10}^{+0.22} \times 10^{24}$ cm$^{-2}$ and $\Gamma = 1.50^{+0.07}$. Our results are consistent with the \textit{NuSTAR} results by \cite{Gandhi17}, who obtained $N_{\mathrm{H}}^{\mathrm{LOS}} = 0.13_{-0.03}^{+0.03} \times 10^{24}$ cm$^{-2}$ and $\Gamma = 1.40^{+0.08}$ with the decoupled MYTorus model.

\begin{figure*}
\gridline{\fig{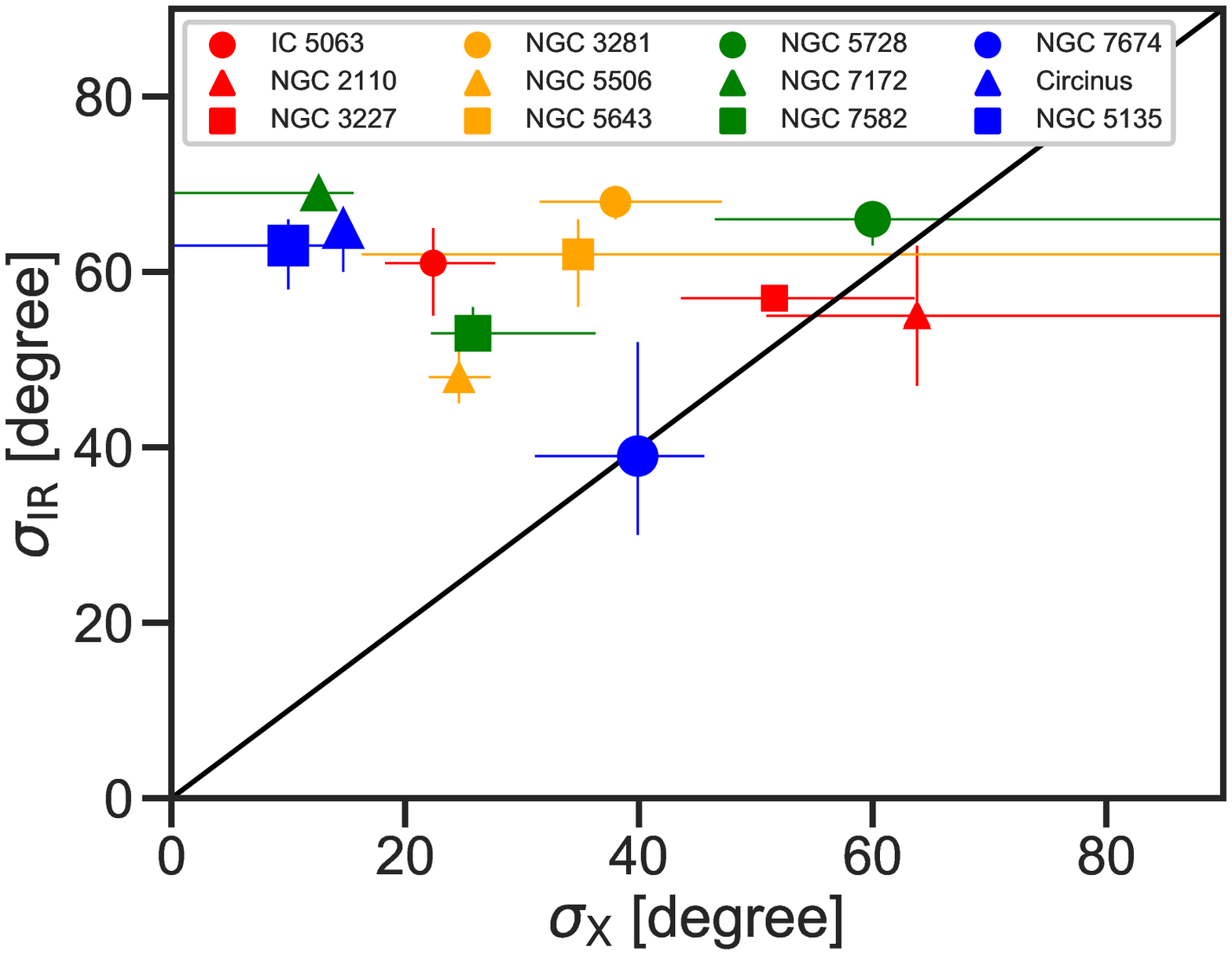}{0.5\textwidth}{(a)}\fig{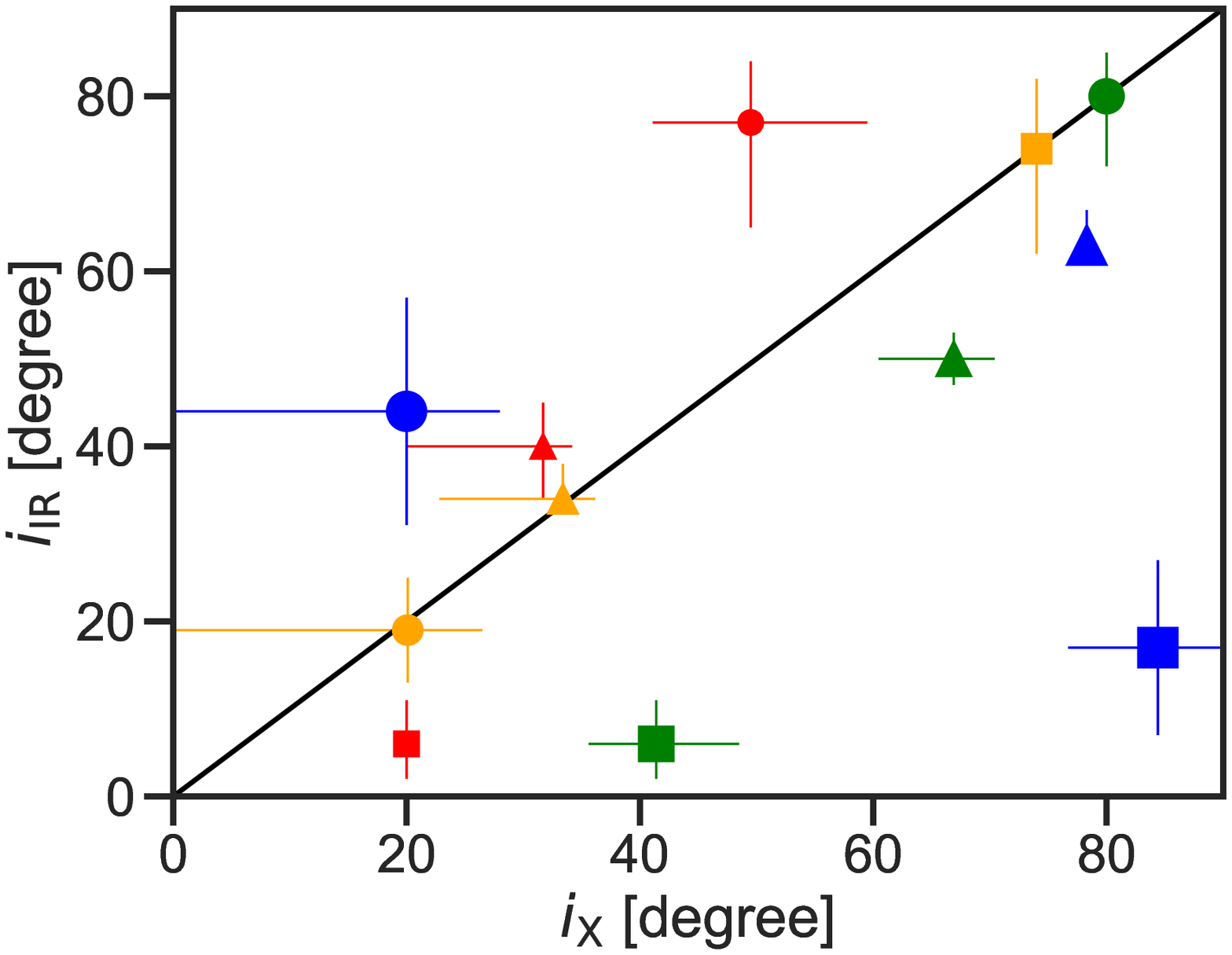}{0.5\textwidth}{(b)}}
\gridline{\fig{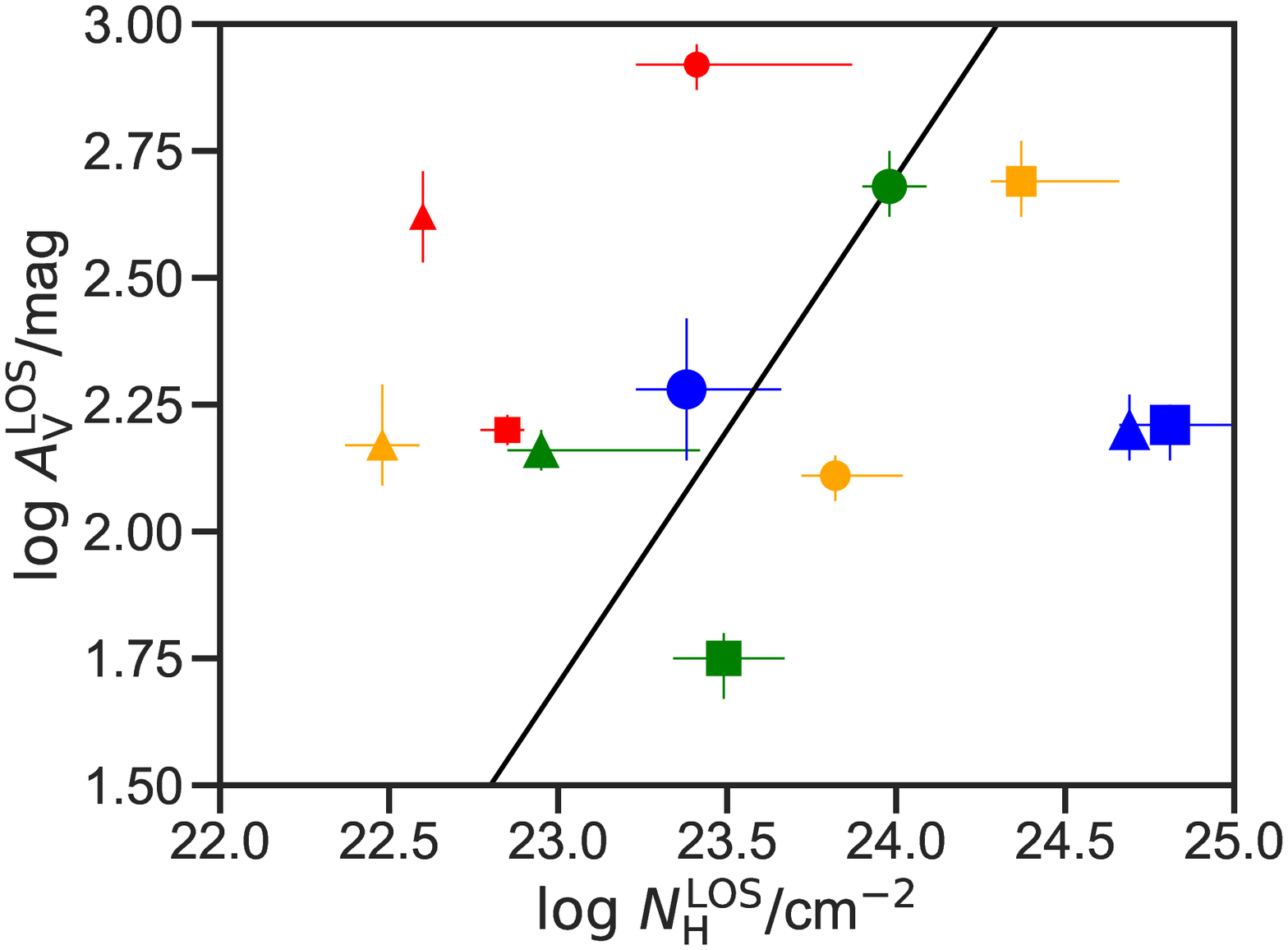}{0.5\textwidth}{(c)}\fig{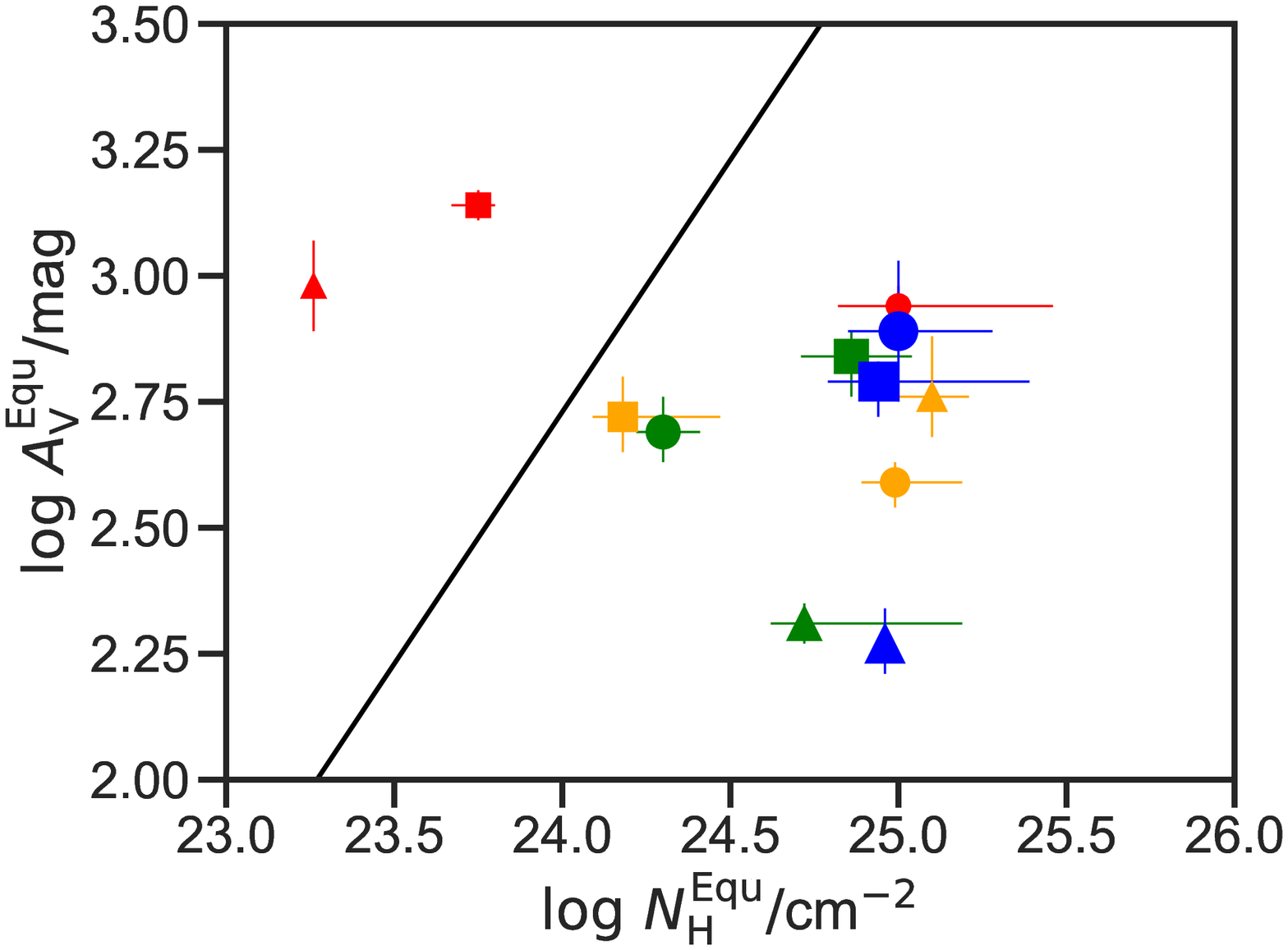}{0.5\textwidth}{(d)}}
\caption{(a) Correlation between the torus angular width obtained from the X-ray spectrum ($\sigma_{\mathrm{X}}$) and that from the infrared one ($\sigma_{\mathrm{IR}}$). The black line shows $\sigma_{\mathrm{X}} = \sigma_{\mathrm{IR}}$. (b) Correlation between the inclination angle obtained from the X-ray spectrum ($i_{\mathrm{X}}$) and that from the infrared one ($i_{\mathrm{IR}}$). The black line shows $i_{\mathrm{X}} = i_{\mathrm{IR}}$. (c) Correlation between the hydrogen column density along the line of sight obtained from the X-ray spectrum ($N_{\mathrm{H}}^{\mathrm{LOS}}$) and the V-band extinction along the line of sight from the infrared one ($A_{\mathrm{V}}^{\mathrm{LOS}}$). (d) Correlation between the hydrogen column density along the equatorial plane obtained from the X-ray spectrum ($N_{\mathrm{H}}^{\mathrm{Equ}}$) and the V-band extinction along the equatorial plane from the infrared one ($A_{\mathrm{V}}^{\mathrm{Equ}}$). The black line corresponds to the Galactic value: $N_{\mathrm{H}}/A_{\mathrm{V}} = 1.87 \times 10^{21}$ cm$^{-2}$ mag$^{-1}$ \citep{Draine03}. Note that we fix $i_{\mathrm{X}}$ to $i_{\mathrm{IR}}$ for NGC 3227, NGC 5643, and NGC 5728. For NGC 3227, we fix $i_{\mathrm{X}}$ at $20\degr$, the lower boundary value in the XCLUMPY model,  whereas $i_{\mathrm{IR}} = 6\degr$.}
\end{figure*}

\section{Discussion}
\subsection{Self-consistency of Our X-ray Spectral Model}
We have shown that the broadband X-ray spectra of the 10 nearby obscured AGNs can be well reproduced by employing the XCLUMPY model. XCLUMPY has only 3 free torus parameters: the hydrogen column density along the equatorial plane $N_{\mathrm{H}}^{\mathrm{Equ}}$, the torus angular width $\sigma_{\mathrm{X}}$, and the inclination $i_{\mathrm{X}}$. To make the spectral model self-consistent, the line of sight column density of the transmitted component has been linked to the torus parameters through equation (3).

Here we have introduced a correction factor $\alpha$, within a range of 0.5--2.0, to take into account a possible fluctuation in the line of sight absorption. We find that in 7 out of the 10 sources the 90\% confidence range of $\alpha$ contains unity (Table~3), meaning that the torus geometry assumed in XCLUMPY is consistent with the data. In the remaining 3 sources, NGC 2110, NGC 5643 and NGC 5728, $\alpha$ is pegged at either of the boundary values (0.5 or 2.0). The small sample size makes it difficult to judge if it is purely a result of statistical fluctuation in the clump number in the line of sight. Rather, it would be possible that the actual matter distribution is not that simple as in the XCLUMPY geometry. For instance, we would observe a large $\alpha$ value when a single optically-thick cloud is present only at the line of sight. A small $\alpha$ value would be expected if matter is more sharply concentrated in the equatorial plane than the Gaussian distribution. These limitations in using the XCLUMPY model must be always bared in mind when interpreting the results.

We have confirmed that our conclusions presented below do not change when we fix $\alpha = 1$ and are robust against parameter coupling among the torus parameters ($N_{\mathrm{H}}^{\mathrm{Equ}}$, $\sigma_{\mathrm{X}}$, and $i_{\mathrm{X}}$). Finally, we note that our spectral model ignores a possible time variability in the line-of-sight column density among different epochs, which may be the case for NGC 7582 (Section~4.9).

\subsection{Comparison of Torus Parameters Obtained from X-Ray and Infrared Spectra}
In this subsection, we compare the torus parameters obtained from the X-ray spectra and those from the infrared ones. Before that, we summarize main driving spectral features that constrain these parameters in the analysis of the infrared and X-ray data. In the CLUMPY model (infrared), the depth of silicate absorption at 9.7$\mu$m mainly determines the line-of-sight extinction ($A_{\mathrm{V}}$). \cite{Ichikawa15} considered the foreground extinction from the host galaxy for some objects \citep[][Table~1]{Ichikawa15}. The SED slope from near to mid infrared wavelengths (the ratio of the mid to near infrared fluxes) constrains $\sigma_{\mathrm{IR}}$; large $\sigma_{\mathrm{IR}}$ results in a steeper slope because emission from inner hot dust becomes more obscured \citep[][Figure~8]{Nenkova08b}. Similarly, a larger inclination also makes the slope steeper, although the dependence is weak at $i_{\mathrm{IR}} < 60$ degree. \cite{Ichikawa15} limited the range of $i_{\mathrm{IR}}$ when an independent constraint on the inclination is available \citep[][Table~1]{Ichikawa15}.

The high-quality broadband X-ray spectra enable us to separate the absorbed transmitted component and the reflection component from the torus. Unfortunately, the parameter dependencies of the reflection component in the XCLUMPY model are not simple \citep[][Figure~2]{Tanimoto19}. Nevertheless, we may roughly understand that (1) the flux ratio between the hard ($>10$ keV) and soft ($<10$ keV) bands mainly determines $N_{\mathrm{H}}^{\mathrm{Equ}}$ and (2) the spectral slope of the reflection component below 7.1 keV constrains $\sigma$. As described above, $N_{\mathrm{H}}^{\mathrm{LOS}}$ gives another constraint to the torus parameters through equation (3). Note that we did not consider the foreground absorption adopted by \cite{Ichikawa15} in the X-ray spectral analysis. Assuming the $N_{\mathrm{H}}/A_{\mathrm{V}}$ ratio of the Galactic ISM, however, its estimated contribution is found to be ignorable in $N_{\mathrm{H}}^{\mathrm{LOS}}$ or smaller than its uncertainty in all cases but NGC 5506.
 
To increase the sample, we include the Circinus galaxy \citep{Tanimoto19}\footnote{We have reanalyzed the same X-ray spectra of the Circinus galaxy as presented in \citet{Tanimoto19} by introducing the $\alpha$ parameter, which was not considered in the original analysis. We have confirmed $\alpha = 0.95_{-0.05}^{+0.13}$.} and NGC 5135 \citep{Yamada20}\footnote{\citet{Yamada20} assume $\alpha = 1$ because it cannot be well determined due to the limited photon statistics of the spectra.} fitted with the XCLUMPY model in the following discussions. Figure~2 plots the relations between (a) the torus angular width obtained from the X-ray spectrum ($\sigma_{\mathrm{X}}$) and that from the infrared one ($\sigma_{\mathrm{IR}}$), and (b) the inclination angle obtained from the X-ray spectrum ($i_{\mathrm{X}}$) and that from the infrared one ($i_{\mathrm{IR}}$), (c) the hydrogen column density along the line of sight obtained from the X-ray spectrum ($N_{\mathrm{H}}^{\mathrm{LOS}}$) and the V-band extinction along the line of sight from the infrared one ($A_{\mathrm{V}}^{\mathrm{LOS}}$), and (d) the hydrogen column density along the equatorial plane obtained from the X-ray spectrum ($N_{\mathrm{H}}^{\mathrm{Equ}}$) and the V-band extinction along the equatorial plane obtained from the infrared one ($A_{\mathrm{V}}^{\mathrm{Equ}}$). Note that we fix the inclination angles of three objects: NGC 3227, NGC 5643, and NGC 5728.

Figure~2(a) indicates that $\sigma_{\mathrm{IR}}$ is systematically larger than $\sigma_{\mathrm{X}}$. Here we recall that the X-ray spectra trace all material including gas and dust in a rather unbiased manner, while the infrared data trace only dust in a temperature dependent way. This means that the apparent dust distribution as seen in the infrared band is effectively more extended in the vertical direction to the equatorial plane than the gas distribution. We infer that this can be explained by contribution to the observed infrared flux from dusty polar outflows, which are commonly observed in nearby AGNs by infrared interferometric observations \citep[e.g.,][]{Tristram14, Lyu18}. Since the CLUMPY model does not include such a polar dust component, this may lead to overestimate the actual angular width of the torus. The mean temperature of polar dust is lower than that of hot dust in the innermost torus region, because of its larger distance from the SMBH. Hence, it works to make the mid- to near-infrared flux ratio larger, leading to a large $\sigma_{IR}$ value (see above). This effect becomes more significant when the infrared flux from the torus is reduced due to extinction by dust in outer cooler regions (such as circumnuclear disks) compared with the prediction by CLUMPY. Such flux reduction is predicted by radiative hydrodynamical simulations, which show a ``shadow'' region around the equatorial plane in the mid infrared image \citep{Wada16}. 

By contrast, the X-ray results are less affected by the polar outflows. This is because the mass carried by the polar outflows is much smaller than that contained in the torus itself \citep{Wada16} and because hard X-rays emitted at the central engine can penetrate through the torus. \cite{Liu19} examined X-ray signatures of the polar outflows with ray-tracing simulations and found that it only contributed to the X-ray spectrum below 2 keV such as Si K$\alpha$ emission lines, not to Fe K$\alpha$ emission lines and hard X-ray continuum above 10 keV. Hence we expect that X-ray results mainly trace the equatorial dusty torus distribution. We recall that a reflection component with a Fe K$\alpha$ line from a extended region of $\gg 1$ pc scale may be contained in our X-ray spectra, whose contribution differs object to object. For instance, a few \% and $\approx$30\% of the Fe K$\alpha$ line fluxes in Circinus \citep{Kawamuro19} and NGC 1068 \citep{Bauer15}, respectively. This would lead to overestimate $\sigma_{\mathrm{X}}$ if the column density of the extended Fe K$\alpha$ region is comparable to that in the torus, whereas the effect is less significant for $\sigma_{\mathrm{IR}}$ because the infrared data are better localized ($< 1''$) than the X-ray data ($>10''$). Hence, the differences between $\sigma_{\mathrm{X}}$ and $\sigma_{\mathrm{IR}}$ would be even enhanced after correcting them for this possible effect.

Figure~2(b) shows that the correlation between $i_{\mathrm{X}}$ and $i_{\mathrm{IR}}$ is not good after excluding the objects for which we have assumed $i_{\mathrm{X}} = i_{\mathrm{IR}}$ in the X-ray spectral analysis. We infer that this is because it is difficult to constrain $i_{\mathrm{IR}}$ from the infrared data because of its small dependence on the SED at low inclinations (see above). Although the X-ray results are inevitably subject to coupling with the $\alpha$ parameter (see equation (3)), we have taken it into account in estimating the uncertainties. In the following discussion, we refer to $i_{\mathrm{X}}$ as an estimator of the inclination.

Since the line of sight extinction is more directly determined from the X-ray and infrared data through photoelectric absorption and silicate absorption, respectively, we mainly focus on the correlation between $N_{\mathrm{H}}^{\mathrm{LOS}}$ and $A_{\mathrm{V}}^{\mathrm{LOS}}$ (Figure~2c). The mean value of $N_{\mathrm{H}}^{\mathrm{LOS}}/A_{\mathrm{V}}^{\mathrm{LOS}}$ is close to that of the Galactic ISM: $N_{\mathrm{H}}/A_{\mathrm{V}} = 1.87 \times 10^{21}$ cm$^{-2}$ mag$^{-1}$ \citep{Draine03}. This is consistent with the results of \cite{Burtscher16}, who investigated $N_{\mathrm{H}}^{\mathrm{LOS}}/A_{\mathrm{V}}^{\mathrm{LOS}}$ utilizing colors of dust. This result would be little affected even if we overestimated $N_{\mathrm{H}}^{\mathrm{LOS}}$ of NGC 5506 due to the possible foreground absorption (see above). We note that Figures 2(c) and 2(d) indicate scatter ($\simeq$1 dex) in the ratio between the hydrogen column density and the V-band extinction, both along the line of sight and along the equatorial plane.

Under the presence of dusty polar outflows, the obtained value of $A_{\mathrm{V}}$ should be a flux-weighted average from two different regions, the polar outflows and the torus. Since the polar outflows are located above the torus, the extinction for them is smaller than that for the torus itself. As discussed in Section~5.2, the relative contribution from dusty polar outflows to the total infrared flux increases with the inclination. Hence the $A_{\mathrm{V}}$ value may be largely underestimated than that toward the torus at high inclination systems, leading to large $N_{\mathrm{H}}^{\mathrm{LOS}}/A_{\mathrm{V}}^{\mathrm{LOS}}$ values, even without invoking a significant amount of dust-free gas inside the dust sublimation radius \cite[e.g.,][]{Davies15, Ichikawa19, Kawakatu19}.

If we exclude the two highest inclination ($i \geq 70^\circ$) objects (Circinus galaxy and NGC 5135) as exceptions, more than half of our sample seem to have ``dust-rich'' circumnuclear environment compared with the Galactic ISM; this conclusion is even strengthened if we correct for the contribution from the polar outflows. \cite{Ogawa19} found such dust-rich AGNs by applying the XCLUMPY to the broadband X-ray spectra of two Seyfert 1 galaxies. This trend is opposite to that reported for some AGNs that show cold absorption in X-rays and optical broad emission lines \citep[e.g.][]{Maiolino01b, Maiolino01a}.

\begin{figure*}
\gridline{\fig{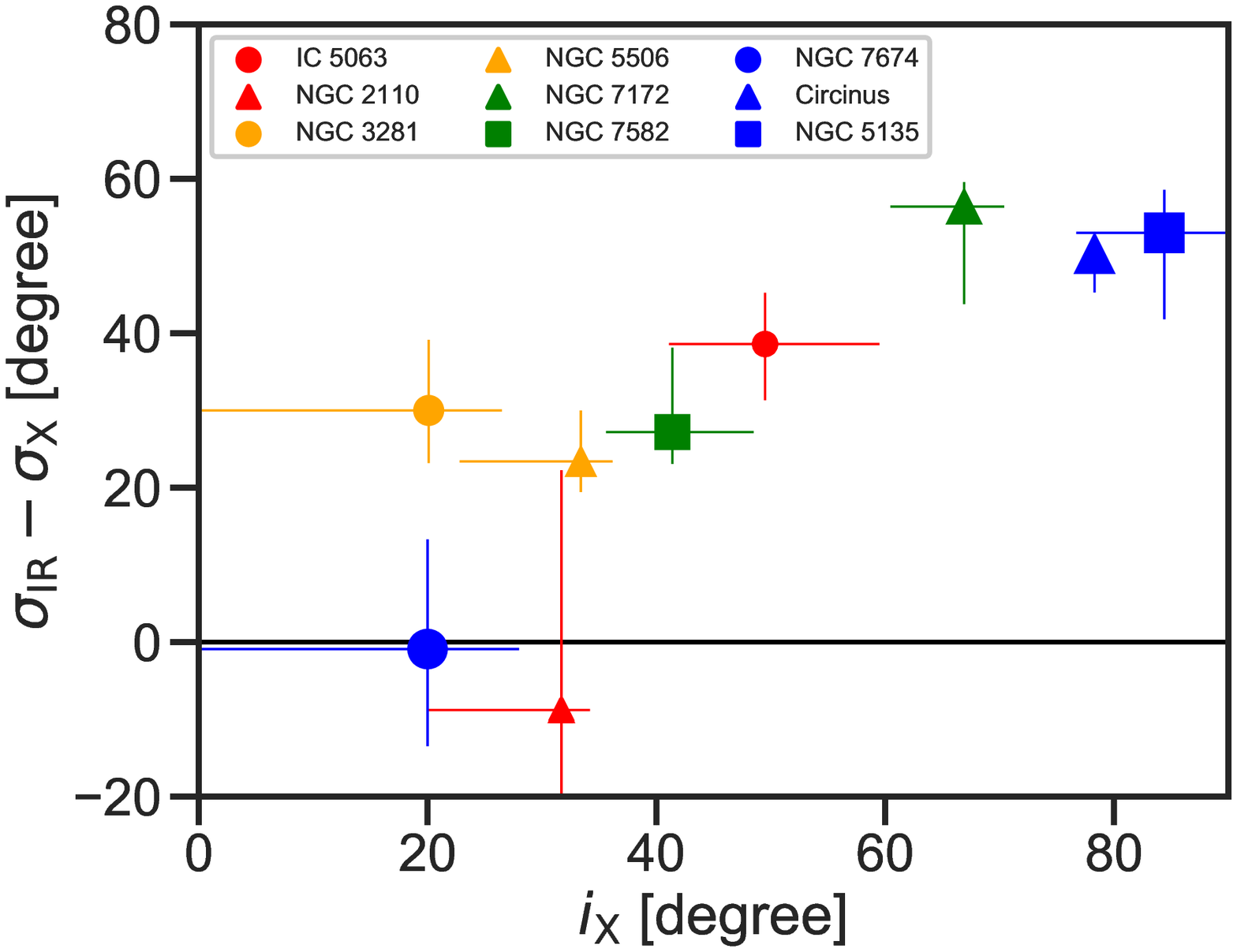}{0.5\textwidth}{(a)}\fig{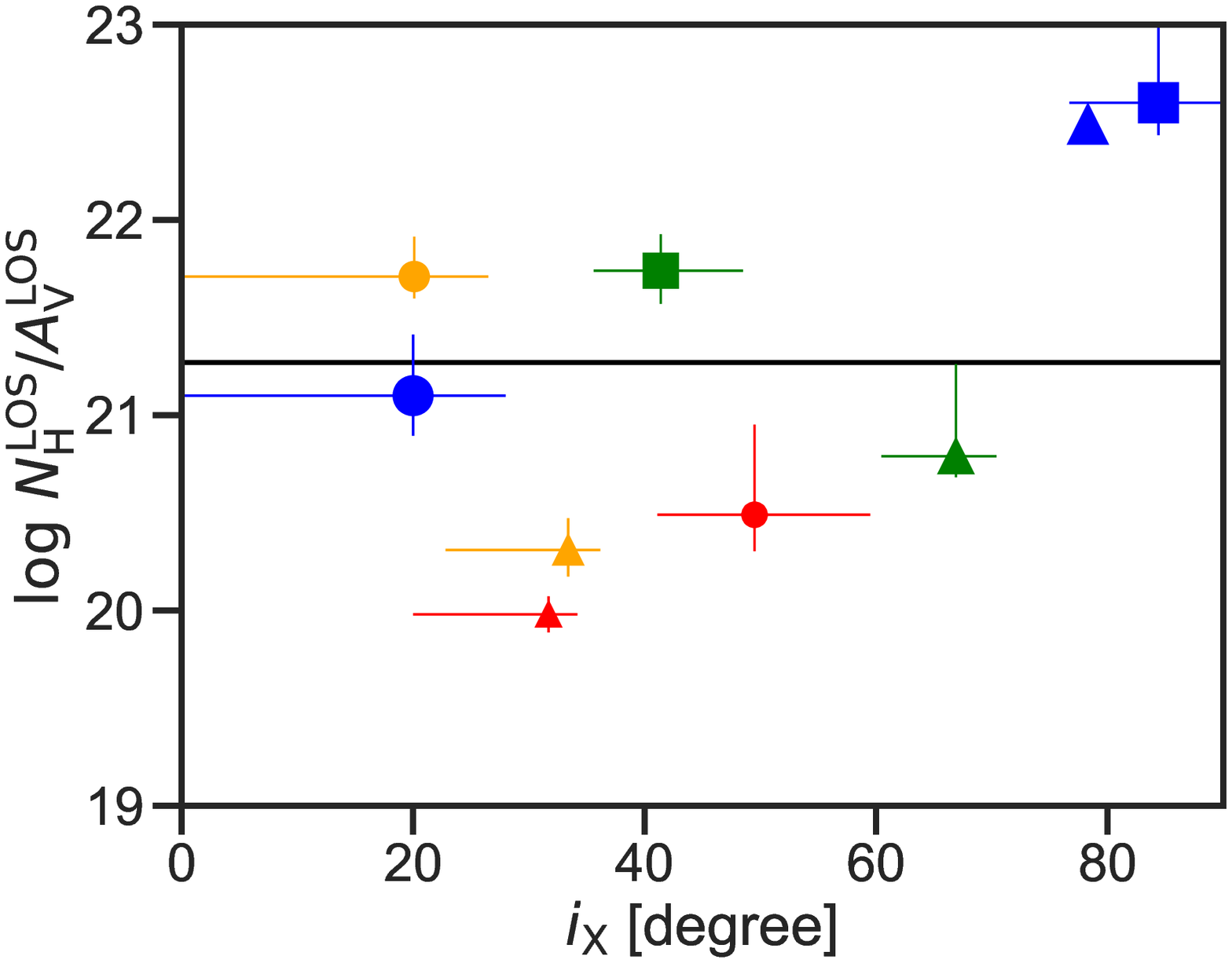}{0.5\textwidth}{(b)}}
\caption{(a) $\sigma_{\mathrm{IR}}-\sigma_{\mathrm{X}}$ against $i_{\mathrm{X}}$. The black line shows $\sigma_{\mathrm{X}} = \sigma_{\mathrm{IR}}$. (b) $\log N_{\mathrm{H}}^{\mathrm{LOS}}/A_{\mathrm{V}}^{\mathrm{LOS}}$ against the inclination angle obtained from the X-ray spectrum ($i_{\mathrm{X}}$). The black line corresponds to the Galactic value: $N_{\mathrm{H}}/A_{\mathrm{V}} = 1.87 \times 10^{21}$ cm$^{-2}$ mag$^{-1}$ \citep{Draine03}.}
\end{figure*}

\subsection{Origin of Torus Parameter Differences between X-ray and IR}
To reinforce our interpretation on the torus-parameter differences between X-ray and infrared, Figure~3 plots $\sigma_{\mathrm{IR}}-\sigma_{\mathrm{X}}$ and $N_{\mathrm{H}}^{\mathrm{LOS}}/A_{\mathrm{V}}^{\mathrm{LOS}}$ as a function of inclination ($i_{\mathrm{X}}$). Here we exclude the three objects whose $i_{\mathrm{X}}$ cannot be well determined from the X-ray spectra (NGC 3227, NGC 5643, and NGC 5728).

Figure~3(a) indicates that $\sigma_{\mathrm{IR}}-\sigma_{\mathrm{X}}$ correlates with $i_{\mathrm{X}}$. We find that the X-ray spectral fit shows a positive degeneracy between $\sigma_{\mathrm{X}}$ and $i_{\mathrm{X}}$, which even strengthens the presence of this correlation. The tendency still exists even if we exclude NGC 3281 and NGC 7674, which show $i_{\mathrm{X}} \sim 20$ degree with reduced $\chi^2$ of $\approx$1.2. The result is expected from radiative hydrodynamics simulations, which showed that relative contribution from the dusty polar outflow to the total observed mid-infrared flux increases with the inclination \citep[][Figure~5b]{Wada16}\footnote{Although the matter distribution assumed in the X/CLUMPY models is too simple compared with that in the \cite{Wada16} model, it would give a good first-order approximation to represent the broad column-density peak around the equatorial plane \citep[][Figure~4]{Wada16}}. This supports $\sigma_{\mathrm{IR}}$ is largely overestimated at high inclination systems.

Figure~3(b) shows that the two highest inclination objects (Circinus Galaxy and NGC 5135) have especially large $N_{\mathrm{H}}^{\mathrm{LOS}}/A_{\mathrm{V}}^{\mathrm{LOS}}$ values. As mentioned above, this is because $A_{\mathrm{V}}$ in the torus is likely underestimated by the contribution of the dusty polar outflows that are subject to smaller extinction. At lower inclinations, this effect becomes less important, and the observed ratio of $N_{\mathrm{H}}^{\mathrm{LOS}}/A_{\mathrm{V}}^{\mathrm{LOS}}$ would correctly represent the gas-to-dust ratio of matter in the line of sight if the travel paths of X-ray and infrared radiation are common. Our results imply a large object-to-object variation ($\sim$1 dex) in the gas-to-dust ratio of an AGN torus with a mean close to the Galactic value.

\newpage
\section{Conclusion}
\begin{enumerate}
\item We apply our X-ray spectral model from a clumpy torus (XCLUMPY: \citealt{Tanimoto19}) to the X-ray spectra of 10 obscured AGNs observed with both \textit{Suzaku} and \textit{NuSTAR}.
\item The torus angular widths obtained from the infrared spectra ($\sigma_{\mathrm{IR}}$) are systematically larger than those from the X-ray ones ($\sigma_{\mathrm{X}}$). Their difference is larger in higher inclination objects. These results can be explained by a significant contribution from dusty polar outflows to the infrared flux, as observed in infrared interferometric observations and predicted by theoretical simulations.
\item The ratios between the line of sight hydrogen column density and V-band extinction ($N_{\mathrm{H}}^{\mathrm{LOS}}/A_{\mathrm{V}}^{\mathrm{LOS}}$) show large scatter ($\simeq$1 dex) around the Galactic ISM value, suggesting that a large fraction of AGNs have dust-rich circumnuclear environments.
\end{enumerate}

\acknowledgements
This work was financially supported by the Grant-in-Aid for JSPS fellows
for young researchers 17J06407 (A.T.) and Scientific Research 17K05384
and 20H01946 (Y.U.) and 18H05861 (H.O.). This research has made use of data and/or software provided by the High Energy Astrophysics Science Archive Research Center (HEASARC), which is a service of the Astrophysics Science Division at NASA/GSFC and the High Energy Astrophysics Division of the Smithsonian Astrophysical Observatory. This research has also made use of the NASA/IPAC Extragalactic Database (NED), which is operated by the Jet Propulsion Laboratory, California Institute of Technology, under contract with the National Aeronautics and Space Administration.
\facilities{\textit{Suzaku}: 701030030, 701079010, 702010010, 702052040, 703022050, 703030010, 703033010, 704010010, 707034010, 708023010. \textit{NuSTAR}: 60001151002, 60061061002, 60061201002, 60061256002, 60061302002, 60061308002, 60061323002, 60061362002, 60201003002, 60202002002.}
\software{HEAsoft 6.26 (HEASARC 2018), XSPEC \citep{Arnaud96}}

\bibliographystyle{aasjournal}
\bibliography{tanimoto}

\begin{thebibliography}{}
\expandafter\ifx\csname natexlab\endcsname\relax\def\natexlab#1{#1}\fi

\bibitem[{{Alonso-Herrero} {et~al.}(2012{\natexlab{a}}){Alonso-Herrero},
  {Pereira-Santaella}, {Rieke}, \& {Rigopoulou}}]{Alonso12a}
{Alonso-Herrero}, A., {Pereira-Santaella}, M., {Rieke}, G.~H., \& {Rigopoulou},
  D. 2012{\natexlab{a}}, \apj, 744, 2

\bibitem[{{Alonso-Herrero} {et~al.}(2011){Alonso-Herrero}, {Ramos Almeida},
  {Mason}, {Asensio Ramos}, {Roche}, {Levenson}, {Elitzur}, {Packham},
  {Rodr{\'\i}guez Espinosa}, {Young}, {D{\'\i}az-Santos}, \&
  {P{\'e}rez-Garc{\'\i}a}}]{Alonso11}
{Alonso-Herrero}, A., {Ramos Almeida}, C., {Mason}, R., {et~al.} 2011, \apj,
  736, 82

\bibitem[{{Alonso-Herrero} {et~al.}(2012{\natexlab{b}}){Alonso-Herrero},
  {S{\'a}nchez-Portal}, {Ramos Almeida}, {Pereira-Santaella}, {Esquej},
  {Garc{\'\i}a-Burillo}, {Castillo}, {Gonz{\'a}lez-Mart{\'\i}n}, {Levenson},
  {Hatziminaoglou}, {Acosta-Pulido}, {Gonz{\'a}lez-Serrano}, {Povi{\'c}},
  {Packham}, \& {P{\'e}rez-Garc{\'\i}a}}]{Alonso12b}
{Alonso-Herrero}, A., {S{\'a}nchez-Portal}, M., {Ramos Almeida}, C., {et~al.}
  2012{\natexlab{b}}, \mnras, 425, 311

\bibitem[{{Alonso-Herrero} {et~al.}(2013){Alonso-Herrero}, {Roche}, {Esquej},
  {Gonz{\'a}lez-Mart{\'\i}n}, {Pereira-Santaella}, {Ramos Almeida}, {Levenson},
  {Packham}, {Asensio Ramos}, {Mason}, {Rodr{\'\i}guez Espinosa}, {Alvarez},
  {Colina}, {Aretxaga}, {D{\'\i}az-Santos}, {Perlman}, \& {Telesco}}]{Alonso13}
{Alonso-Herrero}, A., {Roche}, P.~F., {Esquej}, P., {et~al.} 2013, \apjl, 779,
  L14

\bibitem[{{Anders} \& {Grevesse}(1989)}]{Anders89}
{Anders}, E., \& {Grevesse}, N. 1989, \gca, 53, 197

\bibitem[{{Antonucci}(1993)}]{Antonucci93}
{Antonucci}, R. 1993, Annual Review of Astronomy and Astrophysics, 31, 473

\bibitem[{{Ar{\'e}valo} {et~al.}(2014){Ar{\'e}valo}, {Bauer}, {Puccetti},
  {Walton}, {Koss}, {Boggs}, {Brandt}, {Brightman}, {Christensen}, {Comastri},
  {Craig}, {Fuerst}, {Gandhi}, {Grefenstette}, {Hailey}, {Harrison}, {Luo},
  {Madejski}, {Madsen}, {Marinucci}, {Matt}, {Saez}, {Stern}, {Stuhlinger},
  {Treister}, {Urry}, \& {Zhang}}]{Arevalo14}
{Ar{\'e}valo}, P., {Bauer}, F.~E., {Puccetti}, S., {et~al.} 2014, \apj, 791, 81

\bibitem[{{Arnaud}(1996)}]{Arnaud96}
{Arnaud}, K.~A. 1996, in Astronomical Data Analysis Software and Systems V,
  Vol. 101, 17

\bibitem[{{Audibert} {et~al.}(2017){Audibert}, {Riffel}, {Sales}, {Pastoriza},
  \& {Ruschel-Dutra}}]{Audibert17}
{Audibert}, A., {Riffel}, R., {Sales}, D.~A., {Pastoriza}, M.~G., \&
  {Ruschel-Dutra}, D. 2017, \mnras, 464, 2139

\bibitem[{{Balokovi{\'c}} {et~al.}(2018){Balokovi{\'c}}, {Brightman},
  {Harrison}, {Comastri}, {Ricci}, {Buchner}, {Gandhi}, {Farrah}, \&
  {Stern}}]{Balokovic18}
{Balokovi{\'c}}, M., {Brightman}, M., {Harrison}, F.~A., {et~al.} 2018, \apj,
  854, 42

\bibitem[{{Bauer} {et~al.}(2015){Bauer}, {Ar{\'e}valo}, {Walton}, {Koss},
  {Puccetti}, {Gandhi}, {Stern}, {Alexander}, {Balokovi{\'c}}, {Boggs},
  {Brandt}, {Brightman}, {Christensen}, {Comastri}, {Craig}, {Del Moro},
  {Hailey}, {Harrison}, {Hickox}, {Luo}, {Markwardt}, {Marinucci}, {Matt},
  {Rigby}, {Rivers}, {Saez}, {Treister}, {Urry}, \& {Zhang}}]{Bauer15}
{Bauer}, F.~E., {Ar{\'e}valo}, P., {Walton}, D.~J., {et~al.} 2015, \apj, 812,
  116

\bibitem[{{Bianchi} {et~al.}(2009){Bianchi}, {Piconcelli}, {Chiaberge},
  {Bail{\'o}n}, {Matt}, \& {Fiore}}]{Bianchi09}
{Bianchi}, S., {Piconcelli}, E., {Chiaberge}, M., {et~al.} 2009, \apj, 695, 781

\bibitem[{{Buchner} {et~al.}(2019){Buchner}, {Brightman}, {Nandra}, {Nikutta},
  \& {Bauer}}]{Buchner19}
{Buchner}, J., {Brightman}, M., {Nandra}, K., {Nikutta}, R., \& {Bauer}, F.~E.
  2019, \aap, 629, A16

\bibitem[{{Burtscher} {et~al.}(2016){Burtscher}, {Davies}, {Graci{\'a}-Carpio},
  {Koss}, {Lin}, {Lutz}, {Nandra}, {Netzer}, {Orban de Xivry}, {Ricci},
  {Rosario}, {Veilleux}, {Contursi}, {Genzel}, {Schnorr-M{\"u}ller},
  {Sternberg}, {Sturm}, \& {Tacconi}}]{Burtscher16}
{Burtscher}, L., {Davies}, R.~I., {Graci{\'a}-Carpio}, J., {et~al.} 2016, \aap,
  586, A28

\bibitem[{{Davies} {et~al.}(2015){Davies}, {Burtscher}, {Rosario},
  {Storchi-Bergmann}, {Contursi}, {Genzel}, {Graci{\'a}-Carpio}, {Hicks},
  {Janssen}, {Koss}, {Lin}, {Lutz}, {Maciejewski}, {M{\"u}ller-S{\'a}nchez},
  {Orban de Xivry}, {Ricci}, {Riffel}, {Riffel}, {Schartmann},
  {Schnorr-M{\"u}ller}, {Sternberg}, {Sturm}, {Tacconi}, \&
  {Veilleux}}]{Davies15}
{Davies}, R.~I., {Burtscher}, L., {Rosario}, D., {et~al.} 2015, \apj, 806, 127

\bibitem[{{Draine}(2003)}]{Draine03}
{Draine}, B.~T. 2003, Annual Review of Astronomy and Astrophysics, 41, 241

\bibitem[{{Fabbiano} {et~al.}(2017){Fabbiano}, {Elvis}, {Paggi}, {Karovska},
  {Maksym}, {Raymond}, {Risaliti}, \& {Wang}}]{Fabbiano17}
{Fabbiano}, G., {Elvis}, M., {Paggi}, A., {et~al.} 2017, \apjl, 842, L4

\bibitem[{{Fukazawa} {et~al.}(2009){Fukazawa}, {Mizuno}, {Watanabe}, {Kokubun},
  {Takahashi}, {Kawano}, {Nishino}, {Sasada}, {Shirai}, {Takahashi}, {Umeki},
  {Yamasaki}, {Yasuda}, {Bamba}, {Ohno}, {Takahashi}, {Ushio}, {Enoto},
  {Kitaguchi}, {Makishima}, {Nakazawa}, {Uehara}, {Yamada}, {Yuasa}, {Isobe},
  {Kawaharada}, {Tanaka}, {Tashiro}, {Terada}, \& {Yamaoka}}]{Fukazawa09}
{Fukazawa}, Y., {Mizuno}, T., {Watanabe}, S., {et~al.} 2009, Publications of
  the Astronomical Society of Japan, 61, S17

\bibitem[{{Fukazawa} {et~al.}(2011){Fukazawa}, {Hiragi}, {Yamazaki}, {Mizuno},
  {Hayashi}, {Hayashi}, {Nishino}, {Takahashi}, \& {Ohno}}]{Fukazawa11}
{Fukazawa}, Y., {Hiragi}, K., {Yamazaki}, S., {et~al.} 2011, \apj, 743, 124

\bibitem[{{Fuller} {et~al.}(2016){Fuller}, {Lopez-Rodriguez}, {Packham},
  {Ramos-Almeida}, {Alonso-Herrero}, {Levenson}, {Radomski}, {Ichikawa},
  {Garc{\'\i}a-Bernete}, {Gonz{\'a}lez-Mart{\'\i}n}, {D{\'\i}az-Santos}, \&
  {Mart{\'\i}nez-Paredes}}]{Fuller16}
{Fuller}, L., {Lopez-Rodriguez}, E., {Packham}, C., {et~al.} 2016, \mnras, 462,
  2618

\bibitem[{{Fuller} {et~al.}(2019){Fuller}, {Lopez-Rodriguez}, {Packham},
  {Ichikawa}, {Togi}, {Alonso-Herrero}, {Ramos-Almeida}, {Diaz-Santos},
  {Levenson}, \& {Radomski}}]{Fuller19}
---. 2019, \mnras, 483, 3404

\bibitem[{{Furui} {et~al.}(2016){Furui}, {Fukazawa}, {Odaka}, {Kawaguchi},
  {Ohno}, \& {Hayashi}}]{Furui16}
{Furui}, S., {Fukazawa}, Y., {Odaka}, H., {et~al.} 2016, \apj, 818, 164

\bibitem[{{Gandhi} {et~al.}(2017){Gandhi}, {Annuar}, {Lansbury}, {Stern},
  {Alexander}, {Bauer}, {Bianchi}, {Boggs}, {Boorman}, {Brandt}, {Brightman},
  {Christensen}, {Comastri}, {Craig}, {Del Moro}, {Elvis}, {Guainazzi},
  {Hailey}, {Harrison}, {Koss}, {Lamperti}, {Malaguti}, {Masini}, {Matt},
  {Puccetti}, {Ricci}, {Rivers}, {Walton}, \& {Zhang}}]{Gandhi17}
{Gandhi}, P., {Annuar}, A., {Lansbury}, G.~B., {et~al.} 2017, \mnras, 467, 4606

\bibitem[{{Garc{\'\i}a} {et~al.}(2013){Garc{\'\i}a}, {Dauser}, {Reynolds},
  {Kallman}, {McClintock}, {Wilms}, \& {Eikmann}}]{Garcia13}
{Garc{\'\i}a}, J., {Dauser}, T., {Reynolds}, C.~S., {et~al.} 2013, \apj, 768,
  146

\bibitem[{{Garc{\'\i}a-Bernete} {et~al.}(2015){Garc{\'\i}a-Bernete}, {Ramos
  Almeida}, {Acosta-Pulido}, {Alonso-Herrero}, {S{\'a}nchez-Portal},
  {Castillo}, {Pereira-Santaella}, {Esquej}, {Gonz{\'a}lez-Mart{\'\i}n},
  {D{\'\i}az-Santos}, {Roche}, {Fisher}, {Povi{\'c}}, {P{\'e}rez Garc{\'\i}a},
  {Valtchanov}, {Packham}, \& {Levenson}}]{Garcia15}
{Garc{\'\i}a-Bernete}, I., {Ramos Almeida}, C., {Acosta-Pulido}, J.~A.,
  {et~al.} 2015, \mnras, 449, 1309

\bibitem[{{Garc{\'\i}a-Bernete} {et~al.}(2019){Garc{\'\i}a-Bernete}, {Ramos
  Almeida}, {Alonso-Herrero}, {Ward}, {Acosta-Pulido}, {Pereira-Santaella},
  {Hern{\'a}n-Caballero}, {Asensio Ramos}, {Gonz{\'a}lez-Mart{\'\i}n},
  {Levenson}, {Mateos}, {Carrera}, {Ricci}, {Roche}, {Marquez}, {Packham},
  {Masegosa}, \& {Fuller}}]{Garcia19}
{Garc{\'\i}a-Bernete}, I., {Ramos Almeida}, C., {Alonso-Herrero}, A., {et~al.}
  2019, \mnras, 486, 4917

\bibitem[{{Gonz{\'a}lez-Mart{\'\i}n} {et~al.}(2013){Gonz{\'a}lez-Mart{\'\i}n},
  {Rodr{\'\i}guez-Espinosa}, {D{\'\i}az-Santos}, {Packham}, {Alonso-Herrero},
  {Esquej}, {Ramos Almeida}, {Mason}, \& {Telesco}}]{Gonzalez13}
{Gonz{\'a}lez-Mart{\'\i}n}, O., {Rodr{\'\i}guez-Espinosa}, J.~M.,
  {D{\'\i}az-Santos}, T., {et~al.} 2013, \aap, 553, A35

\bibitem[{{Haan} {et~al.}(2011){Haan}, {Surace}, {Armus}, {Evans}, {Howell},
  {Mazzarella}, {Kim}, {Vavilkin}, {Inami}, {Sanders}, {Petric}, {Bridge},
  {Melbourne}, {Charmandaris}, {Diaz-Santos}, {Murphy}, {U}, {Stierwalt}, \&
  {Marshall}}]{Haan11}
{Haan}, S., {Surace}, J.~A., {Armus}, L., {et~al.} 2011, \aj, 141, 100

\bibitem[{{Harrison} {et~al.}(2013){Harrison}, {Craig}, {Christensen},
  {Hailey}, {Zhang}, {Boggs}, {Stern}, {Cook}, {Forster}, {Giommi},
  {Grefenstette}, {Kim}, {Kitaguchi}, {Koglin}, {Madsen}, {Mao}, {Miyasaka},
  {Mori}, {Perri}, {Pivovaroff}, {Puccetti}, {Rana}, {Westergaard}, {Willis},
  {Zoglauer}, {An}, {Bachetti}, {Barri{\`e}re}, {Bellm}, {Bhalerao},
  {Brejnholt}, {Fuerst}, {Liebe}, {Markwardt}, {Nynka}, {Vogel}, {Walton},
  {Wik}, {Alexander}, {Cominsky}, {Hornschemeier}, {Hornstrup}, {Kaspi},
  {Madejski}, {Matt}, {Molendi}, {Smith}, {Tomsick}, {Ajello}, {Ballantyne},
  {Balokovi{\'c}}, {Barret}, {Bauer}, {Blandford}, {Brandt}, {Brenneman},
  {Chiang}, {Chakrabarty}, {Chenevez}, {Comastri}, {Dufour}, {Elvis}, {Fabian},
  {Farrah}, {Fryer}, {Gotthelf}, {Grindlay}, {Helfand}, {Krivonos}, {Meier},
  {Miller}, {Natalucci}, {Ogle}, {Ofek}, {Ptak}, {Reynolds}, {Rigby},
  {Tagliaferri}, {Thorsett}, {Treister}, \& {Urry}}]{Harrison13}
{Harrison}, F.~A., {Craig}, W.~W., {Christensen}, F.~E., {et~al.} 2013, \apj,
  770, 103

\bibitem[{{Heckman} \& {Best}(2014)}]{Heckman14}
{Heckman}, T.~M., \& {Best}, P.~N. 2014, \araa, 52, 589

\bibitem[{{H{\"o}nig} \& {Beckert}(2007)}]{Honig07}
{H{\"o}nig}, S.~F., \& {Beckert}, T. 2007, \mnras, 380, 1172

\bibitem[{{H{\"o}nig} {et~al.}(2006){H{\"o}nig}, {Beckert}, {Ohnaka}, \&
  {Weigelt}}]{Honig06}
{H{\"o}nig}, S.~F., {Beckert}, T., {Ohnaka}, K., \& {Weigelt}, G. 2006, \aap,
  452, 459

\bibitem[{{H{\"o}nig} {et~al.}(2012){H{\"o}nig}, {Kishimoto}, {Antonucci},
  {Marconi}, {Prieto}, {Tristram}, \& {Weigelt}}]{Honig12}
{H{\"o}nig}, S.~F., {Kishimoto}, M., {Antonucci}, R., {et~al.} 2012, \apj, 755,
  149

\bibitem[{{Ichikawa} {et~al.}(2015){Ichikawa}, {Packham}, {Ramos Almeida},
  {Asensio Ramos}, {Alonso-Herrero}, {Gonz{\'a}lez-Mart{\'\i}n}, {Lopez-
  Rodriguez}, {Ueda}, {D{\'\i}az- Santos}, {Elitzur}, {H{\"o}nig}, {Imanishi},
  {Levenson}, {Mason}, {Perlman}, \& {Alsip}}]{Ichikawa15}
{Ichikawa}, K., {Packham}, C., {Ramos Almeida}, C., {et~al.} 2015, \apj, 803,
  57

\bibitem[{{Ichikawa} {et~al.}(2019){Ichikawa}, {Ricci}, {Ueda}, {Bauer},
  {Kawamuro}, {Koss}, {Oh}, {Rosario}, {Shimizu}, {Stalevski}, {Fuller},
  {Packham}, \& {Trakhtenbrot}}]{Ichikawa19}
{Ichikawa}, K., {Ricci}, C., {Ueda}, Y., {et~al.} 2019, \apj, 870, 31

\bibitem[{{Ikeda} {et~al.}(2009){Ikeda}, {Awaki}, \& {Terashima}}]{Ikeda09}
{Ikeda}, S., {Awaki}, H., \& {Terashima}, Y. 2009, \apj, 692, 608

\bibitem[{{Ishisaki} {et~al.}(2007){Ishisaki}, {Maeda}, {Fujimoto}, {Ozaki},
  {Ebisawa}, {Takahashi}, {Ueda}, {Ogasaka}, {Ptak}, {Mukai}, {Hamaguchi},
  {Hirayama}, {Kotani}, {Kubo}, {Shibata}, {Ebara}, {Furuzawa}, {Iizuka},
  {Inoue}, {Mori}, {Okada}, {Yokoyama}, {Matsumoto}, {Nakajima}, {Yamaguchi},
  {Anabuki}, {Tawa}, {Nagai}, {Katsuda}, {Hayashida}, {Bamba}, {Miller},
  {Sato}, \& {Yamasaki}}]{Ishisaki07}
{Ishisaki}, Y., {Maeda}, Y., {Fujimoto}, R., {et~al.} 2007, Publications of the
  Astronomical Society of Japan, 59, 113

\bibitem[{{Izumi} {et~al.}(2016){Izumi}, {Kawakatu}, \& {Kohno}}]{Izumi16}
{Izumi}, T., {Kawakatu}, N., \& {Kohno}, K. 2016, \apj, 827, 81

\bibitem[{{Kallman} {et~al.}(2014){Kallman}, {Evans}, {Marshall}, {Canizares},
  {Longinotti}, {Nowak}, \& {Schulz}}]{Kallman14}
{Kallman}, T., {Evans}, D.~A., {Marshall}, H., {et~al.} 2014, \apj, 780, 121

\bibitem[{{Kawakatu} {et~al.}(2019){Kawakatu}, {Wada}, \&
  {Ichikawa}}]{Kawakatu19}
{Kawakatu}, N., {Wada}, K., \& {Ichikawa}, K. 2019, arXiv e-prints,
  arXiv:1912.02408

\bibitem[{{Kawamuro} {et~al.}(2019){Kawamuro}, {Izumi}, \&
  {Imanishi}}]{Kawamuro19}
{Kawamuro}, T., {Izumi}, T., \& {Imanishi}, M. 2019, \pasj, 71, 68

\bibitem[{{Kawamuro} {et~al.}(2016{\natexlab{a}}){Kawamuro}, {Ueda}, {Tazaki},
  {Ricci}, \& {Terashima}}]{Kawamuro16a}
{Kawamuro}, T., {Ueda}, Y., {Tazaki}, F., {Ricci}, C., \& {Terashima}, Y.
  2016{\natexlab{a}}, The Astrophysical Journal Supplement Series, 225, 14

\bibitem[{{Kawamuro} {et~al.}(2016{\natexlab{b}}){Kawamuro}, {Ueda}, {Tazaki},
  {Terashima}, \& {Mushotzky}}]{Kawamuro16b}
{Kawamuro}, T., {Ueda}, Y., {Tazaki}, F., {Terashima}, Y., \& {Mushotzky}, R.
  2016{\natexlab{b}}, \apj, 831, 37

\bibitem[{{Kokubun} {et~al.}(2007){Kokubun}, {Makishima}, {Takahashi},
  {Murakami}, {Tashiro}, {Fukazawa}, {Kamae}, {Madejski}, {Nakazawa},
  {Yamaoka}, {Terada}, {Yonetoku}, {Watanabe}, {Tamagawa}, {Mizuno}, {Kubota},
  {Isobe}, {Takahashi}, {Sato}, {Takahashi}, {Hong}, {Kawaharada}, {Kawano},
  {Mitani}, {Murashima}, {Suzuki}, {Abe}, {Miyawaki}, {Ohno}, {Tanaka},
  {Yanagida}, {Itoh}, {Ohnuki}, {Tamura}, {Endo}, {Hirakuri}, {Hiruta},
  {Kitaguchi}, {Kishishita}, {Sugita}, {Takahashi}, {Takeda}, {Enoto},
  {Hirasawa}, {Katsuta}, {Matsumura}, {Onda}, {Sato}, {Ushio}, {Ishikawa},
  {Murase}, {Odaka}, {Suzuki}, {Yaji}, {Yamada}, {Yamasaki}, \&
  {Yuasa}}]{Kokubun07}
{Kokubun}, M., {Makishima}, K., {Takahashi}, T., {et~al.} 2007, Publications of
  the Astronomical Society of Japan, 59, 53

\bibitem[{{Kormendy} \& {Ho}(2013)}]{Kormendy13}
{Kormendy}, J., \& {Ho}, L.~C. 2013, Annual Review of Astronomy and
  Astrophysics, 51, 511

\bibitem[{{Koss} {et~al.}(2017){Koss}, {Trakhtenbrot}, {Ricci}, {Lamperti},
  {Oh}, {Berney}, {Schawinski}, {Balokovi{\'c}}, {Baronchelli}, {Crenshaw},
  {Fischer}, {Gehrels}, {Harrison}, {Hashimoto}, {Hogg}, {Ichikawa}, {Masetti},
  {Mushotzky}, {Sartori}, {Stern}, {Treister}, {Ueda}, {Veilleux}, \&
  {Winter}}]{Koss17}
{Koss}, M., {Trakhtenbrot}, B., {Ricci}, C., {et~al.} 2017, \apj, 850, 74

\bibitem[{{Koyama} {et~al.}(2007){Koyama}, {Tsunemi}, {Dotani}, {Bautz},
  {Hayashida}, {Tsuru}, {Matsumoto}, {Ogawara}, {Ricker}, {Doty}, {Kissel},
  {Foster}, {Nakajima}, {Yamaguchi}, {Mori}, {Sakano}, {Hamaguchi},
  {Nishiuchi}, {Miyata}, {Torii}, {Namiki}, {Katsuda}, {Matsuura}, {Miyauchi},
  {Anabuki}, {Tawa}, {Ozaki}, {Murakami}, {Maeda}, {Ichikawa}, {Prigozhin},
  {Boughan}, {Lamarr}, {Miller}, {Burke}, {Gregory}, {Pillsbury}, {Bamba},
  {Hiraga}, {Senda}, {Katayama}, {Kitamoto}, {Tsujimoto}, {Kohmura}, {Tsuboi},
  \& {Awaki}}]{Koyama07}
{Koyama}, K., {Tsunemi}, H., {Dotani}, T., {et~al.} 2007, Publications of the
  Astronomical Society of Japan, 59, 23

\bibitem[{{Krolik} \& {Begelman}(1988)}]{Krolik88}
{Krolik}, J.~H., \& {Begelman}, M.~C. 1988, \apj, 329, 702

\bibitem[{{Laor} \& {Draine}(1993)}]{Laor93}
{Laor}, A., \& {Draine}, B.~T. 1993, \apj, 402, 441

\bibitem[{{Liu} {et~al.}(2019){Liu}, {H{\"o}nig}, {Ricci}, \&
  {Paltani}}]{Liu19}
{Liu}, J., {H{\"o}nig}, S.~F., {Ricci}, C., \& {Paltani}, S. 2019, \mnras, 490,
  4344

\bibitem[{{Liu} \& {Li}(2014)}]{Liu14}
{Liu}, Y., \& {Li}, X. 2014, \apj, 787, 52

\bibitem[{{Lopez-Rodriguez} {et~al.}(2018){Lopez-Rodriguez}, {Fuller},
  {Alonso-Herrero}, {Efstathiou}, {Ichikawa}, {Levenson}, {Packham},
  {Radomski}, {Ramos Almeida}, {Benford}, {Berthoud}, {Hamilton}, {Harper},
  {Kov{\'a}vcs}, {Santos}, {Staguhn}, \& {Herter}}]{Lopez-Rodriguez18}
{Lopez-Rodriguez}, E., {Fuller}, L., {Alonso-Herrero}, A., {et~al.} 2018, \apj,
  859, 99

\bibitem[{{Lyu} \& {Rieke}(2018)}]{Lyu18}
{Lyu}, J., \& {Rieke}, G.~H. 2018, \apj, 866, 92

\bibitem[{{Magdziarz} \& {Zdziarski}(1995)}]{Magdziarz95}
{Magdziarz}, P., \& {Zdziarski}, A.~A. 1995, \mnras, 273, 837

\bibitem[{{Maiolino} {et~al.}(2001{\natexlab{a}}){Maiolino}, {Marconi}, \&
  {Oliva}}]{Maiolino01b}
{Maiolino}, R., {Marconi}, A., \& {Oliva}, E. 2001{\natexlab{a}}, \aap, 365, 37

\bibitem[{{Maiolino} {et~al.}(2001{\natexlab{b}}){Maiolino}, {Marconi},
  {Salvati}, {Risaliti}, {Severgnini}, {Oliva}, {La Franca}, \&
  {Vanzi}}]{Maiolino01a}
{Maiolino}, R., {Marconi}, A., {Salvati}, M., {et~al.} 2001{\natexlab{b}},
  \aap, 365, 28

\bibitem[{{Marchesi} {et~al.}(2019){Marchesi}, {Ajello}, {Zhao}, {Marcotulli},
  {Balokovi{\'c}}, {Brightman}, {Comastri}, {Cusumano}, {Lanzuisi}, {La
  Parola}, {Segreto}, \& {Vignali}}]{Marchesi19a}
{Marchesi}, S., {Ajello}, M., {Zhao}, X., {et~al.} 2019, \apj, 872, 8

\bibitem[{{Marinucci} {et~al.}(2015){Marinucci}, {Matt}, {Bianchi}, {Lu},
  {Arevalo}, {Balokovi{\'c}}, {Ballantyne}, {Bauer}, {Boggs}, {Christensen},
  {Craig}, {Gandhi}, {Hailey}, {Harrison}, {Puccetti}, {Rivers}, {Walton},
  {Stern}, \& {Zhang}}]{Marinucci15}
{Marinucci}, A., {Matt}, G., {Bianchi}, S., {et~al.} 2015, \mnras, 447, 160

\bibitem[{{Mateos} {et~al.}(2017){Mateos}, {Carrera}, {Barcons},
  {Alonso-Herrero}, {Hern{\'a}n-Caballero}, {Page}, {Ramos Almeida},
  {Caccianiga}, {Miyaji}, \& {Blain}}]{Mateos17}
{Mateos}, S., {Carrera}, F.~J., {Barcons}, X., {et~al.} 2017, \apjl, 841, L18

\bibitem[{{Matt} {et~al.}(2015){Matt}, {Balokovi{\'c}}, {Marinucci},
  {Ballantyne}, {Boggs}, {Christensen}, {Comastri}, {Craig}, {Gandhi},
  {Hailey}, {Harrison}, {Madejski}, {Madsen}, {Stern}, \& {Zhang}}]{Matt15}
{Matt}, G., {Balokovi{\'c}}, M., {Marinucci}, A., {et~al.} 2015, \mnras, 447,
  3029

\bibitem[{{Mitsuda} {et~al.}(2007){Mitsuda}, {Bautz}, {Inoue}, {Kelley},
  {Koyama}, {Kunieda}, {Makishima}, {Ogawara}, {Petre}, {Takahashi}, {Tsunemi},
  {White}, {Anabuki}, {Angelini}, {Arnaud}, {Awaki}, {Bamba}, {Boyce}, {Brown},
  {Chan}, {Cottam}, {Dotani}, {Doty}, {Ebisawa}, {Ezoe}, {Fabian}, {Figueroa},
  {Fujimoto}, {Fukazawa}, {Furusho}, {Furuzawa}, {Gendreau}, {Griffiths},
  {Haba}, {Hamaguchi}, {Harrus}, {Hasinger}, {Hatsukade}, {Hayashida}, {Henry},
  {Hiraga}, {Holt}, {Hornschemeier}, {Hughes}, {Hwang}, {Ishida}, {Ishisaki},
  {Isobe}, {Itoh}, {Iyomoto}, {Kahn}, {Kamae}, {Katagiri}, {Kataoka},
  {Katayama}, {Kawai}, {Kilbourne}, {Kinugasa}, {Kissel}, {Kitamoto}, {Kohama},
  {Kohmura}, {Kokubun}, {Kotani}, {Kotoku}, {Kubota}, {Madejski}, {Maeda},
  {Makino}, {Markowitz}, {Matsumoto}, {Matsumoto}, {Matsuoka}, {Matsushita},
  {McCammon}, {Mihara}, {Misaki}, {Miyata}, {Mizuno}, {Mori}, {Mori}, {Morii},
  {Moseley}, {Mukai}, {Murakami}, {Murakami}, {Mushotzky}, {Nagase}, {Namiki},
  {Negoro}, {Nakazawa}, {Nousek}, {Okajima}, {Ogasaka}, {Ohashi}, {Oshima},
  {Ota}, {Ozaki}, {Ozawa}, {Parmar}, {Pence}, {Porter}, {Reeves}, {Ricker},
  {Sakurai}, {Sanders}, {Senda}, {Serlemitsos}, {Shibata}, {Soong}, {Smith},
  {Suzuki}, {Szymkowiak}, {Takahashi}, {Tamagawa}, {Tamura}, {Tamura},
  {Tanaka}, {Tashiro}, {Tawara}, {Terada}, {Terashima}, {Tomida}, {Torii},
  {Tsuboi}, {Tsujimoto}, {Tsuru}, {Turner}, {Ueda}, {Ueno}, {Ueno}, {Uno},
  {Urata}, {Watanabe}, {Yamamoto}, {Yamaoka}, {Yamasaki}, {Yamashita},
  {Yamauchi}, {Yamauchi}, {Yaqoob}, {Yonetoku}, \& {Yoshida}}]{Mitsuda07}
{Mitsuda}, K., {Bautz}, M., {Inoue}, H., {et~al.} 2007, Publications of the
  Astronomical Society of Japan, 59, S1

\bibitem[{{Murphy} \& {Yaqoob}(2009)}]{Murphy09}
{Murphy}, K.~D., \& {Yaqoob}, T. 2009, \mnras, 397, 1549

\bibitem[{{Nenkova} {et~al.}(2008{\natexlab{a}}){Nenkova}, {Sirocky},
  {Ivezi{\'c}}, \& {Elitzur}}]{Nenkova08a}
{Nenkova}, M., {Sirocky}, M.~M., {Ivezi{\'c}}, {\v{Z}}., \& {Elitzur}, M.
  2008{\natexlab{a}}, \apj, 685, 147

\bibitem[{{Nenkova} {et~al.}(2008{\natexlab{b}}){Nenkova}, {Sirocky},
  {Nikutta}, {Ivezi{\'c}}, \& {Elitzur}}]{Nenkova08b}
{Nenkova}, M., {Sirocky}, M.~M., {Nikutta}, R., {Ivezi{\'c}}, {\v{Z}}., \&
  {Elitzur}, M. 2008{\natexlab{b}}, \apj, 685, 160

\bibitem[{{Netzer}(2015)}]{Netzer15}
{Netzer}, H. 2015, \araa, 53, 365

\bibitem[{{Noda} {et~al.}(2014){Noda}, {Makishima}, {Yamada}, {Nakazawa},
  {Sakurai}, \& {Miyake}}]{Noda14}
{Noda}, H., {Makishima}, K., {Yamada}, S., {et~al.} 2014, \apj, 794, 2

\bibitem[{{Odaka} {et~al.}(2011){Odaka}, {Aharonian}, {Watanabe}, {Tanaka},
  {Khangulyan}, \& {Takahashi}}]{Odaka11}
{Odaka}, H., {Aharonian}, F., {Watanabe}, S., {et~al.} 2011, \apj, 740, 103

\bibitem[{{Odaka} {et~al.}(2016){Odaka}, {Yoneda}, {Takahashi}, \&
  {Fabian}}]{Odaka16}
{Odaka}, H., {Yoneda}, H., {Takahashi}, T., \& {Fabian}, A. 2016, \mnras, 462,
  2366

\bibitem[{{Ogawa} {et~al.}(2020){Ogawa}, {Ueda}, \& {Tanimoto}}]{Ogawa20}
{Ogawa}, S., {Ueda}, Y., \& {Tanimoto}, A. 2020, in prep.

\bibitem[{{Ogawa} {et~al.}(2019){Ogawa}, {Ueda}, {Yamada}, {Tanimoto}, \&
  {Kawaguchi}}]{Ogawa19}
{Ogawa}, S., {Ueda}, Y., {Yamada}, S., {Tanimoto}, A., \& {Kawaguchi}, T. 2019,
  \apj, 875, 115

\bibitem[{{Panessa} {et~al.}(2015){Panessa}, {Tarchi}, {Castangia}, {Maiorano},
  {Bassani}, {Bicknell}, {Bazzano}, {Bird}, {Malizia}, \&
  {Ubertini}}]{Panessa15}
{Panessa}, F., {Tarchi}, A., {Castangia}, P., {et~al.} 2015, \mnras, 447, 1289

\bibitem[{{Protassov} {et~al.}(2002){Protassov}, {van Dyk}, {Connors},
  {Kashyap}, \& {Siemiginowska}}]{Protassov02}
{Protassov}, R., {van Dyk}, D.~A., {Connors}, A., {Kashyap}, V.~L., \&
  {Siemiginowska}, A. 2002, \apj, 571, 545

\bibitem[{{Ramos Almeida} {et~al.}(2014{\natexlab{a}}){Ramos Almeida},
  {Alonso-Herrero}, {Levenson}, {Asensio Ramos}, {Rodr{\'\i}guez Espinosa},
  {Gonz{\'a}lez-Mart{\'\i}n}, {Packham}, \& {Mart{\'\i}nez}}]{Ramos14a}
{Ramos Almeida}, C., {Alonso-Herrero}, A., {Levenson}, N.~A., {et~al.}
  2014{\natexlab{a}}, \mnras, 439, 3847

\bibitem[{{Ramos Almeida} {et~al.}(2011{\natexlab{a}}){Ramos Almeida},
  {Dicken}, {Tadhunter}, {Asensio Ramos}, {Inskip}, {Hardcastle}, \&
  {Mingo}}]{Ramos11b}
{Ramos Almeida}, C., {Dicken}, D., {Tadhunter}, C., {et~al.}
  2011{\natexlab{a}}, \mnras, 413, 2358

\bibitem[{{Ramos Almeida} \& {Ricci}(2017)}]{Ramos17}
{Ramos Almeida}, C., \& {Ricci}, C. 2017, Nature Astronomy, 1, 679

\bibitem[{{Ramos Almeida} {et~al.}(2009){Ramos Almeida}, {Levenson},
  {Rodr{\'\i}guez Espinosa}, {Alonso-Herrero}, {Asensio Ramos}, {Radomski},
  {Packham}, {Fisher}, \& {Telesco}}]{Ramos09}
{Ramos Almeida}, C., {Levenson}, N.~A., {Rodr{\'\i}guez Espinosa}, J.~M.,
  {et~al.} 2009, \apj, 702, 1127

\bibitem[{{Ramos Almeida} {et~al.}(2011{\natexlab{b}}){Ramos Almeida},
  {S{\'a}nchez-Portal}, {P{\'e}rez Garc{\'\i}a}, {Acosta-Pulido}, {Castillo},
  {Asensio Ramos}, {Gonz{\'a}lez-Serrano}, {Alonso-Herrero}, {Rodr{\'\i}guez
  Espinosa}, {Hatziminaoglou}, {Coia}, {Valtchanov}, {Povi{\'c}}, {Esquej},
  {Packham}, \& {Altieri}}]{Ramos11c}
{Ramos Almeida}, C., {S{\'a}nchez-Portal}, M., {P{\'e}rez Garc{\'\i}a}, A.~M.,
  {et~al.} 2011{\natexlab{b}}, \mnras, 417, L46

\bibitem[{{Ramos Almeida} {et~al.}(2011{\natexlab{c}}){Ramos Almeida},
  {Levenson}, {Alonso-Herrero}, {Asensio Ramos}, {Rodr{\'\i}guez Espinosa},
  {P{\'e}rez Garc{\'\i}a}, {Packham}, {Mason}, {Radomski}, \&
  {D{\'\i}az-Santos}}]{Ramos11a}
{Ramos Almeida}, C., {Levenson}, N.~A., {Alonso-Herrero}, A., {et~al.}
  2011{\natexlab{c}}, \apj, 731, 92

\bibitem[{{Ramos Almeida} {et~al.}(2014{\natexlab{b}}){Ramos Almeida},
  {Alonso-Herrero}, {Esquej}, {Gonz{\'a}lez-Mart{\'\i}n}, {Riffel},
  {Garc{\'\i}a-Bernete}, {Rodr{\'\i}guez Espinosa}, {Packham}, {Levenson},
  {Roche}, {D{\'\i}az-Santos}, {Aretxaga}, \& {{\'A}lvarez}}]{Ramos14b}
{Ramos Almeida}, C., {Alonso-Herrero}, A., {Esquej}, P., {et~al.}
  2014{\natexlab{b}}, \mnras, 445, 1130

\bibitem[{{Ricci} {et~al.}(2018){Ricci}, {Ho}, {Fabian}, {Trakhtenbrot},
  {Koss}, {Ueda}, {Lohfink}, {Shimizu}, {Bauer}, {Mushotzky}, {Schawinski},
  {Paltani}, {Lamperti}, {Treister}, \& {Oh}}]{Ricci18}
{Ricci}, C., {Ho}, L.~C., {Fabian}, A.~C., {et~al.} 2018, \mnras, 480, 1819

\bibitem[{{Rivers} {et~al.}(2014){Rivers}, {Markowitz}, {Rothschild}, {Bamba},
  {Fukazawa}, {Okajima}, {Reeves}, {Terashima}, \& {Ueda}}]{Rivers14}
{Rivers}, E., {Markowitz}, A., {Rothschild}, R., {et~al.} 2014, \apj, 786, 126

\bibitem[{{Schoonjans} {et~al.}(2011){Schoonjans}, {Brunetti}, {Golosio},
  {Sanchez del Rio}, {Sol{\'e}}, {Ferrero}, \& {Vincze}}]{Schoonjans11}
{Schoonjans}, T., {Brunetti}, A., {Golosio}, B., {et~al.} 2011, Spectrochimica
  Acta, 66, 776

\bibitem[{{Takahashi} {et~al.}(2007){Takahashi}, {Abe}, {Endo}, {Endo}, {Ezoe},
  {Fukazawa}, {Hamaya}, {Hirakuri}, {Hong}, {Horii}, {Inoue}, {Isobe}, {Itoh},
  {Iyomoto}, {Kamae}, {Kasama}, {Kataoka}, {Kato}, {Kawaharada}, {Kawano},
  {Kawashima}, {Kawasoe}, {Kishishita}, {Kitaguchi}, {Kobayashi}, {Kokubun},
  {Kotoku}, {Kouda}, {Kubota}, {Kuroda}, {Madejski}, {Makishima}, {Masukawa},
  {Matsumoto}, {Mitani}, {Miyawaki}, {Mizuno}, {Mori}, {Mori}, {Murashima},
  {Murakami}, {Nakazawa}, {Niko}, {Nomachi}, {Okada}, {Ohno}, {Oonuki}, {Ota},
  {Ozawa}, {Sato}, {Shinoda}, {Sugiho}, {Suzuki}, {Taguchi}, {Takahashi},
  {Takahashi}, {Takeda}, {Tamura}, {Tamura}, {Tanaka}, {Tanihata}, {Tashiro},
  {Terada}, {Tominaga}, {Uchiyama}, {Watanabe}, {Yamaoka}, {Yanagida}, \&
  {Yonetoku}}]{Takahashi07}
{Takahashi}, T., {Abe}, K., {Endo}, M., {et~al.} 2007, Publications of the
  Astronomical Society of Japan, 59, 35

\bibitem[{{Tanimoto} {et~al.}(2018){Tanimoto}, {Ueda}, {Kawamuro}, {Ricci},
  {Awaki}, \& {Terashima}}]{Tanimoto18}
{Tanimoto}, A., {Ueda}, Y., {Kawamuro}, T., {et~al.} 2018, \apj, 853, 146

\bibitem[{{Tanimoto} {et~al.}(2019){Tanimoto}, {Ueda}, {Odaka}, {Kawaguchi},
  {Fukazawa}, \& {Kawamuro}}]{Tanimoto19}
{Tanimoto}, A., {Ueda}, Y., {Odaka}, H., {et~al.} 2019, \apj, 877, 95

\bibitem[{{Tazaki} {et~al.}(2011){Tazaki}, {Ueda}, {Terashima}, \&
  {Mushotzky}}]{Tazaki11}
{Tazaki}, F., {Ueda}, Y., {Terashima}, Y., \& {Mushotzky}, R.~F. 2011, \apj,
  738, 70

\bibitem[{{Tristram} {et~al.}(2014){Tristram}, {Burtscher}, {Jaffe},
  {Meisenheimer}, {H{\"o}nig}, {Kishimoto}, {Schartmann}, \&
  {Weigelt}}]{Tristram14}
{Tristram}, K. R.~W., {Burtscher}, L., {Jaffe}, W., {et~al.} 2014, \aap, 563,
  A82

\bibitem[{{Urry} \& {Padovani}(1995)}]{Urry95}
{Urry}, C.~M., \& {Padovani}, P. 1995, Publications of the Astronomical Society
  of the Pacific, 107, 803

\bibitem[{{van den Bosch}(2016)}]{Bosch16}
{van den Bosch}, R. C.~E. 2016, \apj, 831, 134

\bibitem[{{Wada} {et~al.}(2016){Wada}, {Schartmann}, \& {Meijerink}}]{Wada16}
{Wada}, K., {Schartmann}, M., \& {Meijerink}, R. 2016, \apj, 828, L19

\bibitem[{{Willingale} {et~al.}(2013){Willingale}, {Starling}, {Beardmore},
  {Tanvir}, \& {O'Brien}}]{Willingale13}
{Willingale}, R., {Starling}, R.~L.~C., {Beardmore}, A.~P., {Tanvir}, N.~R., \&
  {O'Brien}, P.~T. 2013, \mnras, 431, 394

\bibitem[{{Yamada} {et~al.}(2020){Yamada}, {Ueda}, {Oda}, {Tanimoto}, \&
  {Ricci}}]{Yamada20}
{Yamada}, S., {Ueda}, Y., {Oda}, S., {Tanimoto}, A., \& {Ricci}, C. 2020, in
  prep.

\end{thebibliography}
\end{document}